\documentclass[onecolumn, pra, letterpaper, nofootinbib]{revtex4}

\usepackage{epsfig}
\usepackage{graphicx}
\usepackage{subfigure}
\usepackage{amssymb}
\usepackage{amsmath}

\newcommand {\ket}[1] {|#1 \rangle}
\newcommand {\bra}[1] {\langle#1 |}

\newcommand {\rmi}[0] {{\rm i}}
\newcommand{\f}[2]{\hat{\psi}_{#2}(\mathbf{#1})}
\newcommand{\fd}[2]{\hat{\psi}_{#2}^{\dag}(\mathbf{#1})}

\begin{document}

\title{Quantum trajectories and open many-body quantum systems}


\author{Andrew J. Daley\footnote{Email: andrew.daley@strath.ac.uk}}
\affiliation{Department of Physics and SUPA, University of Strathclyde, Glasgow G4 0NG, Scotland, UK\\Department of Physics and Astronomy, University of Pittsburgh, Pittsburgh, Pennsylvania 15260, USA}
\date{May 26, 2014}


\begin{abstract}

The study of open quantum systems - microscopic systems exhibiting quantum coherence that are coupled to their environment - has become increasingly important in the past years, as the ability to control quantum coherence on a single particle level has been developed in a wide variety of physical systems. In quantum optics, the study of open systems goes well beyond understanding the breakdown of quantum coherence. There, the coupling to the environment is sufficiently well understood that it can be manipulated to drive the system into desired quantum states, or to project the system onto known states via feedback in quantum measurements. Many mathematical frameworks have been developed to describe such systems, which for atomic, molecular, and optical (AMO) systems generally provide a very accurate description of the open quantum system on a microscopic level. In recent years, AMO systems including cold atomic and molecular gases and trapped ions have been applied heavily to the study of many-body physics, and it has become important to extend previous understanding of open system dynamics in single- and few-body systems to this many-body context. A key formalism that has already proven very useful in this context is the quantum trajectories technique. This method was developed in quantum optics as a numerical tool for studying dynamics in open quantum systems, and falls within a broader framework of continuous measurement theory as a way to understand the dynamics of large classes of open quantum systems. In this article, we review the progress that has been made in studying open many-body systems in the AMO context, focussing on the application of ideas from quantum optics, and on the implementation and applications of quantum trajectories methods in these systems. Control over dissipative processes promises many further tools to prepare interesting and important states in strongly interacting systems, including the realisation of parameter regimes in quantum simulators that are inaccessible via current techniques. 
\bigskip


\end{abstract}
 
 \maketitle
\tableofcontents

\section{Introduction}
\subsection{Background of open quantum systems}

Coupling between microscopic quantum mechanical systems and their environment is important in essentially every system where quantum mechanical behaviour is observed. No quantum system that has been measured in the laboratory is an ideal, perfectly isolated (or \emph{closed}) system. Rather, coherent dynamics (as described by a Schr\"odinger equation) typically last only over short timescales, before the dynamics become dominated by coupling of the \emph{open} system \cite{Breuer2007,Gardiner2005,Carmichael1993,Carmichael1999,Weiss2012,Muller2012} to its environment, leading to decoherence, and the onset of more classical behaviour. Over the last few decades, quantum mechanical behaviour has been observed and controlled in a diverse range of systems, to the point where many systems can be controlled on the level of a single atom, ion, molecule, photon or electron. As a result, the need to better understand dynamics in open quantum systems has increased. While philosophically it would be possible to extend the boundary of the system and include the environment in a larger system, this is typically impractical mathematically due to the enormous numbers of degrees of freedom involved in describing the environment. Hence, we usually look to find an effective description for dynamics in the approximately isolated system.

\begin{figure}[tb]
\begin{center}
\includegraphics[width=12cm]{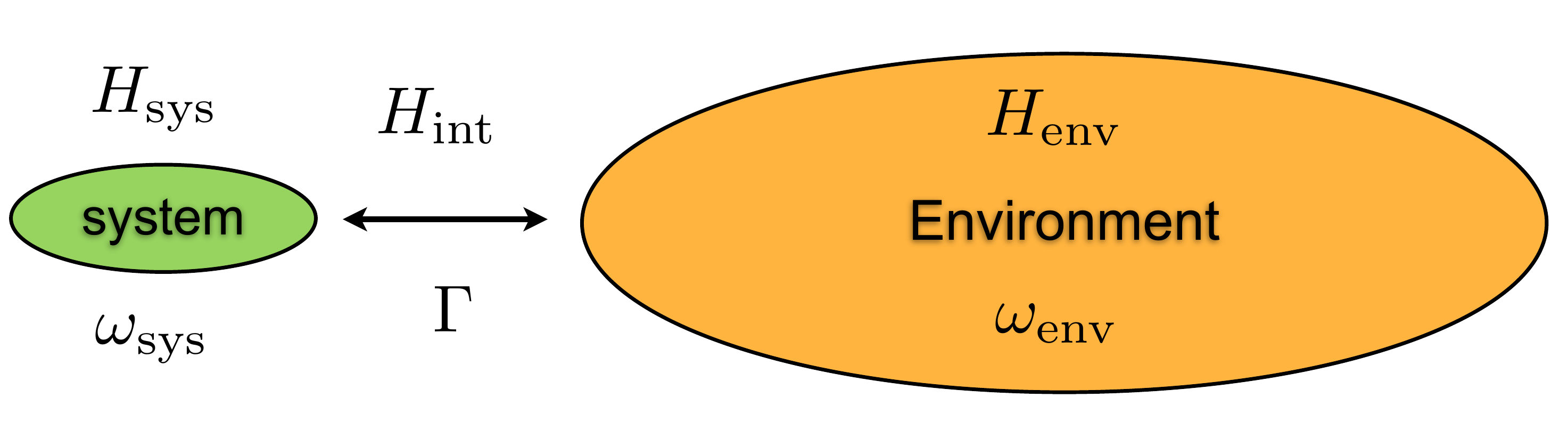}
\end{center} 
\caption{General framework for an open quantum system: A small quantum system interacts with its environment, leading to a combination of coherent and dissipative dynamics for the small system. In quantum optics, and atomic, molecular, and optical (AMO) systems more generally, the resulting dynamics can both be microscopically well understood, as well as controlled to engineer dynamics or specific quantum states. This involves a separation of energy and frequency scales, where the energy and frequency scales of the small system that are directly coupled to the environment ($\omega_{\rm sys}$) and the relaxation rates for relevant correlation functions in the environment ($\omega_{\rm env}$) are much larger than the frequency scales of dynamics induced by the coupling ($\Gamma$).}\label{fig:openquantumsystem}
\end{figure}

With the recent focus on controlling quantum coherence, especially to store and manipulate quantum information \cite{Nielsen2000,Monroe2013,Awschalom2013,Devoret2013}, investigation of open quantum systems has become prominent across various fields of physics. Quantum coherence has been observed, and decoherence studies performed in atomic systems including photon modes in cavities \cite{Kimble2008,Deleglise2008,Gleyzes2007,Guerlin2007,Nogues1999}; single trapped atoms \cite{Bloch2008}, ions \cite{Monroe2013,Haffner2008,Blatt2008,Monz2011,Leibfried2003,Myatt2000}, and molecules \cite{Carr2009,Yan2013}; and in solid state systems \cite{Awschalom2013,Hanson2008} including superconducting junctions \cite{Clarke2008}, and spin systems including quantum dots, colour centres, Cooper pair boxes \cite{Devoret2013,Reed2012}, also in conjunction with microwave stripline cavities. In condensed matter physics, there have been many investigations of dissipative phenomena, a whole class of which began with the Caldeira-Leggett treatment of a spin coupled to a bosonic bath \cite{Caldeira1981,Leggett1987}. The study of open quantum systems also has importance in cosmology \cite{Hu1995}, and quantum-optical approaches have recently been applied to treat pion decay in high-energy physics \cite{Lello2013,Lello2013a}.

While many studies seek to characterise and reduce the destruction of quantum coherence in an open system, a key guiding aim in quantum optics over the last thirty years has been the use of controlled coupling to the environment to control and manipulate quantum coherence with high precision. This includes manipulating the environment in such a way as to drive the system into desired quantum states, increasing quantum coherence in the sample. In the laboratory, this philosophy began with optical pumping in atomic physics \cite{Brossel1952,Hawkins1953}, whereby using laser driving and spontaneous emission processes, atoms can be driven into a single atomic state with extraordinarily high fidelities approaching 100\%. This has lead on to techniques for laser cooling of trapped ions \cite{Leibfried2003,Wineland1978,Neuhauser1978} and of atomic and molecular samples \cite{Metcalf1999}, which has allowed the production of atomic gases with temperatures of the order of $1~\mu$K. This, in turn, set the stage for realising Bose-Einstein condensation \cite{Anderson1995a,Davis1995b,Bradley1995a} and degenerate Fermi gases \cite{DeMarco1999a,Truscott2001a}, as well as providing a level of high-precision control necessary for quantum computing with trapped ions \cite{Haffner2008,Leibfried2003,Cirac1995}. Such driving processes, making use of the coupling of the system to its environment, have been extended to algorithmic cooling in Nuclear Magnetic Resonance (NMR) systems \cite{Boykin2002,Baugh2005}, and are recently being applied to cool the motional modes of macroscopic oscillators to near their quantum ground state \cite{Chan2011}. A detailed understanding of the back-action on the quantum state of the system from coupling to the environment has also been useful in the context of high-fidelity state detection (e.g., in electron shelving \cite{Dehmelt1982,Nagourney1986,Bergquist1986,Sauter1986,Hegerfeldt1992b,Hegerfeldt1995}), or in quantum feedback \cite{Wiseman2010, Handel2005, Wiseman1993, Foster2000, Bushev2006}, and have also been exploited for control over atoms and photons in Cavity QED \cite{Haroche2013,Raimond2001}.

This high level of control is possible because in quantum optics several key approximations can often be made regarding the coupling of a system to its environment, which in combination make the control of this coupling both tractable theoretically and feasible experimentally. This includes the fact that the dynamics are often Markovian (i.e., the environment relaxes rapidly to an equilibrium state and on relevant timescales has no memory of previous interactions with the system), and that the system-environment coupling is typically weak compared with relevant system and environment energy scales. The possibility of making such approximations has resulted in the development of several complementary formalisms, including quantum Langevin equations \cite{Gardiner1985,Yurke1984,Caves1982,Ritsch1988,Parkins1988}, quantum master equations \cite{Gardiner2005,Lax1963,Collett1984,Gardiner1985}, and continuous measurement theory \cite{Wiseman2010,Gardiner2005}. These techniques and related ideas have been applied extensively over the last thirty years to a variety of single-particle and few-particle quantum systems, especially in the context of atomic, molecular, and optical (AMO) experiments. Techniques such as quantum master equations in Lindblad form \cite{Gardiner2005} are sometimes applied in other contexts, especially in a range of solid-state systems. However, many solid-state systems are dominated by non-Markovian aspects of their dissipative dynamics \cite{Breuer2007}, which are consequently not captured by this form. This is in strong contrast to the AMO/quantum optics context, where the approximations involved in deriving these equations are typically good to many orders of magnitude.

Many numerical techniques were developed for solving the equations produced within these formalisms. One simple but highly effective technique in this context is the quantum trajectory technique (also known as the Monte Carlo wavefunction method) \cite{Plenio1998,Carmichael1993,Dalibard1992,Molmer1993,Dum1992}, which allows for the numerical solution of a master equation without propagating a density matrix directly. Instead, pure states are propagated in time, with the dissipative process being described by a modification to the Hamiltonian, combined with quantum jumps - sudden changes in the state that take place at particular times. By taking an appropriate stochastic average over the times and type of quantum jumps, expectation values for the system propagated under the master equation can be faithfully and efficiently reconstructed within well-controlled statistical errors. This technique is particularly appealing, because when combined with continuous measurement theory, it can also help to give a physical intuition into the workings of the dissipative process.

\subsection{Open many-body quantum systems in a AMO context}

Over the last decade and a half, AMO systems (cold atoms and molecules, as well as trapped ions and photons) have been increasingly used to study strongly-interacting many-body systems. This has reached a level where these systems can be used to engineer microscopic Hamiltonians for the purposes of quantum simulation \cite{Cirac2012,Bloch2012,Blatt2012,Aspuru-Guzik2012,Muller2012,Bloch2008b,Lewenstein2007,Lewenstein2012}. As a result, it has become important to explore the application of existing experimental techniques and theoretical formalisms from quantum optics in this new many-body context. Particularly strong motivation in this respect has come from the desire to engineer particularly interesting many-body states, which often require very low temperatures (or entropies) in the system. Understanding dissipative dynamics in the many-body system is then necessary both to control heating processes, and to provide new means to drive the systems to lower temperatures. 

Such studies are particularly facilitated by the fact that many of the same approximations that are made in describing open few-particle systems in quantum optics can also be made in these AMO many-body systems. This means both that the physics of open systems as they appear in quantum optics can be investigated in a wholly new context, and that mathematical techniques such as master equations and quantum Langevin equations, as well as numerical techniques such as quantum trajectory methods can be immediately applied to these systems in a way that faithfully represents the microscopic physics. 
  
This has already lead to many important developments. Some of these have direct practical importance in the experiments - e.g., in order to cool many-body systems to the temperatures required to realise important many-body physics \cite{McKay2011}, it has become important to understand decoherence in a many-body context in order to characterise the related heating processes \cite{Pichler2010,Poletti2012,Poletti2013,Pichler2012,Pichler2013,Schachenmayer2014}. Other studies have involved fundamentally different approaches to state preparation, including studies of how to use controlled dissipation to produce new cooling methods \cite{Griessner2006,Griessner2007,Chen2014,Griessner2005,Daley2004a}, or to drive the system into desired many-body states - including Bose-Einstein condensates (BECs) or metastable states such as $\eta$ pairs \cite{Diehl2008,Kraus2008}, paired states \cite{Diehl2010,Yi2012,Han2009,Sandner2011,Ates2012}, or even states with topological order \cite{Diehl2011,Bardyn2013}. Dissipation can also play a key role in suppressing two-body \cite{Syassen2008,Garcia-Ripoll2009,Zhu2014,Yan2013} and three-body  \cite{Daley2009,Mark2012,Schutzhold2010} loss processes via a continuous quantum Zeno effect, enabling the production of interesting many-body states requiring effective three-body interactions \cite{ChenYC2011,Privitera2011,Bonnes2011,Titvinidze2011,Diehl2010b,Diehl2010d,Lee2010,Diehl2010a,Roncaglia2010,Kantian2009,Capogrosso-Sansone2009,Daley2009,Ottenstein2008,Chen2008,Rapp2008,Rapp2007}, and similar effects are seen in locally induced single-particle losses in cold gases \cite{Barontini2013,Zezyulin2012,Barmettler2011}. Such effects can also be used to manipulate or protect states in quantum computations \cite{Verstraete2009,Kliesch2011} and quantum memories \cite{Pastawski2011}, as well as to prepare entangled states dissipatively \cite{Cho2011,McCutcheon2009,Vacanti2009}, or realise specific quantum gates for group-II atoms \cite{Daley2011,Daley2008b}. It is also possible to generate spin-squeezed states using collisional loss in fermions \cite{Foss-Feig2012} or collisions with background gases \cite{Caballar2014}, as well as to protect states during adiabatic state preparation \cite{Stannigel2013}. 

Studies of dissipation have more recently included investigating the competition between dissipation and coherent dynamics, including dissipative phase transitions \cite{Diehl2010c,Sieberer2013a,Honing2012}, excitation dynamics in clouds of Rydberg atoms \cite{Olmos2012,Lesanovsky2013,Lesanovsky2014,Petrosyan2013,Honing2013,Olmos2013,Ates2012}, and collective spin dynamics on the clock transition in group II atoms \cite{Martin2013}. Other work has begun to characterise dissipative driving processes by identifying universal classes of states and phase transitions that can be realised in this way \cite{Tauber2013,Sieberer2013}. The observation of many-body dissipation in AMO systems extends further to interacting ultracold atoms in optical cavities \cite{Ritsch2013}, beginning with the observation of the Dicke phase transition \cite{Baumann2010,Dimer2007} and continuing to more complex dynamics of many moving atoms interacting with the cavity \cite{Brooks2012,Habibian2013,Brahms2012,Torre2013,Buchhold2013}, as well as to systems of strongly interacting photons \cite{Peyronel2012,Reinhard2012,Petrosyan2011,Sevincli2011,Gorshkov2011}. In addition, dissipative quantum simulation and state preparation \cite{Muller2012} has been observed in trapped ions \cite{Barreiro2011,Schindler2013} and proposed in optomechanical arrays \cite{Tomadin2012,Lechner2013}.

\subsection{Purpose and outline of this review}

As is clear from the previous section, this area of research, combining ideas and techniques from quantum optics with strongly interacting many-body systems, has developed dramatically in the last few years. Control over dissipation promises many new tools to prepare interesting and important many-body states, and to investigate models in quantum simulators that are inaccessible to current methods. 

In this review, we set about facilitating discussion between the AMO and many-body communities by introducing open many-body systems as they are understood in the AMO context. We will focus on real-time dynamics, especially of open quantum system out of thermal equilibrium. We will also build our discussion around the use of quantum trajectory techniques, which provide a numerical method for detailing with dissipative dynamics, and also together with continuous measurement theory, provide a way to understand the dynamics for this class of open AMO quantum systems. We discuss the numerical implementation of quantum trajectories for many-body systems, and address a series of examples highlighting interesting physics and important tools that are already being explored in these dissipative many-body systems.

The review is structured as follows -- we begin with a brief introduction to open quantum systems as they are discussed in quantum optics, noting the key approximations that can be made for many open AMO systems, and introducing notation. We then give an introduction to the quantum trajectories method and its physical interpretation, as well as its integration with time-dependent methods for many-body systems, especially the time-dependent Density Matrix Renormalisation Group (t-DMRG). We then introduce several examples of open many-body AMO systems, including light scattering from strongly interacting atoms in an optical lattice, collisional two-body and three-body losses as well as loss processes with single atoms, and quantum state preparation by reservoir engineering -- including methods to drive systems dissipatively into states with important many-body character. We finally return to further details of the physical interpretation of quantum trajectories by briefly introducing continuous measurement theory, before finishing with an outlook and summary.

\section{Open Quantum Systems in Quantum Optics}

Throughout this review, we will work in the framework of a quantum optics description of open quantum systems. Open quantum systems have been regularly discussed in many areas of physics, but the AMO systems that are studied in quantum optics often enable a series of approximations that allow specific types of understanding and control over the dynamics. In this section we introduce the concept of an open system, and the key approximations that lead to simplified descriptions of system dynamics. We then introduce a key example of those descriptions, specifically the Lindblad form of the master equation. 

\subsection{General framework}
\label{sec:generalqo}

The general framework for an open quantum system is sketched in Fig.~\ref{fig:openquantumsystem} \cite{Breuer2007,Gardiner2005}. We consider a small quantum system\footnote{In a many-body context, a ``small'' quantum system might be relatively large and complex -- it should just be very substantially smaller, e.g., in the scale of total energy, or total number of degrees of freedom, when compared with the environment to which it is coupling}, which is coupled to a large \emph{environment}, which can also be thought of as a \emph{reservoir}. This is a similar relationship to the heat bath and the system in the canonical ensemble of statistical mechanics, and for this reason the environment is also referred to as the ``bath'' in some literature. 

In this setup, the Hamiltonian for the \emph{total system}, including both the system and its environment, consists of three parts, 
\begin{equation}
H_{\rm total} = H_{\rm sys} + H_{\rm env} + H_{\rm int}.
\end{equation}
Here $H_{\rm sys}$ is the system Hamiltonian, describing the coherent dynamics of the system degrees of freedom alone, in the absence of any coupling to the reservoir. In quantum optics, this is often a two-level system (e.g., an atom, a spin, or two states of a Cooper pair box), in which case we typically might have $H_{\rm sys} = \hbar\omega_{\rm sys} \sigma_z$, where $\omega_{\rm sys}$ is a system frequency scale, and $\sigma_z$ is a Pauli operator for a two-level system. Another common system is to have a harmonic oscillator (e.g., a single mode of an optical or microwave cavity, or a single motional mode of a mechanical oscillator), with system Hamiltonian $H_{\rm sys}=\hbar \omega_{\rm  sys} a_{ho}^\dagger a_{ho}$, where $a_{ho}$ is a lowering operator for the harmonic oscillator. The degrees of freedom for the environment and their dynamics are described by the hamiltonian $H_{\rm env}$, and the interaction between the system and the reservoir is described by the hamiltonian $H_{\rm int}$.

In quantum optical systems the reservoir usually consists of a set of bosonic modes, so that
\begin{equation}
H_{\rm res} = \sum_l \int_0^\infty d\omega\, \hbar \omega \, b_l^\dagger (\omega) b_l(\omega), \label{hres}
\end{equation}
where $b_l(\omega)$ is a bosonic annihilation operator for a mode of frequency $\omega$. The index $l$ is convenient for describing multiple discrete modes at a given frequency, e.g., in the case that these are the modes of an external radiation field, $l$ can play the role of a polarisation index. These operators obey the commutation relation $[b_l(\omega),b_{l}^\dagger (\omega')]=\delta(\omega-\omega') \delta_{ll'}$, where $\delta(\omega)$ denotes a Dirac delta function and $\delta_{l l'}$ a Kronecker delta. Note that in general, a reservoir can have a frequency-dependent density of modes $g(\omega)$, which we have assumed for notational convenience is constant in frequency. Below when we make a Markov approximation for the dynamics, we will assume that this density of modes is slowly varying over the frequencies at which the system couples to the reservoir. 

The interaction Hamiltonian $H_{\rm int}$ analogously takes a typical form
\begin{eqnarray}
H_{\rm int} &=& - {\rm i} \hbar \sum_l \int_0^{\infty} d\omega \, \kappa_l (\omega)\left(x^+_l+ x^-_l\right) \left[ b_l(\omega) - b_l^\dagger(\omega)  \right]\\
&\approx& - {\rm i} \hbar \sum_l \int_0^{\infty} d\omega \, \kappa_l (\omega)  \left[x^+_l b_l(\omega) - x^-_l b_l^\dagger(\omega)  \right],\label{hintrwa}
\end{eqnarray}
where $x_l^{\pm}$ is a system operator, and $\kappa(\omega)$ specifies the coupling strength. In connection with the two-level systems listed above, for a two-level system, we might have $x^-_l = \sigma^-$, the spin lowering operator, and $x^+_l=\sigma^+$, the spin raising operator. In the case that the system is a Harmonic oscillator, we would typically have $x_l^{-} = a_{ho}$. In the second line of this expression, we have explicitly applied a rotating wave approximation, which we will now discuss. 

\subsection{Key approximations in AMO systems}
\label{sec:approxamo}

A key to the microscopic understanding and control that we have of open AMO quantum systems is the fact that we can usually make three approximations in describing the interaction between the system and the reservoir that are difficult to make in other systems \cite{Gardiner2005}:

\begin{enumerate}
\item{\emph{The rotating wave approximation} - In the approximate form eq.~\eqref{hintrwa} of the interaction Hamiltonian we gave above, we neglected energy non-conserving terms of the form $x^+_l b_l^\dagger (\omega)$ and $x^-_l b_l (\omega)$. In general, such terms will arise physically in the couplings, and will be of the same order as the energy conserving terms. However, if we transform these operators into a frame rotating with the system and bath frequencies (an interaction picture where the dynamics due to $H_{\rm sys}$ and $H_{\rm env}$ are incorporated in the time dependence of the operators), we see that the energy-conserving terms will be explicitly time-independent, whereas these energy non-conserving terms will be explicitly time-dependent, rotating at twice the typical frequency scale $\omega_{\rm sys}$. This is shown explicitly below in eq.~\eqref{inthamcmeq} in section \ref{sec:inthamcm}. In the case that this frequency scale is much larger than the important frequency scales for system dynamics (or the inverse of the timescales for which we wish to compute the dynamics), the effects of these energy non-conserving terms will average to zero over the relevant dynamical timescales for the system, and so we can neglect their effects in describing the dynamics.}
\item{\emph{The Born approximation} - We also make the approximation that the frequency scales associated with dynamics induced by the system-environment coupling is small in scale compared with the relevant system and environment dynamical frequency scales. That is, if the frequency scales $\omega_{\rm sys}$ at which the system couples are much larger than the frequency scales $\Gamma$ corresponding to the system dynamics induced by the environment, then we can make a Born approximation in time-dependent perturbation theory. }
\item{\emph{The Markov approximation} - We assume that the system-environment coupling is frequency/time-independent over short timescales, and that the environment returns rapidly to equilibrium in a manner essentially unaffected by its coupling to the system, so that the environment is unchanged in time, and the dynamics of the system are not affected by its coupling to the environment at earlier times, i.e., the time evolution of the system does not depend on the history of the system.}
\end{enumerate}

\begin{figure}[tb]
\begin{center}
\includegraphics[width=10cm]{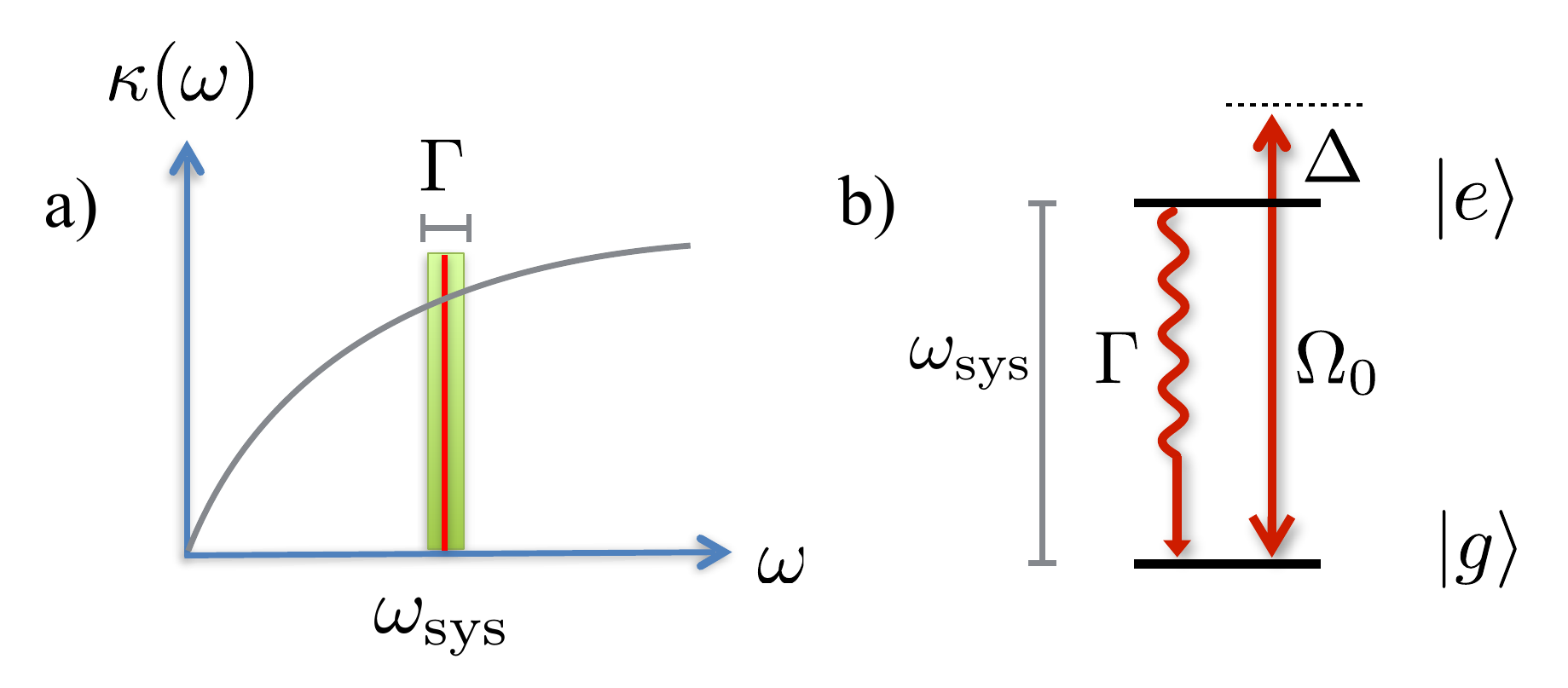}
\end{center} 
\caption{(a) Coupling constant as a function of frequency, $\kappa(\omega)$, illustrating a slowly varying coupling strength around the central system frequency $\omega_{\rm sys}$. This is a key to several of the approximations made for open quantum systems in quantum optics. (b) Comparison of the scales for a two-level system. The system frequency is the optical frequency, $\omega_{\rm sys}$, associated with the energy difference between the ground and excited levels. In order to make the standard approximations, we require the timescale for the decay of the excited state, $\Gamma^{-1}$, i.e., the timescale for the dynamics induced by coupling to the environment, to be much longer than the optical timescale associated with the transition, or in freqeuncy units, $\Gamma^{-1} \gg \omega_{\rm sys}^{-1}$. We also work in a limit where any remaining system dynamics are associated with much smaller frequency scales. In this case, the coupling strength $\Omega_0$, associated with the frequency of Rabi oscillations between the states when they are coupled strongly on resonance, as well as the frequency of detuning from resonance $\Delta$, satisfy $\Omega_0, \Delta \ll \omega_{\rm sys}$.}\label{fig:tls}\label{fig:largefrequency} 
\end{figure}

Each of these three approximations are normally exceptionally well justified, with the neglected terms in equations of motion being many orders of magnitude smaller than those that are retained. This results from the existence of a large frequency/energy scale that dominates all other such scales in the system dynamics described by $H_{\rm sys}$, and in the system-environment interaction $H_{\rm  int}$. Specifically, we usually have a situation where $[H_{\rm sys}, x^\pm_l ]\approx \pm \hbar \omega_{l} x_l^\pm$, and the frequencies $\omega_l$, which are all of the order of some system frequency $\omega_{\rm sys}$, (1) dominate any other processes present in the system dynamics, and (2) are much larger than the frequency scales of dynamics induced by coupling to the reservoir, $\Gamma$. The first of these two conditions allows us to make rotating wave approximations in the interaction Hamiltonian, without which we could not make the Markov approximation. The second of these allows us to make the Born approximation. 

The Markov approximation can be physically interpreted in two parts: it implies firstly that the system-environment coupling should be independent of frequency, to remove any back-action of the environment on the system that is not local in time, and secondly, that any correlation functions in the reservoir should retain no long-term memory of the interaction with the system. The first of these two pieces is again justified by the large frequency scale for the energy being transferred between the system and the reservoir relative to the effective coupling strength. As depicted in Fig.~\ref{fig:largefrequency}, the system and environment couple in a range of frequencies that is of the order of the effective coupling strength, $\Gamma$, and if $\Gamma \ll \omega_{\rm sys}$, then the variation in coupling strength $\kappa(\omega)$ over the range of frequencies where the modes couple significantly to the system is very small. For a single index $l$, we can write $\kappa(\omega) \approx \sqrt{\Gamma/(2\pi)}$. Note that we implicitly made this assumption for the density of modes in the reservoir $g(\omega)$ when we didn't include such a density of modes  in eqs.~\eqref{hres} and \eqref{hintrwa} above. The second requirement depends on the physical details of the reservoir, and justifying this part of the approximation usually requires comparing the typical coupling timescales $\Gamma^{-1}$ with environment relaxation timescales $\omega_{\rm env}^{-1}$, which are in general different to $\omega_{\rm sys}^{-1}$. However, for many typical cases, we also find that $\omega_{\rm env} \sim \omega_{\rm sys}$.

In the case of decaying two-level atoms, or loss of photons from an optical cavity, these timescales are both related to the relevant optical frequency. On a physical level, when a two-level atom decays and emits a photon of wavelength $\lambda_{\rm opt}$ and frequency $\omega_{\rm opt}$, then the relevant system timescale $\omega_{\rm sys}^{-1}=\omega_{\rm opt}^{-1}$, and we have a requirement that the decay time, which is the relevant system-environment coupling timescale $\Gamma^{-1}$, is much longer than $\omega_{\rm opt}^{-1}$. In terms of the reservoir relaxation time, physically the time it takes for the photon to propagate away from the atom within the photon wavelength $\lambda$ is also of the order $1/\omega_{\rm opt}$. In that sense, the single frequency $\omega_{\rm opt}$ is the large frequency scale for each of the three approximations. The ratio of this frequency scale to $\Gamma$ is, for typical optical systems $\sim 10^{15}/10^8=10^7$, so that the approximations made here are good to many orders of magnitude.

\subsection{Master Equation}

The approximations given above make it possible to derive an equation of motion for the behaviour of the system alone in the presence of the dissipation due to coupling to the environment. Because of the combination of dissipative and coherent dynamics, this is expressed in terms of the density operator $\rho$ describing the state of the \emph{system}, which can be defined based on the total density operator for the system and the environment $\rho_{\rm total}$ as $\rho={\rm Tr}_{\rm env} \{\rho_{\rm total}\}$, where the trace is taken over the environment degrees of freedom. There are many routes to derive such an equation of motion (see, e.g., \cite{Lax1963,Collett1984,Gardiner1985,Gardiner2005,Breuer2007}), one of which is outlined in appendix \ref{sec:continuousmeasurement}, via a continuous measurement formalism.  

The resulting markovian \emph{master equation} takes a Lindblad form \cite{Lindblad1976} (see, e.g., \cite{Gardiner2005,Collett1984,Gardiner1985})
\begin{eqnarray}
\dot \rho &=&-\frac{\rmi}{\hbar} [H,\rho] -\frac{1}{2} \sum_m  \gamma_m\left[\tilde c_m^\dag \tilde c_m \rho + \rho \tilde c_m^\dag \tilde c_m - 2\tilde c_m \rho \tilde c_m^\dag  \right]. \label{mastereqversion1}
\end{eqnarray}
Here, $H$ is the remaining system Hamiltonian, after the frequency scale $\omega_{\rm sys}$ of the coupling between the system and the environment has been transformed away. For a two-level system, this is a Hamiltonian describing the coupling between levels and a detuning in the rotating frame (see section \ref{sec:twolevel} below), i.e., in an interaction picture we we transform the operators so that we make the dominant coupling terms in the Hamiltonian time-independent. The operators $\tilde c_m$ are sometimes called Lindblad operators (or jump operators, as described below), and describe dissipative dynamics (including decoherence and loss processes) that occur at characteristic rates $\gamma_m$. 

Note that the $\tilde c_m$ are all system operators, and the time-dependence of the environment does not appear in this equation. For the interaction Hamiltonian in Eq.~\eqref{hintrwa}, if we assume that all of the reservoir modes that couple to the system are unoccupied, we will have $\tilde c_l = x^-_l$. One example of $\tilde c_m$ would be a transition operator $\sigma^-$ from an excited state to a ground state in the two-level system depicted in Fig.~\ref{fig:largefrequency}b, describing spontaneous emissions, and further examples are given below in sections \ref{sec:physicalinterp}, \ref{sec:illustrative} and \ref{amosys}. We also note at this point that in general, the $\tilde c_m$ are non-Hermitian, although they can always be interpreted as resulting from continuous measurement of Hermitian operators acting only on the environment, as discussed in appendix \ref{sec:continuousmeasurement}. In the case that all $\tilde c_m$ are Hermitian, we can think of this master equation as representing a continuous measurement of the system operators $c_m$. 

The general form in eq.~\eqref{mastereqversion1} also demonstrates explicitly the trace-preserving property of the master equation (${\rm Tr}\{ \dot \rho\}=0$). This form of dissipative dynamics is sometimes used as a toy model for dissipation because of its relatively simple interpretation. In the AMO / quantum optics context, it is particularly important because it describes the microscopic behaviour under the well-controlled approximations described in section \ref{sec:approxamo}.

In order to reduce the notational complexity, we will combine the rate coefficients with the operators in the form $c_m=\sqrt{\gamma_m}\tilde c_m$, and we will also equate energy and frequency scales setting $\hbar\equiv 1$, to obtain 
\begin{eqnarray*}
\dot \rho &=&-\rmi  [H,\rho] -\frac{1}{2} \sum_m \left[c_m^\dag c_m \rho + \rho c_m^\dag c_m - 2c_m \rho c_m^\dag  \right]\nonumber,
\end{eqnarray*}
Note that we can also express this in the convenient alternative form
\begin{eqnarray}
\dot \rho &=&-\rmi  (H_{\rm eff}\rho - \rho H_{\rm eff}^\dag) + \sum_m c_m \rho c_m^\dag, \label{mastereq}
\end{eqnarray}
where we refer to 
\begin{eqnarray*}
H_{\rm eff}=H - \frac{i}{2} \sum_m c_m^\dag c_m
\end{eqnarray*}
as the \emph{effective Hamiltonian} for the dissipative system. in this form, the term $\sum_m c_m \rho c_m^\dag$ is often called the \emph{recycling term}, as it recycles the population that is lost from certain states due to the non-Hermitian effective Hamiltonian, placing it in other states. For a two level system with decay from an excited state to the ground state, one may think of the non-Hermitian part as removing amplitude from the excited state, and the recycling term as reinstating this in the ground state (see the many-body examples for a more detailed discussion of this).

For some simple systems, master equations can be solved analytically, to obtain expectation values of a given physical observable $A$ at particular times, $\langle A(t) \rangle = {\rm Tr}\{ A \rho(t)\}$ or also two-time correlation functions of the form $\langle A(t+\tau) B(t) \rangle$. In the next section we will discuss how the dynamics described by the master equation can be both computed and physically interpreted using quantum trajectories techniques. We then give some specific examples of small physical systems described by master equations. In the following two sections, we then discuss how master equations of the form eq.~\eqref{mastereq} describing many-body systems can be solved by combining quantum trajectories techniques with many-body numerical methods, and then discuss a number of examples of recent work on such physical systems.

\section{Quantum Trajectories}

Quantum trajectory techniques were developed in Quantum Optics in the early 1990s \cite{Dalibard1992,Molmer1993,Dum1992,Dum1992a,Carmichael1993,Plenio1998} as a means of numerically simulating dissipative dynamics,
 and can be applied to any system where the time evolution of the density operator is described via a master equation (in Lindblad
 form). These techniques involve rewriting the master equation as a stochastic average over individual \emph{trajectories}, which can be
  evolved in time numerically as pure states. These techniques avoid the need to
 propagate a full density matrix in time (which is often numerically
 prohibitive), and replace this complexity with stochastic sampling. The key advantage this gives is that if the Hilbert space has dimension $N_H$, then propagating the density matrix means propagating an object of size $N_H^2$, whereas stochastic sampling of states requires propagation of state vectors of size $N_H$ only. Naturally, the penalty that is paid is the need to collect many samples for small statistical errors, and it is important that the number of samples required remains smaller than the size of the Hilbert space in order to make this efficient.
 
In quantum optics, these techniques were developed in parallel by a number of groups, as they arose out of studies of different open-quantum system phenomena. The versions of these techniques that most closely resemble what we present here were developed by Dalibard, Castin and M{\o}lmer \cite{Dalibard1992,Molmer1993} as a Monte Carlo method for simulation of laser cooling; by Dum et al. \cite{Dum1992,Dum1992a}, arising from studies of continuous measurement; by Carmichael \cite{Carmichael1993}, in studying of the generation of non-classical states of light; and by Hegerfeldt and Wilser \cite{Hegerfeldt1992}, in modelling a single radiating atom. The term \emph{quantum trajectories} was coined by Carmichael, whereas the other approaches were referred to as either a \emph{quantum jump approach}, or the \emph{Monte Carlo wavefunction method}. This later term should not be confused with so-called \emph{quantum Monte Carlo} techniques - the Monte Carlo treatment performed here is classical, in the sense that there is no coherent sum of amplitudes involved in evaluating expectation values from the samples we obtain. Mathematically, all of these approaches are essentially equivalent, but there are differences in the numerical implementations.

In quantum optics these methods have been applied to a wide variety of problems, including laser cooling \cite{Molmer1993,Marte1993,Castin1995,Marte1993b} and coherent population trapping \cite{Bardou1994}, the behaviour of cascaded quantum systems \cite{Carmichael1993b,Gardiner1993,Kochan1994}, the continuous quantum Zeno effect \cite{Itano1990,Power1996,Beige1996}, two-photon processes \cite{Garraway1994}, decoherence in atom-field interactions \cite{Moya-Cessa1993}, description of quantum non-demolition measurements \cite{Imamoglu1993} and decoherence in the atom-optics kicked rotor \cite{Ammann1998,Doherty2000,Daley2002,Wimberger2003}. For further examples, see the review by Plenio and Knight \cite{Plenio1998} and references therein. Note that complementary stochastic approaches were developed at the same time, especially by Gisin and his collaborators \cite{Gisin1984,Gisin1989,Gisin1992}. These and other related quantum state diffusion approaches \cite{Goetsch1994, Garraway1994b} involve continuous stochastic processes, but can be directly related back to the quantum trajectories formalism by considering homodyne detection of the output of a quantum system (see, e.g.,\cite{Wiseman1993b,Carmichael1993}). 

In the remainder of this section we introduce the quantum trajectories method and illustrate its use and physical interpretation. We begin by presenting the first-order Monte Carlo wavefunction method put forward by Dalibard et al. \cite{Dalibard1992} and Dum et al. \cite{Dum1992}, and discuss statistical errors and convergence in this method. We then discuss the physical interpretation of quantum trajectories, before continuing with generalisations to calculations that are higher-order in the timestep. We then give two illustrative examples. The first is a simple example involving the optical Bloch equations for a two-level atom, whereas the second example gives a preview of our discussion of open many-body systems by treating dephasing of a hard-core Bose gas on a lattice. 

\subsection{First-order Monte Carlo wavefunction method}
\label{sec:firstorder}
The simplest form of the quantum trajectory method involves expanding the master equation to first order in a time step $\delta t$, and was first described in this form by Dalibard et al. \cite{Dalibard1992} and Dum et al. \cite{Dum1992}. It involves the evolution of individual trajectories $\ket{\phi(t)}$, over which we average values of observables. 

For a single trajectory, the initial state $\ket{\phi(t=0)}$ should be sampled from the density operator at time $t=0$, $\rho(t=0)$. For some systems, $\rho(t=0)$ may be a pure state, and this is the case for a number of our many-body examples. In that case, the initial state is always the same state, $\rho(t=0)=|\phi(t=0)\rangle\langle \phi(t=0)|$. Once we have an initial state, we propagate forward in time, making use of the following procedure in each time step:
\begin{enumerate}
\item Taking the state at the beginning of the time step, $|\phi(t)\rangle$, we first compute the evolution under the effective Hamiltonian. This will give us one candidate for the new state at time $t+\delta t$,
 \begin{eqnarray}
|\phi^{(1)}(t+\delta t)\rangle=\left(1-\rmi H_{\rm eff}\delta t \right)|\phi(t)\rangle,
\end{eqnarray}
and we compute the norm of the corresponding state, which will be less than one because $H_{\rm eff}$ is non-Hermitian:
 \begin{eqnarray}
\langle \phi^{(1)}(t+\delta t)|\phi^{(1)}(t+\delta t)\rangle&=&\langle \phi(t) | \left(1+\rmi H^\dagger_{\rm eff}\delta t \right)\left(1-\rmi H_{\rm eff}\delta t \right)|\phi(t)\rangle\\
&=&1-\delta p .
\end{eqnarray}
Here, we can consider $\delta p$ as arising from different potential decay channels, corresponding to the action of different Lindblad operators $c_m$, 
\begin{eqnarray}
\delta p&=&\delta t\langle \phi(t) | \rmi(H_{\rm eff}-H_{\rm eff}^\dag)|\phi(t)\rangle\\ &=&\delta t \sum_m \langle \phi(t) | c^\dag_m c_m |\phi(t) \rangle \equiv\sum_m \delta p_m.
\end{eqnarray}
We can effectively interpret $\delta p_m$ as the probability that the action described by the operator $c_m$ will occur during this particular time step.

\item Then, we choose the propagated state stochastically in the following manner. We would like to assign probabilities to different outcomes so that:
\begin{itemize}
\item  With probability $1-\delta p$
 \begin{eqnarray}
|\phi(t+\delta t) \rangle =\frac{ |\phi^{(1)}(t+\delta t)\rangle }{\sqrt{1-\delta p}}
\end{eqnarray}
 
\item  With probability $\delta p$
 \begin{eqnarray}
|\phi (t+\delta t) \rangle = \frac{ c_m |\phi(t) \rangle }{\sqrt{\delta p_m / \delta t}}
\end{eqnarray}
where we choose one \emph{particular} $m$, which is taken from all of the possible $m$ with probability
\begin{eqnarray}
\Pi_m=\delta p_m / \delta p
\end{eqnarray}
\end{itemize}
In a practical numerical calculation, this action requires drawing a uniform random number $r_1$ between 0 and 1, and comparing it with $\delta p$. If $r_1>\delta p$, then no jump occurs, and the first option arrising from propagation under $H_{\rm eff}$, i.e., $|\phi (t+\delta t) \rangle \propto |\phi^{(1)}(t+\delta t)\rangle$ is chosen. If $r_1<\delta p$, then a jump occurs, and we must choose the particular $c_m$ operator to apply. We therefore associate each $m$ with an interval of real numbers, with the size of the interval being proportional to $\delta p_m$. Normalising the total interval length to one so that every $m$ corresponds uniquely to a range between 0 and 1, we then choose a second random number $r_2$, also uniformly distributed between 0 and 1, and choose the associated $c_m$ for which the assigned interval contains $r_2$. 

\end{enumerate}

In order to see that this stochastic propagation is equivalent to the master equation, we can form the density operator,
\begin{eqnarray}
\sigma(t)=|\phi(t)\rangle\langle \phi(t) | .
\end{eqnarray}
From the prescription above, the propagation of this density operator in a given step is:
\begin{eqnarray}
\overline{\sigma(t+\delta t)} = (1-\delta p) \frac{| \phi^{(1)} (t+\delta t) \rangle}{\sqrt{1-\delta p}} \frac{\langle \phi^{(1)} (t+\delta t) |}{{\sqrt{1-\delta p}}} 
+ \delta p \sum_m \Pi_m \frac{c_m |\phi(t)\rangle}{\sqrt{\delta p_m / \delta t}}\frac{\langle\phi(t)| c_m^\dag}{\sqrt{\delta p_m / \delta t}},
\end{eqnarray}
where $\overline X$ for any $X$ denotes a statistical average over trajectories, as opposed to the quantum mechanical expectation values or mathematically exact averages, which we will denote $\langle X \rangle$. Rewriting the terms from the above definitions, we obtain
\begin{eqnarray}
\overline{\sigma(t+\delta t)}&=& \sigma(t) -\rmi \delta t (H_{\rm eff} \sigma(t) - \sigma(t) H_{\rm eff}^\dag) +\delta t \sum_m c_m \sigma(t) c_m^\dag,
\end{eqnarray}
which holds whether $\sigma(t)$ corresponds to a pure state or to a mixed state. In this way, we see that taking a stochastic average over trajectories is equivalent to the master equation
\begin{eqnarray}
\dot \rho =-\rmi (H_{\rm eff}\rho - \rho H_{\rm eff}^\dag) + \sum_m c_m \rho c_m^\dag.
\end{eqnarray}
Note that this equivalence doesn't require a particular choice of $\delta t$, and in particular $\delta t$ can be chosen to be small in evaluating this evolution. Naturally, choosing very large $\delta t$ would strongly compromise the accuracy of the method when propagating the state in time. In the subsection \ref{sec:higherorder}, we will deal with how to improve upon this first-order method numerically.

\subsubsection{Computing single-time expectation values}

In order to compute a particular quantity at time $t$, i.e., $\langle A \rangle_t={\rm Tr} \{A\rho(t) \}$, we simply compute the expectation value of $A$ for each of our stochastically propagated trajectories, $\langle \phi(t) | A |\phi(t) \rangle$, and take the average of this quantity over all of the trajectories,
\begin{equation}
\langle A \rangle_t \approx \overline{ \langle \phi(t) | A  | \phi(t) \rangle}. 
\end{equation}

Provided our random number generators are well behaved, the trajectories are statistically independent, allowing for simple estimate of statistical errors in the computation of $\langle A \rangle_t$, as discussed in section \ref{sec:staterror}.

\subsubsection{Computing two-time correlation functions}
Two-time correlation functions of the form $C(t,\tau)=\langle A(t+\tau) B(t)\rangle$ appear in many contexts, including spectral functions \cite{Cohen-Tannoudji1998}, and also current autocorrelation functions in many-body physics \cite{Mahan2000}. These functions are a little more complicated to compute than single-time expectation values. To compute such correlation functions from a master equation, we typically apply the quantum regression theorem \cite{Lax1968,Gardiner2005}, which can be applied generally to an operator $X_{ij}=|i\rangle\langle j|$, where $|i\rangle$ and $|j\rangle$ belong to an orthonormal set of basis states for the Hilbert space. If we write $C_{ij}(t,\tau)=\langle X_{ij}(t+\tau) B(t)\rangle$, then we note that $C_{ij}(t,0)$ are one-time averages that can be calculated directly from the density matrix at a single time, and the $\tau$ dependence can be computed as
\begin{equation}
\frac{\partial C_{ij}}{\partial \tau} (t,\tau) = \sum_{kl} M_{ijkl} C_{kl}(t,\tau),
\end{equation}
where the $M_{ijkl}$ are the same matrix elements that appear in the equation of motion for one-time averages \cite{Lax1968,Gardiner2005}
\begin{equation}
\frac{d \langle X_{ij} (t)\rangle}{dt}=\sum_{kl} M_{ijkl}\langle X_{kl} (t)\rangle.
\end{equation}

To reproduce these values from quantum trajectories, we follow a simple procedure \cite{Molmer1993,Plenio1998}. We propagate a sample trajectory to time $t$, and then generate four helper states:
\begin{eqnarray}
|\chi^R_\pm (0) \rangle&=&\frac{1}{\sqrt{\mu^R_\pm}} (1\pm B) |\phi(t)\rangle,\\
|\chi^I_\pm (0) \rangle&=&\frac{1}{\sqrt{\mu^I_\pm}} (1\pm \rmi B) |\phi(t)\rangle,
\end{eqnarray}
where $\mu^R_\pm$ and $\mu^I_\pm$ normalise the resulting helper states. We then evolve each of these four states using the quantum trajectories procedure, and compute the correlation functions
\begin{eqnarray}
c^R_\pm(\tau)= \langle \chi^R_\pm (\tau)|A|\chi^R_\pm (\tau)\rangle, \\
c^I_\pm(\tau)= \langle \chi^I_\pm (\tau)|A|\chi^I_\pm (\tau)\rangle.
\end{eqnarray}
We can then reconstruct a sample for 
\begin{equation}
C(t,\tau)=\frac{1}{4} \left[\mu^R_+c^R_+(\tau)-\mu^R_-c^R_-(\tau)-\rmi \mu^I_+c_+^I(\tau)+\rmi \mu^I_-c_-^I(\tau)  \right],
\end{equation}
and average this over both the evolution up to time $t$ and the propagation of helper states to time $\tau$.

\subsection{Statistical Errors and convergence}
\label{sec:staterror}

\subsubsection{Estimating statistical errors}

Using the above procedure or generalisations that are higher-order in the timestep $\delta t$ (see section \ref{sec:higherorder}), we generate $N$ sample trajectories. Under the assumption that the random numbers used in the numerical implementation are statistically random and uncorrelated\footnote{See, e.g., Ref.~\cite{Press1995} for a general discussion of random number generators. For typical calculations, only a few tens of thousands of random numbers must be generated, and good quality pseudorandom number generators suffice. See Ref.~\cite{Frauchiger2013} for a recent discussion of improvements to that are possible using quantum random number generators.}, these trajectories are statistically independent, and we can  estimate the correct mean $\langle X \rangle$ of any operator of interest $\hat X$ as 
\begin{equation}
\overline X (t)=  \frac{1}{N} \sum_i X_i \equiv \frac{1}{N} \sum_i \langle \phi_i (t) | \hat X |\phi_i(t) \rangle.
\end{equation}
The central limit theorem implies that for sufficiently large $N$, the probability distribution for $\overline X$ will be well approximated by a Gaussian distribution with mean $\langle X \rangle$. The statistical error in this mean is the standard deviation of that distribution, which in turn can be estimated based on the variance of the values $X_i$ in the following way. We consider
\begin{eqnarray}
{\rm Var}[\overline X]&=& \left \langle \left( \overline X- \langle X \rangle \right)^2 \right \rangle = \left\langle \left( \frac{1}{N} \sum_{i=1}^N X_i \right)^2  - 2 \overline X \langle X \rangle  + \langle X \rangle^2 \right\rangle\nonumber\\
&=& \frac{1}{N^2} \sum_{i=1}^N\sum_{j=1}^N \left\langle X_i X_j \right\rangle-\langle X \rangle^2 
=  \frac{1}{N^2} \sum_{i=1}^N \left\langle X_i^2 \right\rangle+ \frac{N-1}{N}\langle X \rangle^2-\langle X \rangle^2 \nonumber\\
&=&\frac{1}{N} (\langle X^2 \rangle - \langle X \rangle ^2)= \frac{1}{N} {\rm Var}[ X]. \nonumber\\
\end{eqnarray}
In deriving this standard result, we use both the independence of the trajectories $\langle X_i X_{j\neq i} \rangle =\langle X_i \rangle \langle X_{j\neq i} \rangle$, and the replacement that $\langle \overline X \rangle=\langle X \rangle$, i.e., the variance written in the last line is the true variance of the distribution for $X$. As a result, it can be shown (see, e.g., Ref.~\cite{Press1995}) that we should use the approximator
\begin{equation}
{\rm Var}[ X] \approx \frac{\sum_{i =1}^N(X_i - \overline X)^2 }{N-1},
\end{equation}
if we would like to estimate this based on the samples we obtain in the calculation. 

In this sense, we can always calculate the statistical error $\sigma_A$ in our estimate of a quantity $\langle A \rangle$ by taking the \emph{estimate of the population standard deviation} $\Delta A$ from our $N$ samples, and \emph{dividing by $\sqrt{N}$},
\begin{equation}
\sigma_A = \frac{\Delta A}{\sqrt{N}}.
\end{equation}

How many trajectories we will require for good convergence will depend both on the details of the dynamics and on the quantity being calculated. For variables with non-zero mean, we would typically like to have $\sigma_A/\langle A \rangle \ll 1$, which implies that $\sqrt{N} \gg \Delta A / \langle A \rangle$. As the sample estimate overestimates the population standard deviation (see e.g., \cite{Molmer1993}), we can consider the sample estimate for the standard deviation $\Delta A$ here. 

\subsubsection{Global quantities vs. local quantities}
\label{sec:globallocal}

In M{\o}lmer et al. \cite{Molmer1993}, there is a discussion of the different number of trajectories required for global quantities vs. local quantities to be straight-forwardly estimated. This is done for single-particle systems in a Hilbert space dimension $N_H$. 

For \emph{global quantities} such as the total energy $A_G=E_{\rm tot}$, there is usually a fixed relationship between the estimate $\Delta A_G$ and $\langle A_G \rangle$ that is independent of the dimension of the Hilbert space $N_H$. For example, in the Brownian motion of a particle thermalised with a reservoir at temperature $T$, we would expect $\langle A_G \rangle\approx (3/2) k_B T$, where $k_B$ is the Boltzmann constant, and $\Delta A_G\approx \sqrt{3/2} k_B T$. The requirement on the number of trajectories is then simply $N\gg 1$, and we should expect that the relative error is well estimated by $1/\sqrt{N}$ (so that for 10\% relative statistical error, we would require $N\sim 100$ trajectories. These same arguments apply to many-particle systems, we also expect global averages involving all of the particles in a system (or all of the spins or lattice sites in a spin or lattice system) to be efficiently treatable with quantum trajectories methods.

For \emph{local quantities} in a single-particle calculation, the opposite is true. If we try to determine the population of a given eigenstate (of the Hamiltonian, or of the momentum operator), then we expect that in a Hilbert space of dimension $N_H$, our local quantities $A_L$ and their statistical variance will behave as $\langle A_L \rangle \sim 1/N_H$ and $\sigma_A^2 \sim 1/N_H$. As a result, we see that we require $N\gg N_H$ for good statistical convergence, and there is no advantage in using a quantum trajectories technique over direct integration of the master equation.

However, for \emph{local quantities in many-body systems}, the situation is not as clear-cut. If we attempt to compute the population of a given many-body eigenstate out of $N_H$ possible many-body eigenstates, then the previous conclusion for local quantities in a single-particle system still applies. However, single-particle or few-particle quantities in a finite-size system that are local in space or momentum will often scale simply as the size of the system, $L$, whereas for large systems with fixed particle density, $N_H\propto \exp(L)$. As a result, quantities such as the local density on one site in a lattice of $L$ sites, or individual elements of a single-particle density matrix on such a lattice will tend to scale as $\langle A_S \rangle \sim 1/L$ and $\sigma_A^2 \sim 1/L$. We therefore require $N\gg L$ for good statistical convergence, and not $N 
\gg N_H$. Although this will require more trajectories for small relative statistical error than in the case of global quantities (by a factor of $\sqrt{L}$), there are often still large advantages in using quantum trajectory techniques to calculate these quantities when the size of the Hilbert space $N_H$ becomes large.

We will discuss this again below, when we give two illustrative examples for the use of quantum trajectories in section \ref{sec:illustrative}. Specifically, in section \ref{sec:hcb}, we treat the dephasing of hard-core bosons on a lattice, and consider calculation of the total energy, local single-particle correlations, and local densities using quantum trajectory techniques.

\subsection{Physical interpretation}
\label{sec:physicalinterp}
One of the greatest strengths of the quantum trajectories approach to dissipative systems, but also one of the most subtle points in its usage, is that it gives us a simple physical interpretation for the physics induced by the environment on the system. The method is also known as the method of \emph{quantum jumps}, and the Lindblad operators $c_m$ are also called \emph{jump operators}, inviting the picture of a system that evolves under the non-Hermitian effective Hamiltonian $H_{\rm eff}$ and then undergoes \emph{quantum jumps} at certain points in time. The master equation is then an appropriately weighted stochastic average over all of the different times at which the jumps could occur, and all of the different types of jumps that can occur. 

To see this working in practice, consider a two-level system like that depicted in Fig.~\ref{fig:tls}b. If we drive the system, coupling the excited and ground states ($\ket{e}$ and $\ket{g}$) respectively with an effective Rabi frequency $\Omega_0$ and a detuning $\Delta$, then the system will undergo coherent dynamics, interrupted at particular times by spontaneous emission events, which return the atom to the ground state $\ket{g}$, due to the action of a jump operator $c=\sqrt{\Gamma}|g\rangle\langle e|$. If photons can be scattered in multiple directions, then the different directions constitute different channels $c_m$, and we can average over the probability distribution of emission directions. Note that this is presented carefully below in section \ref{sec:lightscattering}. The stochastically chosen times for the jumps are the times at which spontaneous emissions occur, and stochastically sampling different $m$ values amounts to a stochastic sampling of the direction in which the photon is emitted. 

This is a very appealing intuitive picture of the dynamics, and connects strongly with an intuition of what would happen if we are actually able to measure the environment. Indeed, quantum trajectories techniques were originally developed by certain groups from studies of photon counting \cite{Hegerfeldt1992}, or continuous measurement \cite{Dum1992}. If we are able to make perfect measurements and we see a photon appear in the time window $\delta t$, then we know that a jump has occurred, and that the state of the atom should be projected on the ground state. On the other hand, if we know that no jump has occurred, then the corresponding evolution of the system is an evolution under the effective Hamiltonian $H_{\rm eff}$. Already here, we see a key piece of physics that will recur multiple times: \emph{knowing} that no jump has occurred means that we gain information about the system, just as knowing that a jump has occurred gives us information that the atom is projected into the ground state. 

To see this in a simple example, consider preparing the system in a state $\ket{\psi(t)}=\alpha \ket{g}+ \beta \ket{e}$, where $\alpha$ and $\beta$ are complex coefficients, and set $\Omega_0=\Delta=0$. We can then ask what we expect to happen in a single step, depending on whether we observe a spontaneously emitted photon or not. As we are not concerned about the motion of the atom, or therefore about the direction of spontaneously emitted photons, we can consider a single jump operator $c=\sqrt{\Gamma}|g\rangle\langle e|$, and an effective Hamiltonian which is simply $H_{\rm eff} = -\rmi (\Gamma/2) |e\rangle\langle e|$.  If a jump occurs in a time step $\delta t$, then the state after this jump is 
\begin{equation}
\ket{\psi(t+\delta t)}=\frac{c\ket{\psi(t)}}{\Vert c\ket{\psi(t)}\Vert }=\ket{g}.
\end{equation}
So if there is a spontaneous emission, then the state is projected onto the ground state, as expected. Consider now what happens if no jump occurs: We then evolve the state as 
\begin{equation}
\ket{\psi(t+\delta t)}=\frac{\exp(-\rmi H_{\rm eff} \delta t) \ket{\psi(t)}}{\Vert \exp(-\rmi H_{\rm eff} \delta t) \ket{\psi(t)}\Vert }=\frac{ \alpha \ket{g}+ \beta{\rm e}^{-\Gamma \delta t/2} \ket{e}}{\Vert \alpha \ket{g}+ \beta{\rm e}^{-\Gamma \delta t/2} \ket{e}\Vert }=\frac{ \alpha \ket{g}+ \beta{\rm e}^{-\Gamma \delta t/2} \ket{e}}{\sqrt{|\alpha|^2 + |\beta|^2{\rm e}^{-\Gamma \delta t} }}.
\end{equation}
In this way, the probability of finding the system in the excited state decreases relative to the probability of the system being in the ground state, provided $\alpha\neq 0$. Essentially, through the lack of a spontaneously emitted photon, we have gained the knowledge that the system is somewhat more likely to be in the ground state than in the excited state, and this gain in knowledge is reflected in the relative probabilities for occupation of these states at time $t+\delta t$ relative to time $t$. Thus, the dynamics are affected by coupling to the environment even in the absence of actual spontaneously emitted photons.

There are two important cautions to over-interpreting this intuitive physical picture. Firstly, the master equation can typically be expanded in different sets of jump operators. 
If there exists a unitary transformation $T$ in the Hilbert space of the system such that for the dissipative part of the master equation
\begin{equation}
\mathcal{L}\rho = -\frac{1}{2} \sum_m \left[c_m^\dag c_m \rho + \rho c_m^\dag c_m - 2c_m \rho c_m^\dag  \right],
\end{equation}
the operator $T$ satisfies the condition
\begin{equation}
T[\mathcal L \rho]T^\dagger = \mathcal L \left(T\rho T^\dagger\right),
\end{equation}
then we can rewrite each jump operator $c_m$ such that we obtain new operators $d_m=T^\dagger c_m T$, and show \cite{Molmer1993} that
\begin{equation}
\mathcal{L}\rho = -\frac{1}{2} \sum_m \left[d_m^\dag d_m \rho + \rho d_m^\dag d_m - 2d_m \rho d_m^\dag  \right].
\end{equation}
In this sense, the operators $d_m$ are equally good choices for the jumps as the operators $c_m$. In general, for the application of quantum trajectories as a numerical method, we should simply choose the jump operators that provide most rapid convergence of sampling. This has been investigated in a number of contexts \cite{Molmer1993,Holland1998,Breslin1995,Wiseman2000,Atkins2005}, including quantum feedback \cite{Wiseman2005}. However, to ascribe physical meaning to an individual trajectory it is important to be able to associate the specific jumps with measurable properties of the environment after the jump. This detector dependence should even produce measurable differences under certain conditions \cite{Wiseman2012}.

In addition, ascribing a pure state to the system as a function of time implies knowledge of measurement results from the environment. Although each trajectory implies a particular physical picture of the dissipation, it is the average over trajectories that properly specifies our knowledge of the system if we do not measure the environment. At the same time, there are sometimes means to make a measurement on the system and infer the measurement result for the environment. A simple case of this type of \emph{postselection} appears when the dissipative process we deal with is dominated by particle loss, as we will treat in section \ref{sec:particleloss} below. There, if we know that no particles are lost by making a measurement at the end of a dynamical process, then we can project the state of the system on the state simply evolved under the effective Hamiltonian $H_{\rm eff}$, as this is an equivalent mechanism to measure that no jumps occurred during the evolution.   

In order to address how the approximations that we can make in AMO systems give rise to the behaviour exhibited by the system, we will re-derive the quantum trajectory approach in appendix \ref{sec:continuousmeasurement}. There we will make explicit this connection between measurement of the environment and inference of the state of the system.

\subsection{Alternate formulation for higher-order integration in time}

\label{sec:higherorder} 

The downside of the method presented in section \ref{sec:firstorder} is that it is only first order in the time step $\delta t$. Under some conditions, where the effect of dissipation is much slower than other dynamical timescales for the system, it might be possible to continue to apply this first order time step for the dissipative dynamics, but take a step towards higher-order expansions by propagating the state under the effective Hamiltonian more accurately, i.e., replacing the step $|\phi^{(1)}(t+\delta t)\rangle\propto\left(1-\rmi H_{\rm eff}\delta t \right)|\phi(t)\rangle$ with a higher-order version, or exact computation of $|\phi^{(1)}(t+\delta t)\rangle\propto\exp(-\rmi H_{\rm eff}\delta t)|\phi(t)\rangle$ (and corresponding normalisation of the resulting state). 

However, this still doesn't alleviate the problem that the jump takes an entire time step $\delta t$, and effective results in an underestimation of the typical time between jumps of the order of $\delta t$. In this way, there is a systematic first-order overestimation of the rate at which jumps occur. Direct higher-order adjustments to the method presented in section \ref{sec:firstorder} were made by Steinbach et al., \cite{Steinbach1995}, allowing calculations of a Runge-Kutta type up to fourth order. 

A general way to improve the method was originally proposed by Dum et al., \cite{Dum1992a}, in which they approach the problem from the point of view of continuous measurement, and essentially take the limit $\delta t \rightarrow 0$ in thinking about the occurrences of jumps within individual trajectories\footnote{We can't completely take the limit $\delta t \rightarrow 0$ physically, as we have made important approximations regarding the fact that we consider dynamics over long timescales compared with $\omega_{\rm sys}^{-1}$. However, we can take the limit of vanishing $\delta t$ from the point of view of numerical implementations after the other approximations have been implemented. See appendix \ref{sec:continuousmeasurement} for a more detailed discussion.}. 

The revised version of the scheme takes the following form:

\begin{enumerate}
\item{Sample the initial state and begin the propagation under the effective Hamiltonian as in the scheme from section \ref{sec:firstorder}.}
\item{Sample a random number $r$, uniformly distributed between $0$ and $1$.}
\item Numerically solve the equation
\begin{equation}
\Vert\exp(-\rmi H_{\rm eff} t_1) \ket{\phi(t_0)}\Vert^2=r
\end{equation}
in order to find the time $t_1$ at which the next jump occurs, given that the previous jump or the start of the calculation was at time $t_0$. This can be solved using higher order integration methods, including Runge-Kutta.
\item $\ket{\phi(t)}$ is then computed numerically in the time interval $t\in [t_0,t_1]$ as
\begin{equation}
\ket{\phi(t)}=\frac{\exp[-\rmi H_{\rm eff} (t-t_0)] \ket{\phi(t_0)}}{\Vert \exp[-\rmi H_{\rm eff} (t-t_0)] \ket{\phi(t_0)}\Vert }.
\end{equation}
\item At time $t_1$, a quantum jump is applied, with probabilities for application of each $c_m$ determined as in step 2 of the method in section \ref{sec:firstorder}. That is, we choose a particular $m$ based on the probabilities
\begin{equation}
\delta p_m \propto \langle \phi(t_1)|c_m^\dagger c_m|\phi(t_1)\rangle,
\end{equation}
and apply the jump as
\begin{equation}
\ket{\phi(t_1^+)}=\frac{c_m \ket{\phi(t_1^-)}}{\Vert c_m \ket{\phi(t_1^-)}\Vert }.
\end{equation}
Here, $|\phi(t_1^-)\rangle$ is the state obtained in step 4 by propagating in time under $H_{\rm eff}$ up to time $t_1$, and $|\phi(t_1^+)\rangle$ is the state after the jump, i.e., the state we use to continue the time evolution.

\item We now continue the time evolution from step 2, choosing a new random number $r$.
\end{enumerate}

In this method, jumps occur at a particular point in time, rather than taking a fixed length of evolution time, and both the times of the jumps and the evolution under the effective Hamiltonian between the jumps can be solved numerically to arbitrary precision. In this way, we remove the reliance on a first-order Euler expansion.

\subsection{Illustrative examples}
\label{sec:illustrative}

We now give two example applications for quantum trajectories. The first is the simple case of the optical Bloch equations, which describe a driven two-level atom undergoing spontaneous emissions. While the underlying master equation is easily tractable, this gives a good basis for our intuition on the meaning of individual trajectories, and a good example for the calculation of statistical errors. The second example then provides a lead-in to thinking about many-body dissipative systems by considering the dephasing of a hard-core Bose gas on a lattice. This system is somewhat more complicated, though analytical experessions can be found for certain quantities, against which we can straight-forwardly benchmark the statistical error estimates discussed in section \ref{sec:staterror}.

\subsubsection{Optical Bloch Equations}
\label{sec:twolevel}

The optical Bloch equations describe a two-level atom that is driven by a classical laser field \cite{Cohen-Tannoudji1998}, with a detuning $\Delta$ between the frequency of the laser field and the atomic transition frequency, as depicted in Fig.~\ref{fig:largefrequency}b. In the absence of dissipation, this system undergoes Rabi oscillations at a frequency $\Omega$ that depends on the intensity and polarization of the classical laser field, and the dipole matrix elements between the two internal states $\{\ket{g}$ and $\ket{e}\}$. This gives rise to the well-known Rabi Hamiltonian for a spin-1/2 system. In the presence of damping, where an atom can undergo spontaneous emissions, decaying from the excited state to the ground state, we obtain the master equation 
\begin{equation}
\frac{d}{dt} \rho = -{\rm i}[H_{\rm opt},\rho] - \frac{\Gamma}{2} \left( \sigma_+\sigma_- \rho + \rho  \sigma_+\sigma_-  -2 \sigma_- \rho \sigma_+  \right),
\end{equation}
where the Hamiltonian
\begin{equation}
H_{\rm opt} = - \frac{\Omega}{2} \sigma_x - \Delta \sigma_+\sigma_- .
\end{equation}
Here, in the basis $\{\ket{e},\ket{g}\}$, we denote the system density matrix $\rho$ and the Pauli matrices as 
\begin{equation}
\rho=\left(\begin{array}{cc}
\rho_{ee} & {\rho}_{ge}\\
{\rho}_{eg} &\rho_{gg}
\end{array}\right), \,\,\,\sigma_x=\left(\begin{array}{cc}
0 & 1\\
1 & 0
\end{array}\right), \,\,\, \sigma_+=\left(\begin{array}{cc}
0 & 1\\
0 & 0
\end{array}\right), \,\,\,\sigma_- =[\sigma_+]^\dagger .
\end{equation}
We assume $\Omega$ is real for simplicity of notation. 

This master equation (the so-called \emph{optical Bloch equations}) can be expressed in terms of the matrix elements of the system density operator as
\begin{equation}
\frac{d}{dt}\left[\begin{array}{c}
{\rho}_{eg}\\
{\rho}_{ge}\\
\rho_{ee}\\
\rho_{gg}
\end{array}\right]=\left[\begin{array}{cccc}
i\Delta-\Gamma/2 & 0 & -{\rm i}\Omega/2 & {\rm i}\Omega/2\\
0 & -{\rm i}\Delta-\Gamma/2 & {\rm i}\Omega/2 & -{\rm i}\Omega/2\\
-{\rm i}\Omega/2 & {\rm i}\Omega/2 & -\Gamma & 0\\
{\rm i}\Omega/2 & -{\rm i}\Omega/2 & \Gamma & 0
\end{array}\right]\left[\begin{array}{c}
{\rho}_{eg}\\
{\rho}_{ge}\\
\rho_{ee}\\
\rho_{gg}
\end{array}\right].
 \end{equation}
These equations can be solved exactly, and describe damped oscillations of the system between the two internal states, damping to an excited state population $P_e \equiv \rho_{ee}$
\begin{equation}
\rho_{ee}=\frac{\frac{1}{4}|\Omega|^{2}}{\Delta^{2}+\frac{1}{4}\Gamma^{2}+\frac{1}{2}|\Omega|^{2}}.
\end{equation} 

\begin{figure}
\begin{center}
\includegraphics[width=6.8cm]{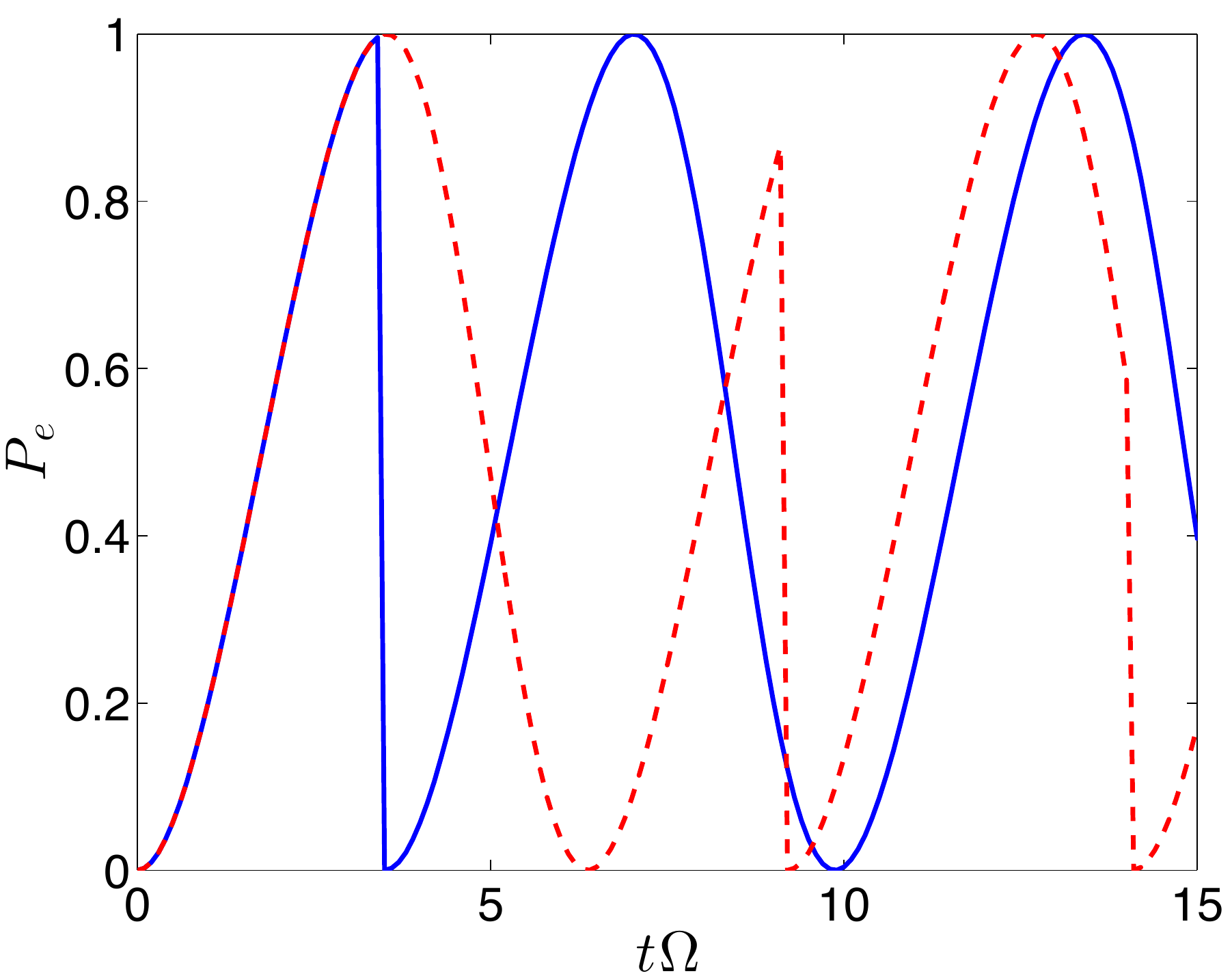}
\includegraphics[width=6.8cm]{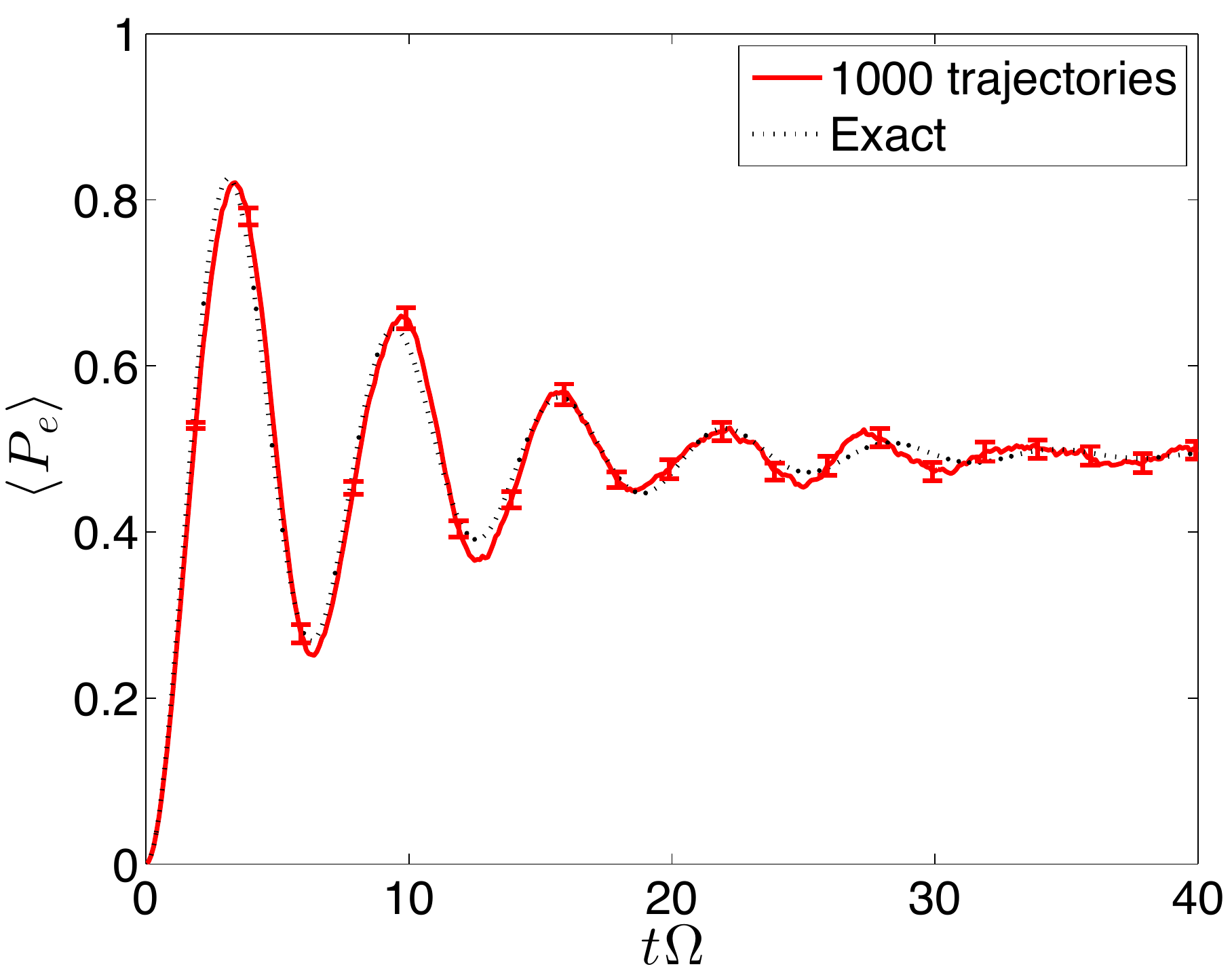}
\end{center}
\caption{Illustrative example of quantum trajectories averaging for a two-level system. (left) Probability of finding the atom in the excited state $P_e$ as a function of time $t\Omega$ for two example trajectories (with blue solid and red dashed lines showing different random samples). We see the effect of quantum jumps, where the atom is projected on the ground state. Here the detuning $\Delta=0$, and $\Gamma=\Omega/6$. (right) Values for $P_e$ averaged over 1000 sample trajectories (Solid line), compared with the exact result from direct integration of the master equation (dotted line). The quantum trajectories results agree with the exact results within the statistical errors, which are shown here as error bars calculated as described in section \ref{sec:staterror}.  } \label{fig:twolevelexample}
\end{figure}

It is straight-forward to formulate the quantum trajectories approach for this master equation, which contains a single jump operator, $c\equiv \sigma_-$, and we can write the corresponding effective Hamiltonian as 
\begin{equation}
H_{\rm eff} = H_{\rm opt} - {\rm i}\frac{\Gamma}{2} \sigma_+\sigma_-.
\end{equation}
The states can be propagated straight-forwardly in time numerically, and we show two example trajectories in the left panel of Fig.~\ref{fig:twolevelexample}. Each of these trajectories exhibits Rabi oscillations, which are interrupted by spontaneous emissions at points in time that are randomly chosen, and thus vary from trajectory to trajectory. After each spontaneous emission, the Rabi oscillations in an individual trajectory reappear with their original amplitude, implying that if we know exactly at which time(s) the atom was reset to its ground state, then we could predict the exact form of the system state at any point in time. In the absence of knowledge of these times, the excited state population damps to its steady-state value, which is marginally below $1/2$ for the case of $\Delta=0$, $\Gamma \ll \Omega$, which we have here. This damping comes from the incoherent averaging over different trajectories, and averaging over 1000 trajectories reliably predicts $P_e$ as a function of time up to statistical errors that are less than a few percent of $P_e$ throughout most of the evolution. This is shown in the right-hand panel of Fig.~\ref{fig:twolevelexample}, where the dotted line represents the exact solution for $P_e$ from the master equation, and the solid line represents the average over trajectories. The statistical error bars shown here are calculated as discussed in section \ref{sec:staterror}, i.e., from the sample of $N_{\rm traj}=1000$ trajectories we compute the population variance of $P_e$, $\Delta P_e^2$, and compute the error as $\sigma_{P_e}=\sqrt{\Delta P_e^2} / \sqrt{N_{\rm traj}}$.

\begin{figure}
\begin{center}
\includegraphics[width=12cm]{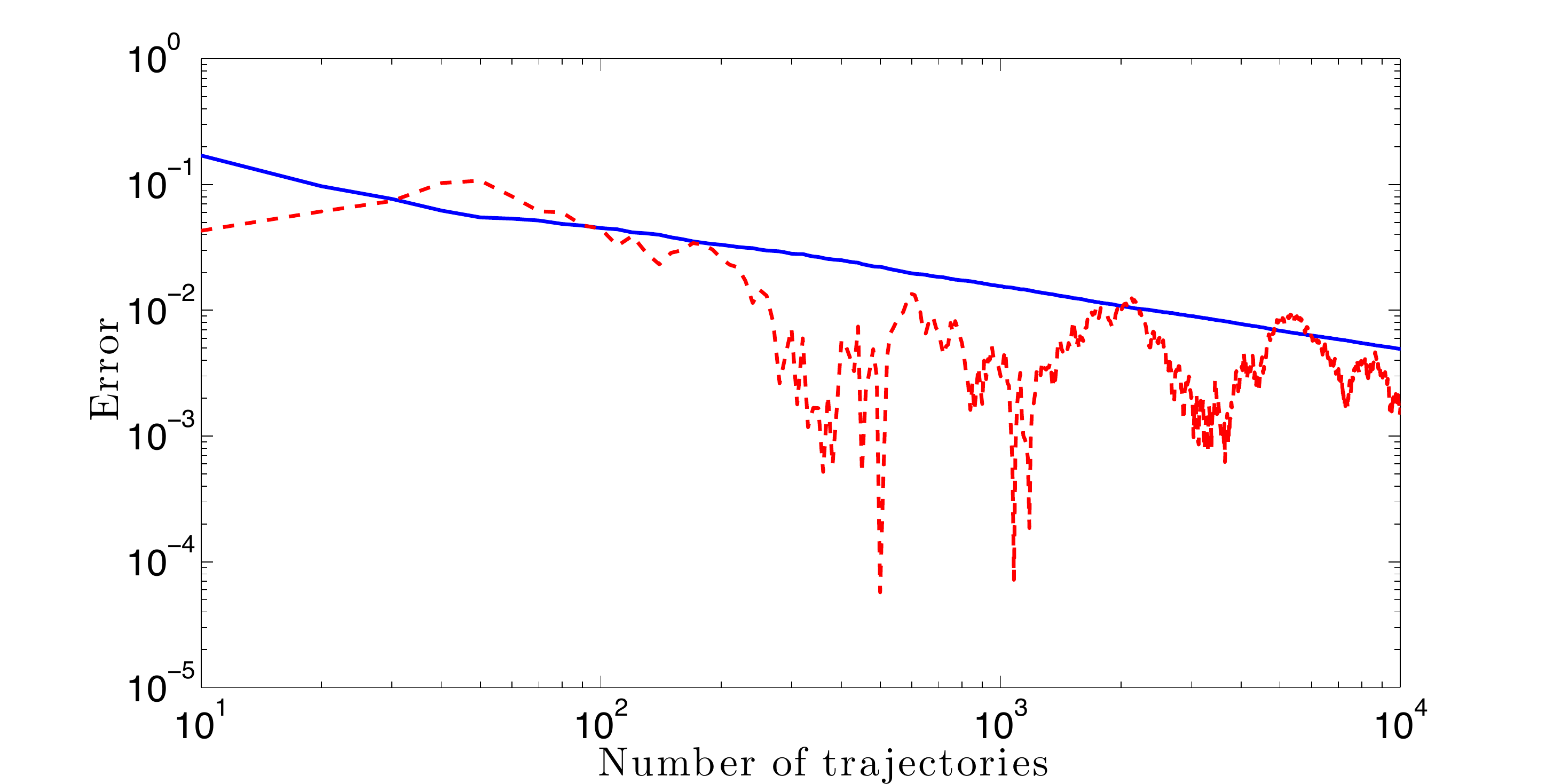}
\end{center}
\caption{Statistical errors in the quantum trajectories computation for a two-level system. Here we compare the statistical error estimate (blue solid line) to the absolute error in the averaged value of $P_e$ when compared with the exactly computed value (red dashed line) at a time $t\Omega=40$. We show these values as a function of the number of trajectories over which we average, with all quantities plotted on a logarithmic scale. As expected, the behaviour of both the statistical error and the discrepancy is somewhat erratic for small numbers of trajectories, but decrease steadily $\propto 1/\sqrt{N}$ for larger numbers. The absolute error is mostly below the statistical error estimate, consistent with a gaussian distribution of possible errors with a standard deviation equal to our statistical error estimate. Note that these values depend on the particular random sample of trajectories obtained in the calculation. From the central limit theorem and the properties of Gaussian distributions that for any given computed values, we expect the absolute error to be smaller than the statistical error as it is quoted here for $68.2\%$ of all possible samples. } \label{fig:twolevelerror}
\end{figure}

In order to look into the statistical errors in more detail, we fix the time, and investigate how the statistical error and the discrepancy between the value of $P_e$ calculated from the trajectory average and the exact value obtained from the master equation vary with $N_{\rm traj}$. This comparison is shown in Fig.~\ref{fig:twolevelerror}, where the solid blue line represents the statistical error, and the dashed red line the discrepancy between the trajectory average and the exactly computed value as a function of the number of trajectories. Each of these lines is shown on a double logarithmic scale to make the scaling $\propto 1/\sqrt{N}$ clearer. The lines shown in the figure represent a particular sample of trajectories, and were chosen so that we simply grew the sample as we increased $N_{\rm traj}$. Naturally, we expect that different samples will produce different plots, but the results shown here constitute a typical example. From the central limit theorem, as described in \ref{sec:staterror}, we expect the means over different samples of trajectories to be approximately normally distributed, and the statistical error represents the standard deviation of that distribution. Thus we expect $68.2\%$ of all possible samples to have discrepancies to exact values that fall below the quoted statistical error. The small ranges of sample sizes where the discrepancy is marginally larger than the estimated statistical errors are perfectly consistent with this analysis.

\subsubsection{Dephasing for hard-core lattice bosons }

\label{sec:hardcorebosons}
\label{sec:hcb}
Our second example focusses on a gas of hard-core bosonic particles moving on a lattice, and introduces dissipation in a simple many-body system. We begin by considering a 1D Hamiltonian of the form  \begin{equation}
H_{\rm bos} = - J \sum_{l} a^\dagger_{l+1} a_l + a^\dagger_{l} a_{l+1},\,\,\, (a_l^\dagger)^2\equiv 0,
\end{equation}
where $a_l$ is a bosonic destruction operator for a particle on lattice site $l$, $J$ is the tunnelling amplitude for a particle moving between sites in a 1D chain, and we have introduced a hard-core constraint, that is, we allow at most one particle on each lattice site. As will be discussed in more detail in sections \ref{sec:numerics} and \ref{sec:amosys} below, bosonic atoms confined to the lowest band of an optical lattice \cite{Jaksch1998,Jaksch2005,Lewenstein2007,Bloch2008b,Lewenstein2012}, the system are described under well-controlled approximations by the Bose-Hubbard model, 
\begin{equation}
H_{BH}=-J\sum_{\langle i,j\rangle}{a}_{i}^{\dag}{a}_{j}+\frac{U}{2}\sum_{i}{a}_{i}^{\dag}{a}_{i}({a}_{i}^{\dag}{a}_{i}-1),\label{HBH}
\end{equation}
where $U$ is the on-site interaction energy shift, and $\langle i,j\rangle$ denotes a sum over nearest-neighbour sites. The model represented by $H_{\rm bos}$ is a limiting case of this when atoms are confined to move along one lattice direction in an optical lattice setup \cite{Jaksch2005,Lewenstein2007,Bloch2008b}, and where the particles are very strongly interacting $U/J\rightarrow \infty$ so that energy conservation strongly disfavours doubly-occupied lattice sites. At less than unit filling of particles in lattice sites, the ground state of $H_{\rm bos}$ is a bosonic superfluid, in which the tails of the momentum distribution decrease algebraically with increasing momentum. The properties of this gas can be derived straight-forwardly via a Jordan-Wigner transformation \cite{Sachdev2011}, which allows the solution to be expressed in terms of the properties of a gas of non-interacting Fermions.

Below, in section \ref{sec:lightscattering}, we will discuss how undergoing spontaneous emissions, i.e., incoherently scattering light from the lattice lasers or another source, tends to localise particles in the system, providing the environment with information about the location of the particles. In the typical limit for optical lattice experiments, this localisation essentially projects a particle onto a single lattice site (although we typically cannot or do not measure which site this was). This localisation in space delocalises the atom in quasi-momentum space across the first Brillouin zone, leading to a broadening of the quasi-momentum distribution and an increase in the kinetic energy beginning from the ground state. For hard-core bosons in 1D, these processes can be approximately described by the master equation
\begin{equation}
\frac{d}{dt} \rho = -{\rm i}[H_{\rm bos},\rho] - \frac{\Gamma}{2} \sum_l \left( a_l^\dagger a_l a_l^\dagger a_l  \rho + \rho  a_l^\dagger a_l a_l^\dagger a_l  -2 a_l^\dagger a_l  \rho a_l^\dagger a_l  \right),
\end{equation}
where $\Gamma$ is the effective rate of spontaneous emission events. 

\begin{figure}
\begin{center}
\includegraphics[width=6.8cm]{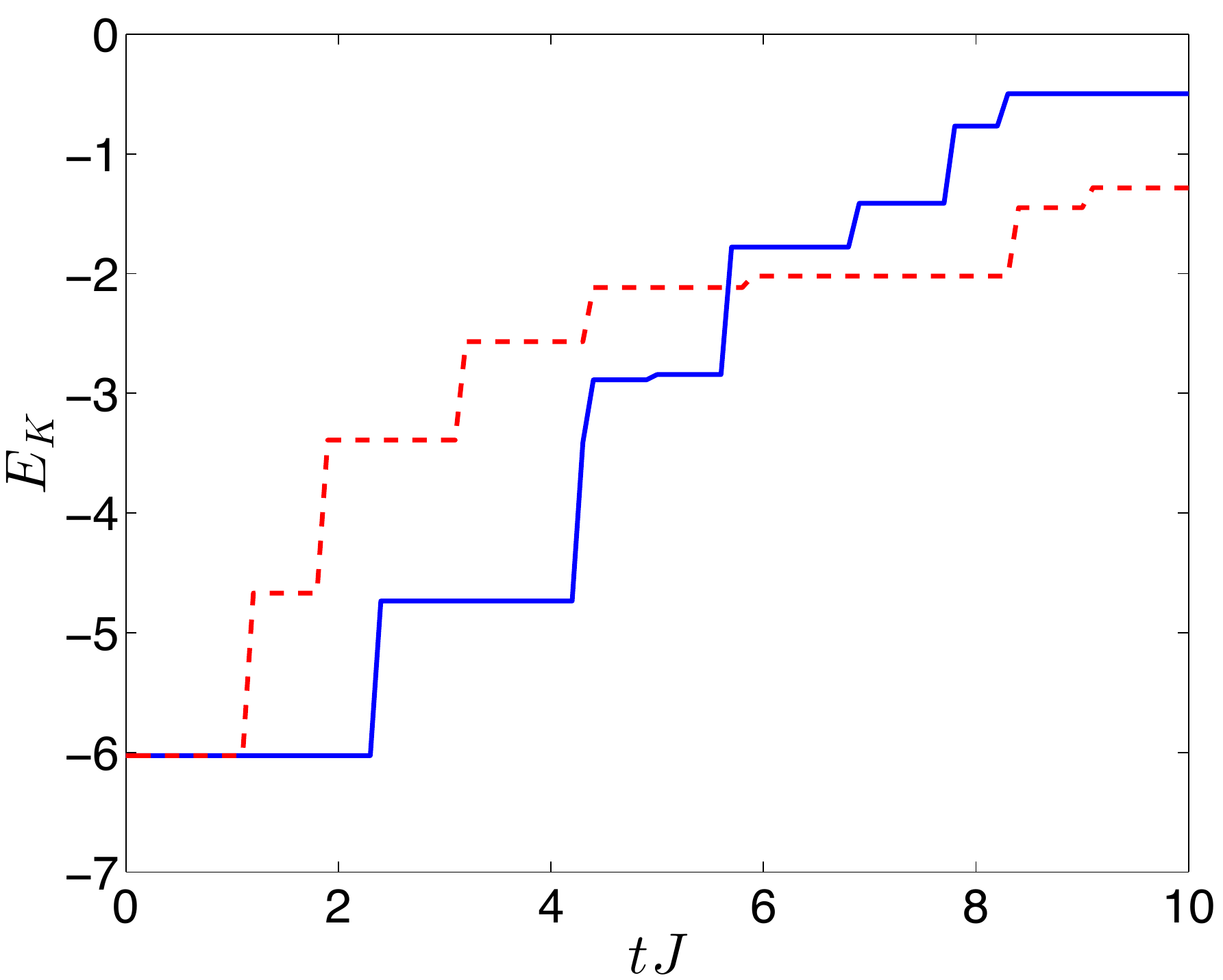}
\includegraphics[width=6.8cm]{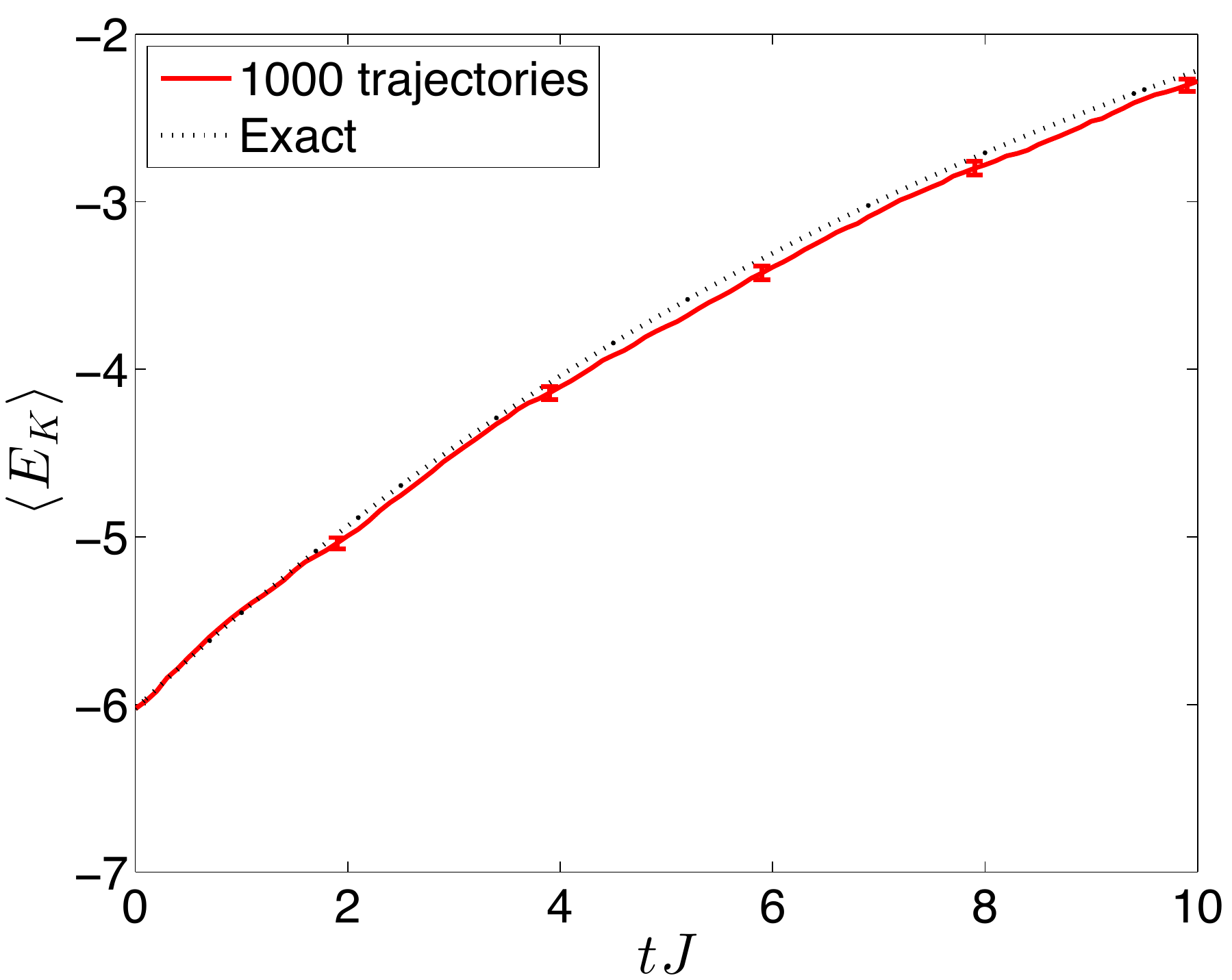}
\end{center}
\caption{Illustrative example of quantum trajectories averaging for a gas of hard-core bosons on a lattice, analogous to the example for a two-level system in Fig.~\ref{fig:twolevelexample}. Here we show results from exact diagonalisation calculations with 5 particles on 10 lattice sites. (left) Kinetic energy of the system of hard-core bosons as a function of time $tJ$ for two example trajectories (with blue solid and red dashed lines showing different random samples). We see the effect of quantum jumps, where the kinetic energy increases as individual atoms are localised in space, and hence spread over the Brillouin zone in quasi-momentum. Here the scattering rate $\Gamma=0.1 J$. (right) Values for the kinetic energy averaged over 1000 sample trajectories, compared with the exact result from eq.~\eqref{hardcorebosonexactke}. As in the case of the two-level system, the quantum trajectories results agree with the exact results within the statistical errors, which are shown here as error bars calculated as described in section \ref{sec:staterror}. } \label{fig:hardcoreexample}
\end{figure}

As in the case of the optical Bloch equations, it is straight-forward to formulate the quantum trajectories approach for this master equation. Now we have a set of jump operators, with as many jump operators as we have lattice sites in the system and $c_m\equiv a_m^\dagger a_m$, i.e., the on-site number operator, and a corresponding effective Hamiltonian
\begin{equation}
H_{\rm eff} = H_{\rm bos} - {\rm i}\frac{\Gamma}{2} \sum_m a_m^\dagger a_m a_m^\dagger a_m.
\end{equation}
Intuitively, this form for the jump operators makes sense, as application of a number operator on a given site, $a_l^\dagger a_l |\psi\rangle/\Vert a_l^\dagger a_l |\psi\rangle \Vert $, leads to the localisation of a single particle on site $l$. The probability that this occurs on a particular site $l$ is proportional to the expectation value $\langle (a_l^\dagger a_l )^2 \rangle$, which reflects the fact that a site that is unoccupied will not give rise to jumps, and for particle numbers greater than one, the rate is enhanced by superradiance \cite{Lehmberg1970,Lehmberg1970a}. We will see below that this enhancement is related to the excitation of a collection of atoms in one site being symmetric. 

The localisation of an atom on a single site corresponds to a spreading of the localised particle over all of the possible states in quasimomentum space, and hence to an increase in kinetic energy. This can be seen in Fig.~\ref{fig:hardcoreexample}, where we propagate states in time numerically using quantum trajectories. As for the previous case of the optical Bloch equations, we show two example trajectories as well as a trajectory average, here for the kinetic energy in the system as a function of time. By direct computation of expectation values from the master equation, we see that  
\begin{equation}
\frac{d}{dt}E_{\rm bos} \equiv \frac{d}{dt}\langle H_{\rm bos} \rangle = - \Gamma \langle H_{\rm bos} \rangle , \label{hardcorebosonexactke}
\end{equation}
so that $E_{\rm bos}(t) = E_{\rm bos}(t=0) \exp (-\Gamma t)$. In the right hand panel of Fig.~\ref{fig:hardcoreexample} we compare this analytical result to the numerical value as a function of time, and observe very good agreement to within the estimated statistical error. 

\begin{figure}
\begin{center}
\includegraphics[width=13cm]{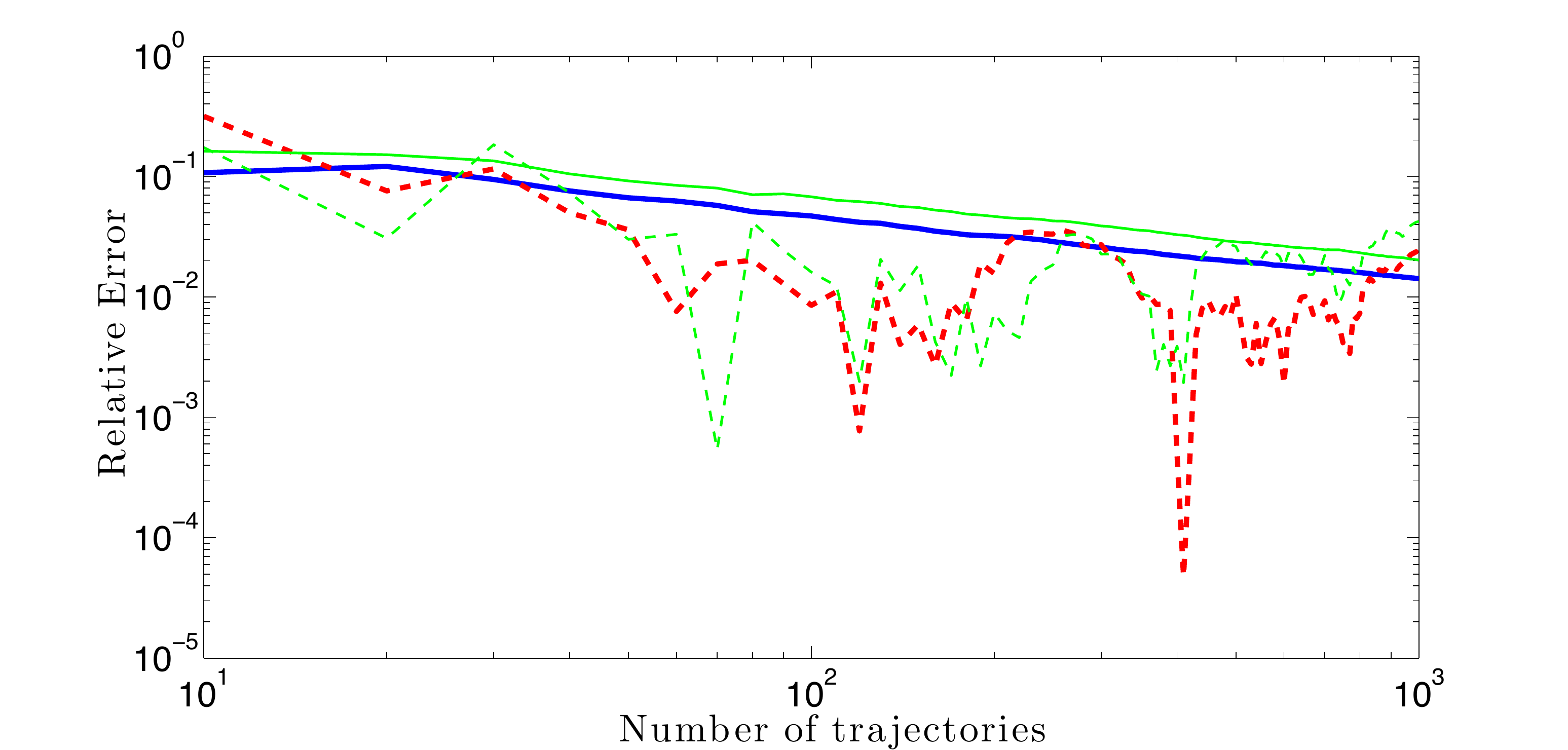}
\end{center}
\caption{Statistical errors in the quantum trajectories computation for hard-core bosons, taken from exact diagonalisation calculations with 5 particles on 10 lattice sites. Analogously to the two-level system case, we compare the statistical error estimate for our estimate of the energy $E_{\rm bos}$ (dark blue solid line) to the absolute error in the averaged value of $E_{\rm bos}$ when compared with the exactly computed value (dark red dashed line) at a time $tJ=8$. We also show the estimate of the error in a local correlation function $\langle (a^\dagger_5 a_6+a^\dagger_6a_5) \rangle$ (light green solid line) compared with the discrepancy between this quantity and the exactly value (light green dashed line) at the same point in time. We show each of these these values as a function of the number of trajectories over which we average, with all quantities plotted on a logarithmic scale. The statistical errors again decrease steadily $\propto 1/\sqrt{N}$ for larger numbers, but the relative error in the local quantity remains larger than the global quantity. } \label{fig:hardcoreerror}
\end{figure}

We investigate this statistical error as a function of the number of trajectories in Fig.~\ref{fig:hardcoreerror}. We fix the time at $tJ=8$, and plot the statistical error and the discrepancy between the value of $E_{\rm bos}$ calculated from the trajectory average and the exact value obtained from the master equation as a function of $N_{\rm traj}$. The solid blue line represents the statistical error, and the dashed red line the discrepancy between the trajectory average and the exactly computed value. From the double logarithmic scale we see the scaling $\propto 1/\sqrt{N}$. As in the example of the optical Bloch equations, we expect the means over different samples of trajectories to be approximately normally distributed, and therefore $68.2\%$ of all possible samples should have discrepancies to exact values that fall below the estimated statistical error. The values shown here are again consistent with that analysis. In addition to this data, we also show using lighter (green) solid and dashed lines the same analysis for a local correlation function $\langle (a^\dagger_5 a_6+a^\dagger_6 a_5) \rangle$. We see that the relative error is higher than for the globally averaged value, but the behaviour of the local correlation function with the number of trajectories is very similar. 

\begin{figure}
\begin{center}
\includegraphics[width=6.67cm]{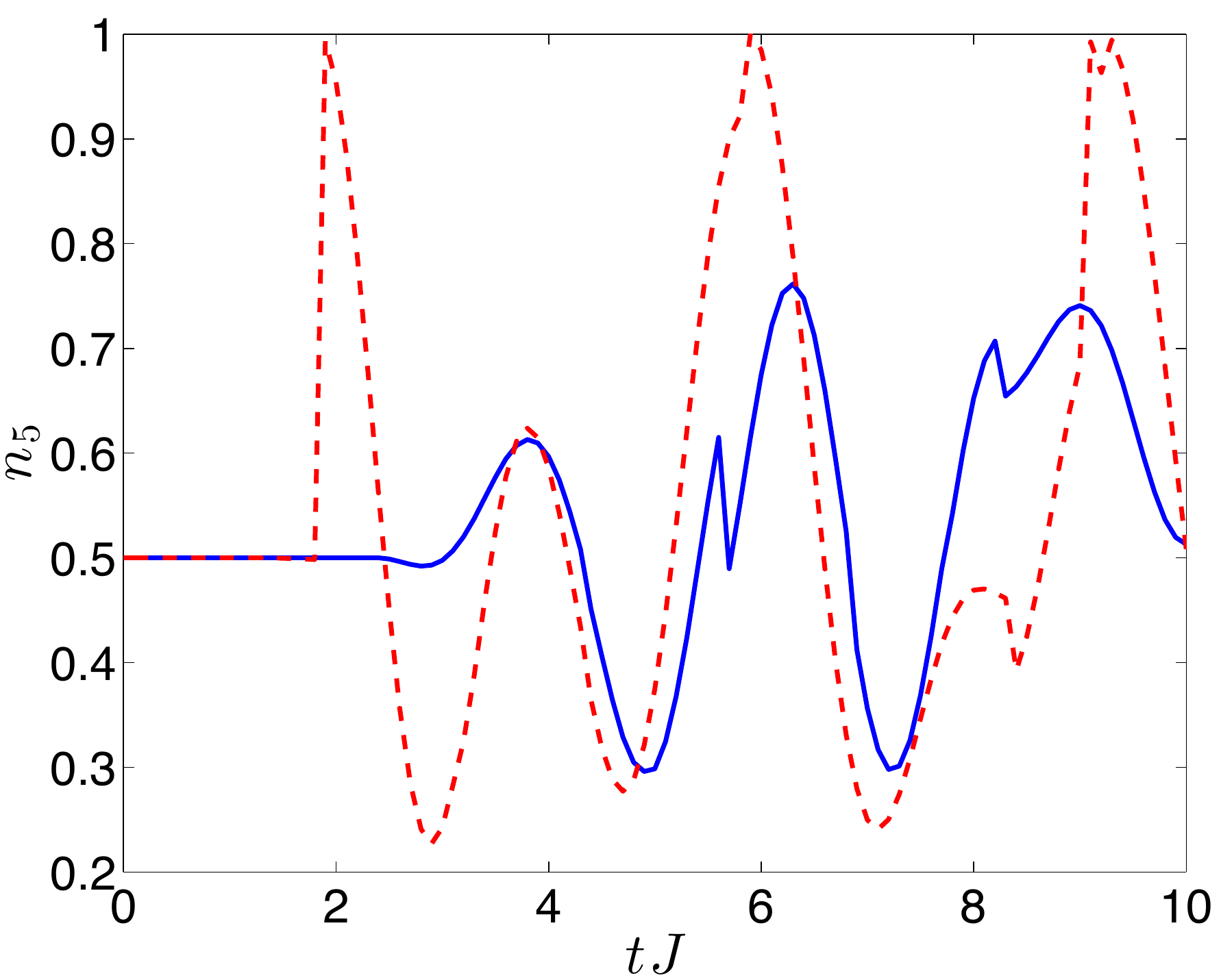}
\includegraphics[width=6.93cm]{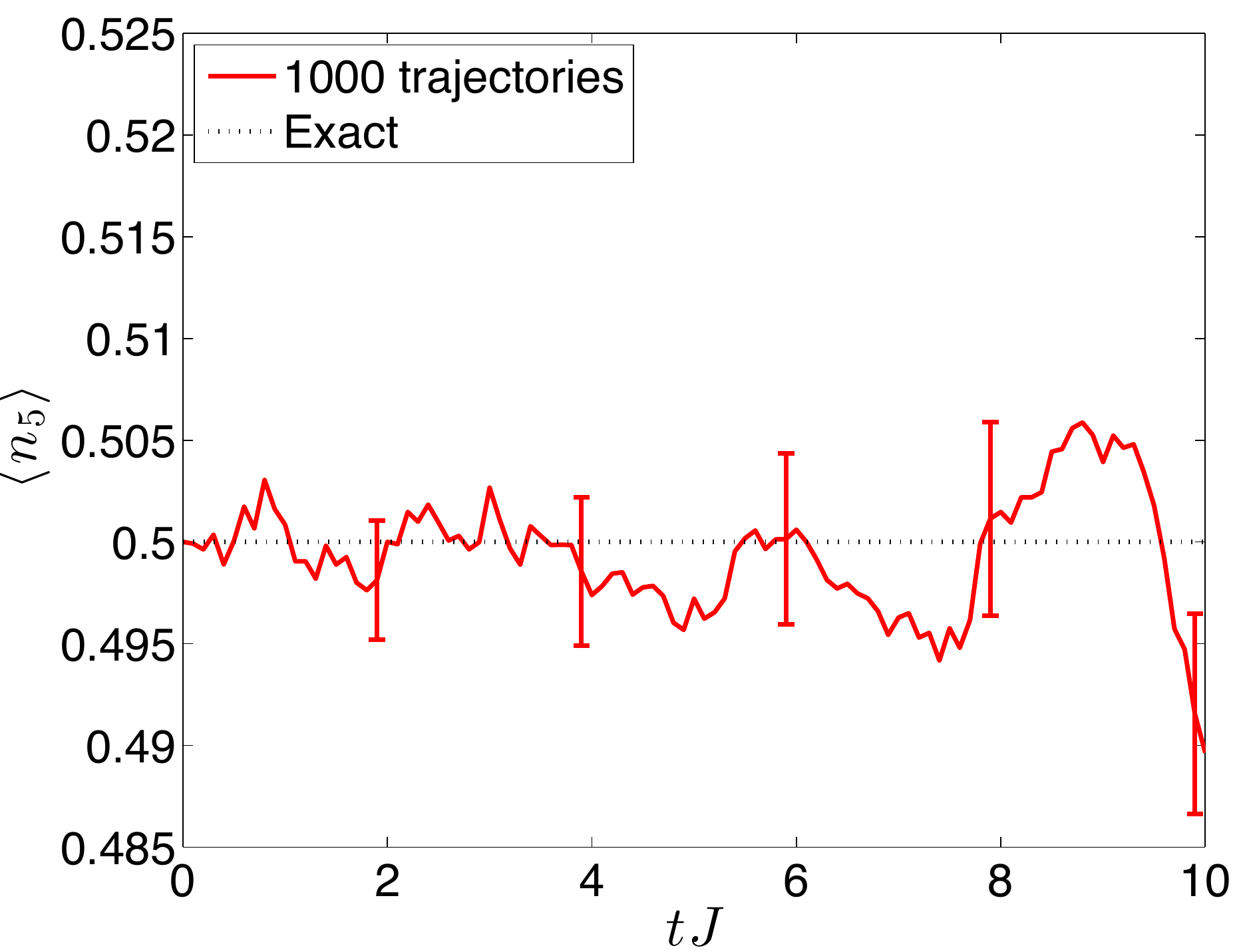}
\end{center}
\caption{Illustrative example of quantum trajectories averaging for a gas of hard-core bosons on a lattice, showing results from exact diagonalisation calculations with 5 particles on 10 lattice sites, as in Fig.~\ref{fig:hardcoreexample}. (left) Density on site 5 of the lattice as a function of time $tJ$ for two example trajectories (with blue solid and red dashed lines showing different random samples). We see the effect of quantum jumps, where nearby jumps give rise to fluctuations either up or down of the local mean density. (right) Values for $\langle n_5\rangle$  averaged over 1000 sample trajectories, and compared with the exact result, which is $\langle n_5 \rangle\equiv \langle a_5^\dagger a_5 \rangle=0.5$, because the equations of motion are homogeneous. We see that the deviation from this value is well approximated by the statistical errors, shown here as error bars calculated as described in section \ref{sec:staterror}. } \label{fig:hardcoredensity}
\end{figure}

In Fig.~\ref{fig:hardcoredensity} we show additional data for another local correlation function, namely the on-site density. This is quite instructive, because it shows how individual trajectories need not exactly enforce symmetries that are present in the master equation as a whole. Specifically, for a system with periodic boundary conditions and an initial density that is uniform across the system, we expect that
\begin{equation}
\frac{d}{dt}\langle a^\dagger_l a_l \rangle=0.
\end{equation}
However, because the individual jumps are local, the spatial invariance in the expectation value of the density is not followed by individual trajectories, as we see clearly in the left panel of Fig.~\ref{fig:hardcoredensity}. This reflects an analogous lack of spatial invariance that might occur in specific runs of an experiment if spontaneous emission events are taking place locally (see appendix \ref{sec:continuousmeasurement} for an interpretation of this in terms of physical processes occurring on individual lattice sites). In the right panel of Fig.~\ref{fig:hardcoredensity}, we see that the expectation value of the density remains constant to within the statistical errors, as is expected. 

In Sec.~\ref{sec:lightscattering} we return to this problem of spontaneous emissions, outlining the derivation of this master equation for many bosons in an optical lattice, also away from the approximations that are made here. We also explain the additional dynamics (including transfer of particles to higher Bloch bands) that are exhibited in that case, and we explain the limits in which the simplified master equation treated here arises.

\section{Integration of quantum trajectories with many-body numerical methods}
\label{sec:numerics}

Over the last few years, several groups have begun applying these techniques to
 describe dissipative dynamics of many-body systems as they arise in AMO systems. A number of examples of such master equations will be given in section \ref{sec:amosys}, beginning with a general version of the light scattering master equation discussed in section \ref{sec:hardcorebosons}. In order to facilitate the solution of the corresponding master equations, quantum trajectories techniques have been combined with many-body methods. In this section we give an overview of what has been done so far, beginning with exact diagonalisation and mean-field methods, and then describing in detail the integration of quantum trajectories with time-dependent density matrix renormalisation group (t-DMRG) techniques. 
 
\subsection{Integration with exact diagonalisation}
\label{sec:ed}

For systems that are not too large, we can store the many-body states exactly, and apply quantum trajectories methods in much the same way as they would be applied to single-particle systems. We only need a method to propagate a state under the effective Hamiltonian $H_{\rm eff}$, and apply jump operators. The main difference in applying this method to many-body systems, as opposed to single-particle systems, is that quantum trajectories can be an efficient way to compute quantities that are local in space or momentum in these systems, because the Hilbert space is large compared with the system size. This was discussed in more detail in section \ref{sec:globallocal} above. In most quantum trajectories involving t-DMRG techniques, the dynamics are first studied for a small system that is tractable to exact diagonalisation (see, e.g., \cite{Daley2009,Kantian2009}). In the case of the Bose-Hubbard model eq.~\eqref{HBH}, such systems are of the order of 10 particles on 10 lattice sites for rapid computations, although recent calculations with the Bose-Hubbard model (without quantum trajectories) have extended to 14 particles on 14 sites \cite{Lux2013,Bauer2011}.

One important optimisation that quantum trajectories can allow involves conserved symmetries of the effective Hamiltonian. Specifically, if the Hamiltonian commutes with a unitary transformation $\hat T$, $[H_{\rm eff},\hat T]=0$, it is usually possible to optimise a calculation by working in one symmetry sector of $\hat T$, reducing the number of basis states that must be used for the Hilbert space. In the case that $[c_m,\hat T]\neq 0$, this is not typically the case for a full master equation simulation. However, in a trajectories calculation, if $[c_m , \hat T] \propto c_m$, it is still possible to make use of the symmetry in reducing the relevant basis size for the Hilbert space, because the application of a jump simply switches the state from one symmetry sector to another. A simple example is the case of particle loss in a many-body system \cite{Daley2009} (see section \ref{amosys} below). Often, the effective Hamiltonian commutes with the total particle number operator, and we can work in a basis with fixed particle number. Loss events reduce this number by a well controlled amount, allowing a different number conserving basis to be used for propagation after the jump \cite{Daley2009}. 

\subsection{Integration with the time-dependent Gutzwiller ansatz}
\label{sec:gutzwiller}

This discussion applies equally to mean-field methods that make it possible to rewrite the dynamics so that it is exactly computable. An important example of this for bosons on a lattice is the possibility to perform time-dependent calculations with a Gutzwiller ansatz \cite{Snoek2007,Wernsdorfer2010,Zakrzewski2005,Jaksch1998}. In calculating ground states of the Bose-Hubbard Hamiltonian, this ansatz takes the form of a product state over different lattice sites,
\begin{equation}
\ket{\psi_{G}}=\prod_{l}\ket{\psi_G^{l}}_{[l]}=\prod_l\sum_{n}f_n^{(l)}\ket{n}_{[l]}. \label{eqgutzwiller}
\end{equation}
Here, $\ket{n}_{[l]}$ is a state of $n$ particles locally on site $l$, and $f_n^{(l)}$ can be used as variational parameters to minimise the energy. For the Bose-Hubbard model $H_{BH}$, this is equivalent to using a mean-field approximation on the operators $a_l$ in the tunnelling part of the Hamiltonian, $a_i^\dagger a_j \rightarrow a_i^\dagger \langle a_j \rangle + \langle a_i^\dagger \rangle a_j - \langle a_i^\dagger \rangle \langle a_j \rangle$, which leads to a Gutzwiller mean-field Hamiltonian that can be expressed as a sum over terms involving only local operators on site $l$,
\begin{equation}
H_{MF}=\sum_l\left[-J\left({\psi_l a_i^{\dag}+\psi_l^{\ast}a_l}\right)+\frac{1}{2} U(a_l^{\dag})^2a_l^2\right].
\end{equation}
Here, $\psi_l\equiv \sum_{j|l}\langle \psi_G^j |a_j|\psi_G^j\rangle$ involves a sum over the sites $j$ that are immediate neighbours of site $l$, and in variational calculations these parameters must be found self-consistently. As a method of determining the ground state, this method faithfully reproduces states for weak interactions \cite{Rokhsar1991}, and is exact in the limit of infinite dimensions\cite{Byczuk2008,Hubener2009,Anders2011}.  In this form, the superfluid phase of the Bose-Hubbard model, which is characterised by off-diagonal long-range order \cite{Sachdev2011}, is represented by a nonzero value of $\psi_l$, whereas for the Mott Insulator state, $\psi_l=0$. For unit filling on the lattice, this transition from the superfluid phase to the Mott Insulator phase in this ansatz occurs at $U/(zJ)\approx5.8$ \cite{Sachdev2011}, where $z$ is the number of neighbouring sites. This value is accurate for a 3D cubic systems to about 15\%, when compared with results from Quantum Monte Carlo calculations \cite{Pollet2013}.

These methods can be directly generalised to time-dependent calculations \cite{Jaksch1998}, and have been seen to give reasonable qualitative results in a number of contexts \cite{Snoek2007,Wernsdorfer2010,Zakrzewski2005}. Because these calculations involve direct evolution of on-site wavefunctions under the mean-field Hamiltonian, they are efficient to compute numerically. Care must be taken, as there are some cases where the method clearly fails - for example, no dynamics can be generated when the initial state is a Mott Insulator state, as in this case the terms in $H_{MF}$ that couple sites are exactly zero. This also leads to artefacts and unphysical slowing down in studying time-dependent transitions from the superfluid to Mott Insulator phase. However, these methods can be very useful in developing understanding of systems in a tractable manner.

Because dynamics are efficiently computable, these methods can be used directly together with quantum trajectory techniques. The averages are then performed over an ensemble of stochastically evolved states, where each state of the ensemble has the Gutzwiller ansatz form of eq.~\eqref{eqgutzwiller}. 

An alternative Gutzwiller ansatz form for solving a master equation was used by Diehl et al. in Ref.~\cite{Diehl2010c}. This involved a factorisation of the system density operator, 
\begin{equation}
\rho_G=\bigotimes_i\rho_i\;\; \rho_l={\rm Tr}_{i\neq l}\{\rho\}
\end{equation}
 which gives rise to a nonlinear set of coupled master equations, each for one site $l$. This ansatz is particularly useful in certain cases where it can be treated analytically \cite{Diehl2010c}, and can also be useful as a numerical method to obtain certain qualitative effects (see, e.g., \cite{Pichler2010}, where this method was used to study transfer of particles to higher bands of an optical lattice in spontaneous emission events).
 
It is important to note that where sampling an ensemble of trajectories in the Gutzwiller ansatz form $|\psi_G\rangle$ can actually contain somewhat more information than $\rho_G$. While neither ansatz can represent quantum entanglement between different sites, by making a product state of local density operators, $\rho_G$ also does not capture the development of non-trivial classical correlations between different parts of the system. However, classical correlations can be captured by an ensemble of trajectories in the form $|\psi_G\rangle$ (each trajectory has no classical correlations, but the ensemble of trajectories can represent these). This can be important not only in treating dissipative dynamics with quantum mechanical reservoirs, but also when sampling over classical noise within such an ansatz, as was observed in Ref.~\cite{Pichler2013}.

\subsection{Time-dependent density matrix renormalization group methods}

An important development in treating dissipative dynamics of many-body systems has been the advent of time-dependent methods for dealing with 1D many-body systems. We will briefly summarise the key features of these methods in this section, and then discuss their integration with quantum trajectories methods\footnote{For a more detailed review of these methods, see, e.g., Ref.~\cite{Schollwock2011}.}. 

Over the last ten years, a range of methods based on matrix product states \cite{Schollwock2011} have been developed, which effectively build on the success of density matrix renormalization group (DMRG) methods \cite{Schollwock2005,White1992} as the most powerful numerical method for generic lattice models in one dimension. Early development of time-dependent methods has included the time-evolving block decimation (TEBD) algorithm \cite{Vidal2003,Vidal2004}, and its integration with DMRG to produce an adaptive time-dependent DMRG (t-DMRG) algorithm \cite{White2004,Daley2004}. Similar methods based on matrix product operators have been developed \cite{McCulloch2007,Verstraete2008,Crosswhite2008,Pirvu2010,Verstraete2004}, which allow for both the study of time-dependent dynamics with long-range interactions \cite{Crosswhite2008}, and the direct study of dissipative dynamics described by a master equation, with matrix product states used to represent a density matrix \cite{Verstraete2004,Zwolak2004}. Such density matrix representations can be used to represent finite-temperature states \cite{Verstraete2004,Verstraete2008}, though finite temperature dynamics alone (without dissipative processes) can also be represented by use of ancilla states \cite{Feiguin2005}, or by sampling minimally entangled states \cite{Stoudenmire2010}. Generalisations have been proposed to higher dimensions \cite{Verstraete2004b,Vidal2008}, although these are typically very computationally intensive.

\begin{figure}
\begin{center}
\includegraphics[width=\textwidth]{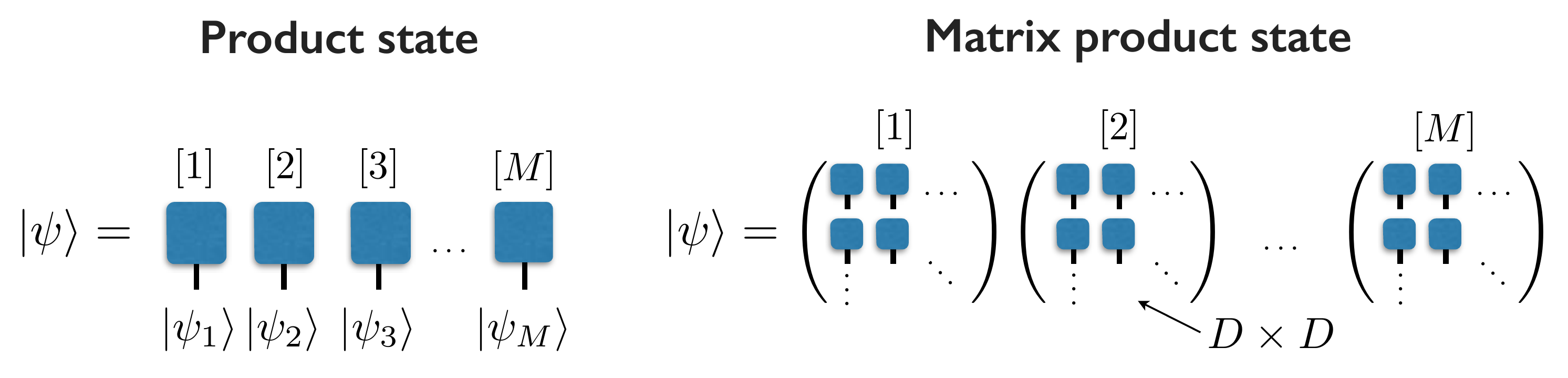}
\end{center}
\caption{Diagram comparing a product state and a matrix product state. Each square represents a state $\ket{\psi_l}_{[l]}$ in a local Hilbert space $\mathcal{H}_l$, e.g., a single spin or lattice site, and the vertical leg on each square box indicates an index for the states of the local Hilbert space. The Hilbert space representing the whole system is the tensor product $\mathcal{H}=\mathcal{H}_1 \otimes \mathcal{H}_2 \otimes \ldots  \otimes \mathcal{H}_L$. A product state can be written as $\ket{\psi_1}_{[1]}\otimes \ket{\psi_2}_{[2]} \otimes \ldots \otimes \ket{\psi_L}_{[L]}$, and represents a quantum state with no entanglement between any two local Hilbert spaces. By multiplying out $D\times D$ matrices, a matrix product state makes it possible to represent superpositions of product states, and this superposition contains entanglement between the local Hilbert spaces. For sufficiently large $D$, any state can be represented in this way, and the ground states of large classes of local 1D many-body Hamiltonians can be represented faithfully for small $D$ in this fashion. This is the basis of the time-dependent density matrix renormalization group method, as discussed in the main text. The maximum amount of entanglement we can represent across any bipartite splitting in the system is bounded in terms of the von Neumann entropy by $S_{\rm vN}\leq \log_2 D$. Note that while each square box here is a similar notation to that used in Ref.~\cite{Schollwock2011} and similar references, in the notation of those references, contractions between matrices are represented by connecting boxes with horizontal lines.} \label{fig:mps}
\end{figure}

Matrix product state methods can be applied to any system for which we can write the Hilbert space as a product of local Hilbert spaces, $\mathcal{H}=\mathcal{H}_1\otimes \mathcal{H}_2\otimes \cdots \otimes \mathcal{H}_L$. This works well for spin chains, where each local Hilbert space $\mathcal{H}_i$ corresponds to the states of a single spin, as well as for bosons or fermions on a lattice, where each local Hilbert space corresponds to the different possible occupations of particles on a single lattice site. If we write the $d$ basis states of the local space $\mathcal{H}_l$ as $|i_l\rangle_{[l]}$, $i_l=1\ldots d$, then any state $\ket{\psi}$ of the complete system can be expressed in the form  
\begin{equation}
|\psi\rangle=\sum_{i_1,i_2,\ldots,i_L = 1}^d c_{i_1i_2\ldots i_L} |i_1\rangle_{[1]} \otimes |i_2\rangle_{[2]} \otimes \ldots.
\otimes |i_L\rangle_{[L]}. \label{mpssystemstate}
\end{equation}
The key to matrix product state methods is a decomposition of the state in which we write $c_{i_1i_2\ldots i_L}$ from this equation as the multiplication of a series of matrices $A_l [i_l]$,
\begin{equation}
c_{i_1i_2\ldots i_L}=A_1 [i_1] \, A_2[i_2] \ldots A_L [i_L] \label{mpseq},
\end{equation}
where the boundary matrices are assumed to be one-dimensional in order to produce coefficients from the matrix multiplication. 

This state is represented in the diagram of Fig.~\ref{fig:mps}, where we visualise the basic concept of a matrix product state. If the matrices were just complex coefficients (i.e., matrices of dimension $D=1$), then this state would be precisely the same as the Gutzwiller ansatz state eq.~\eqref{eqgutzwiller} described in the previous section. It would be a product state, with a simple physical interpretation, and no entanglement between any of the local Hilbert spaces. However, by using matrices of dimension $D\times D$, where $D>1$, we can produce a superposition of different product states, in so doing, allowing entanglement between different local Hilbert spaces. For sufficiently large $D$, any state can be represented in this form, but in general $D$ must grow exponentially with the system size in order to represent an arbitrary state of the Hilbert space $D\propto \exp(L)$. In fact, to represent an arbitrary state of the form given in eq.~\eqref{mpssystemstate}, and $L$ is even, then we must have matrices at the centre of the system of dimension $D=d^{L/2}$.

The key to the success of these methods comes in the understanding of entanglement in many-body systems, which has developed extensively in the past few years \cite{Eisert2010,Amico2008,Calabrese2009,Calabrese2009b,Peschel2009,Cardy2010}. For a system where the total state of the system is a pure state $\psi$, the entanglement between two parts $A$ and $B$ can be quantified by computing the von Neumann entropy $S_{\rm vN}[\rho_A]=-{\rm Tr}[\rho_A \log_2 \rho_A]$ of the reduced density operator for one part of the system $\rho_A={\rm Tr}_B [|\psi\rangle\langle \psi |]$. If the two parts are in a product state, then the reduced density operator $\rho_A$ will represent a pure state, and we have $S_{\rm vN}[\rho_A]=0$. However, if the parts of the system are entangled, $\rho_A$ will represent a mixed state, and its entropy gives the quantification of the amount of entanglement between $A$ and $B$. 

In a matrix product state with matrix dimension $D$, the maximum entanglement across any possible bipartite splitting in the system that can be represented by this state is $S_{\rm vN}=\log_2 (D)$. However, if we take a bipartite splitting of a gapped lattice model or spin chain in 1D, then we expect $S_{\rm vN}$ to be bounded as a function of the size of each part of the system $A$ and $B$, even as we extrapolate to unbounded block size in an infinite-sized system \cite{Eisert2010,Amico2008,Calabrese2009}. If, on the other hand, we have a 1D system that is critical, then we expect $S_{\rm vN}\propto \log(L_s)$, where $L_s$ is the size of the smaller block, $A$ or $B$ \cite{Eisert2010,Calabrese2009}. In either case, the entanglement is in some sense weak, and for finite-size systems it is reasonable to expect that ``slightly-entangled'' matrix product states will represent the states well. While the scaling of the von Neumann entropy is a practical guide, and in fact the scaling of R\'{e}nyi entropies $S_R\equiv (1-R)^{-1} {\rm Tr} [\rho^R]$ of order $R<1$ should be considered in order to prove that a state can be represented in this form \cite{Schuch2008,Schuch2008b}, it is found that these methods work very well under a wide range of circumstances. In fact, it has been shown \cite{Verstraete2006} that ground states of 1D spin systems are expected to be represented faithfully in this form, even at criticality, and the success of DMRG as an essentially exact method for the computation of ground states of 1D systems is practical confirmation of this \cite{Schollwock2011,Schollwock2005}. 

Time-dependent methods can be readily developed, as it is straight-forward to apply either local operators or even long-range operators that are representable in matrix operator form to these states. In particular, applying an operator to two neighbouring sites can be done while only updating the matrices associated with those sites. Then, a simple method to compute time evolution for a local Hamiltonian $\hat H=\sum_{l} \hat H_{l,l+1}$ is to take a Trotter decomposition of the time evolution operator,
\begin{equation}
{\rm e}^{-{\rm i} \hat H (2\delta t) /\hbar}={\rm e}^{-{\rm i} \sum_l \hat H_{l, l+1} \delta t /\hbar}\approx \prod_{l=1}^{L-1} {\rm e}^{-{\rm i} \hat H_{l,l+1} \delta t /\hbar} \prod_{l=L-1}^{1}{\rm e}^{-{\rm i} \hat H_{l,l+1} \delta t /\hbar}+\mathcal{O}(\delta t^3),\label{trotter}
\end{equation}
where each of these operators can be applied to the state by updating only the matrices associated with those sites. This can be straight-forwardly generalised to higher-order Trotter decompositions \cite{Sornborger1999}, and a variety of other methods for efficiently propagating the state, especially Arnoldi methods \cite{Garcia-Ripoll2006,McCulloch2007} are also available. In each case, the numerical cost of propagating the state scales as $D^3$.

\begin{figure}
\begin{center}
\includegraphics[width=7cm]{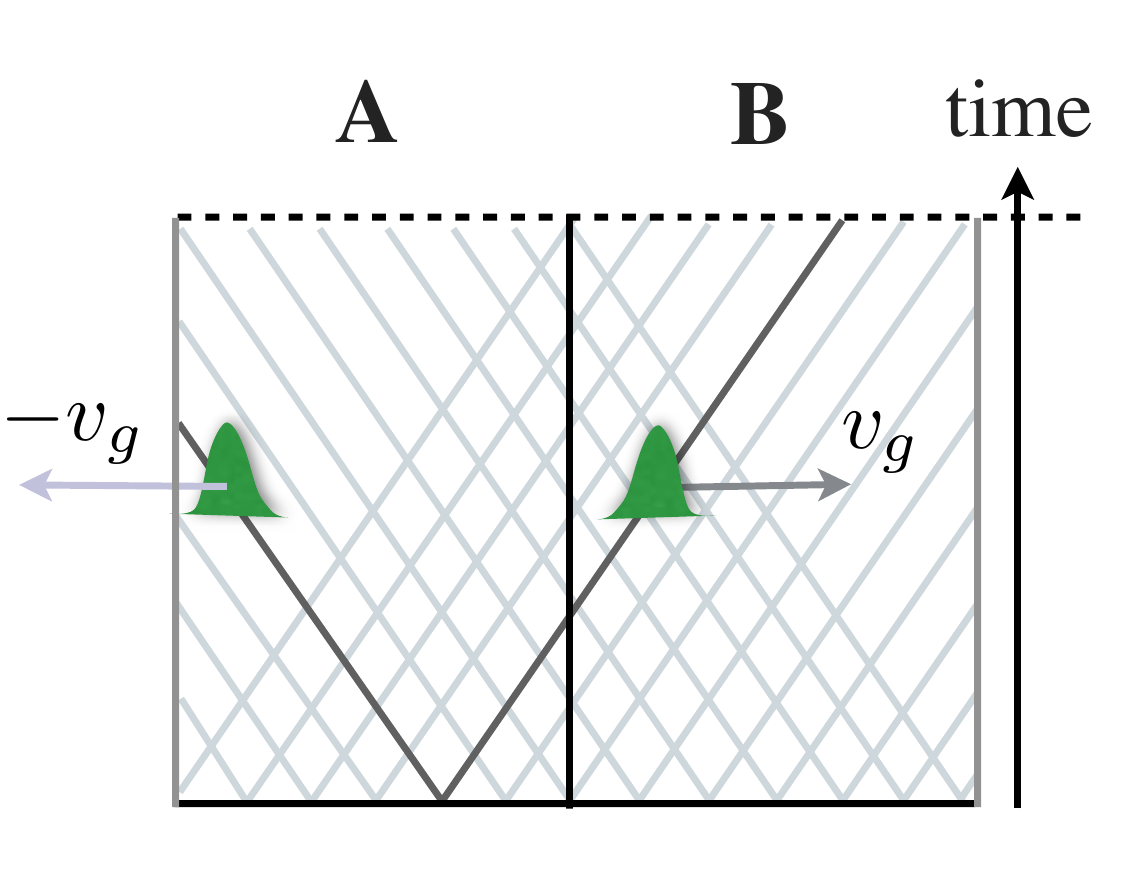}
\end{center}
\caption{After a local or global quench in a 1D system with short-range interactions, quasiparticle excitations are produced that propagate through the system. The rate of information propagation in the system is restricted by the Lieb-Robinson velocity. When we consider a bipartite splitting of the system (here into parts A and B), propagation of quasiparticles across the boundary leads to an increase in spatial entanglement between parts A and B of the system.} \label{fig:quasiparticlediagram}
\end{figure}

For how long a state can be accurately propagated in time using these methods depends on the rate of growth of entanglement during the time-evolution. For cases where the evolution is somewhat close to equilibrium, dynamics on very long timescales can be computed. For example, near-adiabatic state preparation has been studied in Hubbard models on timescales longer than $2000 J^{-1}$ \cite{Kantian2010}. However, general time evolution under a Hamiltonian for which the matrix product state is not a near-equilibrium state can only be computed over short timescales \cite{Gobert2005,Prosen2007,Schuch2008,Schuch2008b,Trotzky2012}. This is directly related to the growth of entanglement after a quantum quench \cite{Calabrese2005,Prosen2007,Lauchli2008,Unanyan2010,Gobert2005}. A simple picture for how this works, which can be clearly seen in the case of nearest-neighbour transverse Ising model \cite{Calabrese2005} is depicted in Fig.~\ref{fig:quasiparticlediagram}. There, we show the propagation in time of quasiparticle excitations, which carry information at a rate that is limited by the Lieb-Robinson velocity \cite{Lieb1972}. Propagation of quasiparticles across a boundary between two parts of the system leads to an increase in spatial entanglement between those parts of the system. In the generic case \cite{Calabrese2005,Schuch2008,Schuch2008b}, this leads to a linear growth in the entanglement entropy as a function of time. This has been investigated in a variety of different models, including the Bose-Hubbard model \cite{Lauchli2008,Daley2012}. 

\begin{figure}
\begin{center}
\includegraphics[width=8cm]{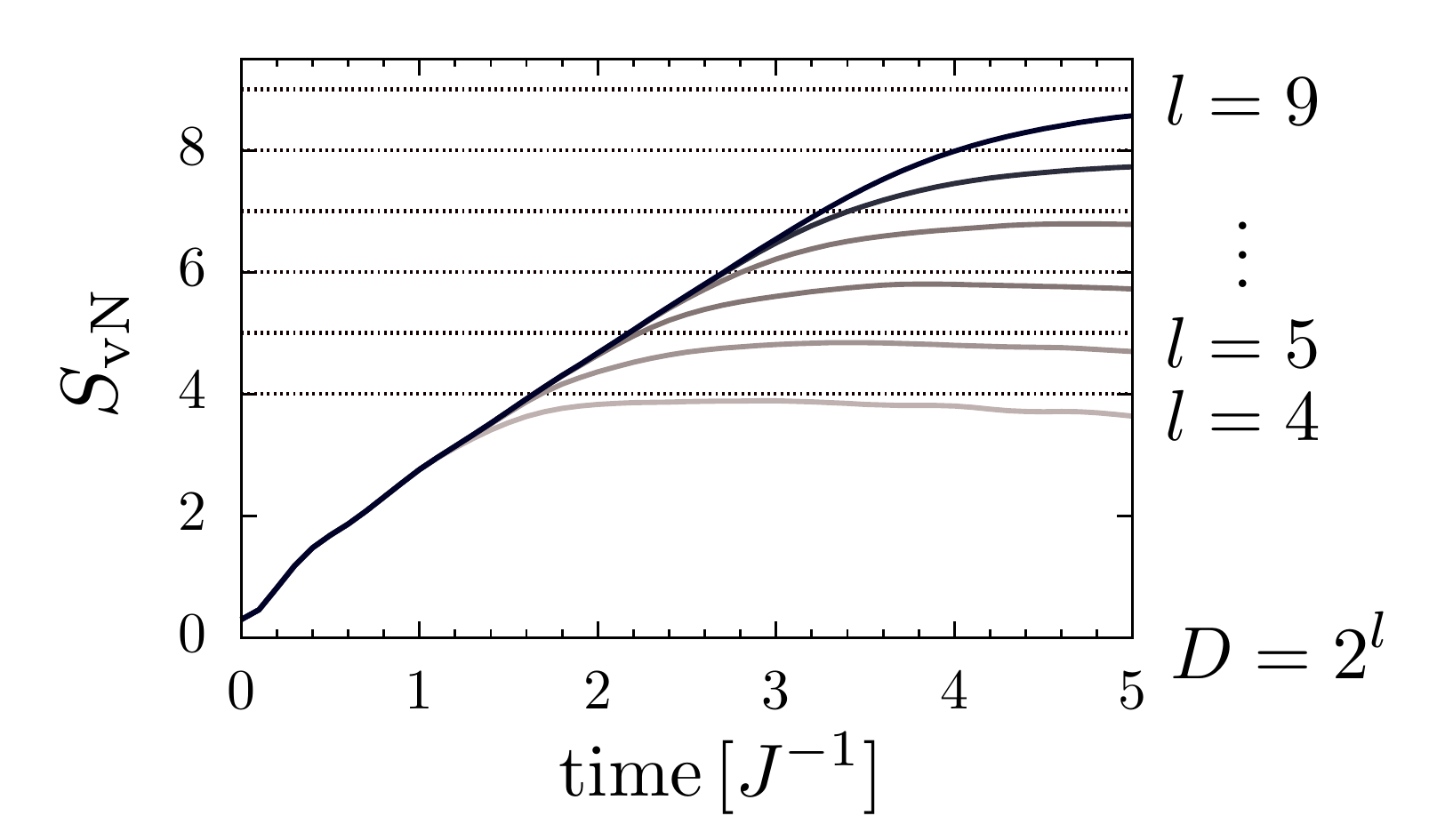}
\end{center}
\caption{Increase in entanglement as a function of time after a quantum quench in the Bose-Hubbard model. Here we consider a quench in the Bose-Hubbard model eq.~\eqref{HBH} from $U=10J$ to $U=J$ in a system with 30 particles on 30 lattice sites, initially in the ground state with $U=10J$. Making a bipartite splitting in the centre of the system (with 15 sites on each side), we compute the von Neumann entropy increase as a function of time using t-DMRG methods with varying values of $D$. We see clearly the bounds in $S_{\rm vN}$ imposed by different values of $D$, which are indicated with dotted horizontal lines, as well as clear convergence of the results at short times to the exact value as we increase $D$. Adapted from A.~J.~Daley, H.~Pichler, J.~Schachenmayer, and P.~Zoller, Phys. Rev. Lett. {\bf 109}, 020505 (2012).} \label{fig:vnegrowth}
\end{figure}

As an example of this, in Fig.~\ref{fig:vnegrowth}, we plot the entanglement across a bipartite splitting the middle of the system as a function of time after a quench in the Bose-Hubbard model. The quench is from a Mott Insulator state in 1D at unit filling, with an initial ground state at $U/J=10$, and a quench to $U/J=1$. The linear growth of the entanglement entropy is clearly seen, as is the ability of t-DMRG methods to capture the quench dynamics only so long as the entanglement remains less than $\log_2 D$. Because we require a matrix size $D> 2^{S_{\rm vN}}$ to represent the entanglement in the system, and because the numerical cost scales as $D^3$, we see that the cost of the algorithm grows exponentially in order to simulate this further in time. For reference, in this computation, a computation on a single CPU core required several days for $D=512$. It has been recently discussed that measurement of such entanglement growth in experiments could be a good way to characterise the behaviour of a quantum simulator in a regime that goes beyond what is currently computable on classical computers  \cite{Daley2012,Cardy2011,Moura-Alves2004,Palmer2005,Guhne2009,Abanin2012,Schachenmayer2013}.

In practice, the practical procedure employed when using t-DMRG methods is to fix an allowed truncation error value $\epsilon_T$ for the propagation of the state in each time step. We then ensure that $D$ is chosen to be large enough so that this error is small in each time step, in an analogous manner to the control of other such errors in numerical computations. The error $\epsilon_T$ we define by considering the exact state $|\Psi\rangle$, that would be produced via application of the time evolution operator exactly on the matrix product state we had at the beginning of the time step, and comparing this with the nearest matrix product state representation with dimension $D$ to this state, $|\Psi_{{\rm MPS},D}\rangle$. Specifically, we define $|\langle \Psi |\Psi_{{\rm MPS},D}\rangle |^2\equiv1-\epsilon_T$, where we compute the square of the inner product between $|\Psi\rangle$ and $|\Psi_{{\rm MPS},D}\rangle$. We then require that $1-\epsilon_T\approx 1$, and we can perform systematic convergence tests in $\epsilon_T$, or in some cases where it is well controlled, even extrapolation to $\epsilon_T \rightarrow 0$. This naturally involves performing calculations with increasing matrix size $D$. 

On a practical level, the size of the the matrix $D$ depends on the particular system being investigated, and the length of time over which the dynamics are computed. Typical calculations often involve several hundred to several thousand time steps, with $D\sim$ 200--2000. A calculation with a small local Hilbert space of 2--5 states and $D\sim500$ takes of the order of a day on a single Intel Ivy Bridge processor core. A significant computational advantage of quantum trajectory techniques is that with access to a large computational cluster, the methods can be immediately parallelised as each trajectory can be run independently.

\subsection{Integration of quantum trajectories and t-DMRG methods}

Integration of quantum trajectories with t-DMRG methods is somewhat straight-forward in so far as t-DMRG provides a direct means to propagate the state under the effective Hamiltonian $H_{\rm eff}$, or to apply jump operators $c_m$ to the state. Initial states can be directly constructed or computed if they are pure states, or known mixtures of pure states, and finite temperature states could also potentially be sampled as minimally entangled thermal states \cite{Stoudenmire2010}. We can then implement the protocol from section \ref{sec:higherorder} directly within this formalism \cite{Daley2009}. In computing the norm of the state as a function of time, it is usually best to ensure that the state stored in a matrix product state is always normalised. If a Trotter decomposition is used to compute the time evolution, for example, it is often useful to renormalize the state stored in memory after every application of local evolution operators \eqref{trotter} to the state. The required normalisation factors can be stored separately in memory and used to compute the norm of the state for the purposes of the protocol in \ref{sec:higherorder}.

A key test for these out of equilibrium calculations is to perform convergence tests in the truncation error $\epsilon_{T}$ \emph{for each individual trajectory}. It is typically sufficient to perform convergence tests for a small representative sample of the trajectories, because the behaviour of the entanglement growth in the system and the related growth in $\epsilon_T$ generally depends on the type of dynamics being induced by the quantum jumps. Care should be taken with this, because jumps effectively produce quantum quenches (often local quantum quenches \cite{Schachenmayer2014}), which can lead to an increase in entanglement.

\begin{figure}
\begin{center}
\includegraphics[width=10cm]{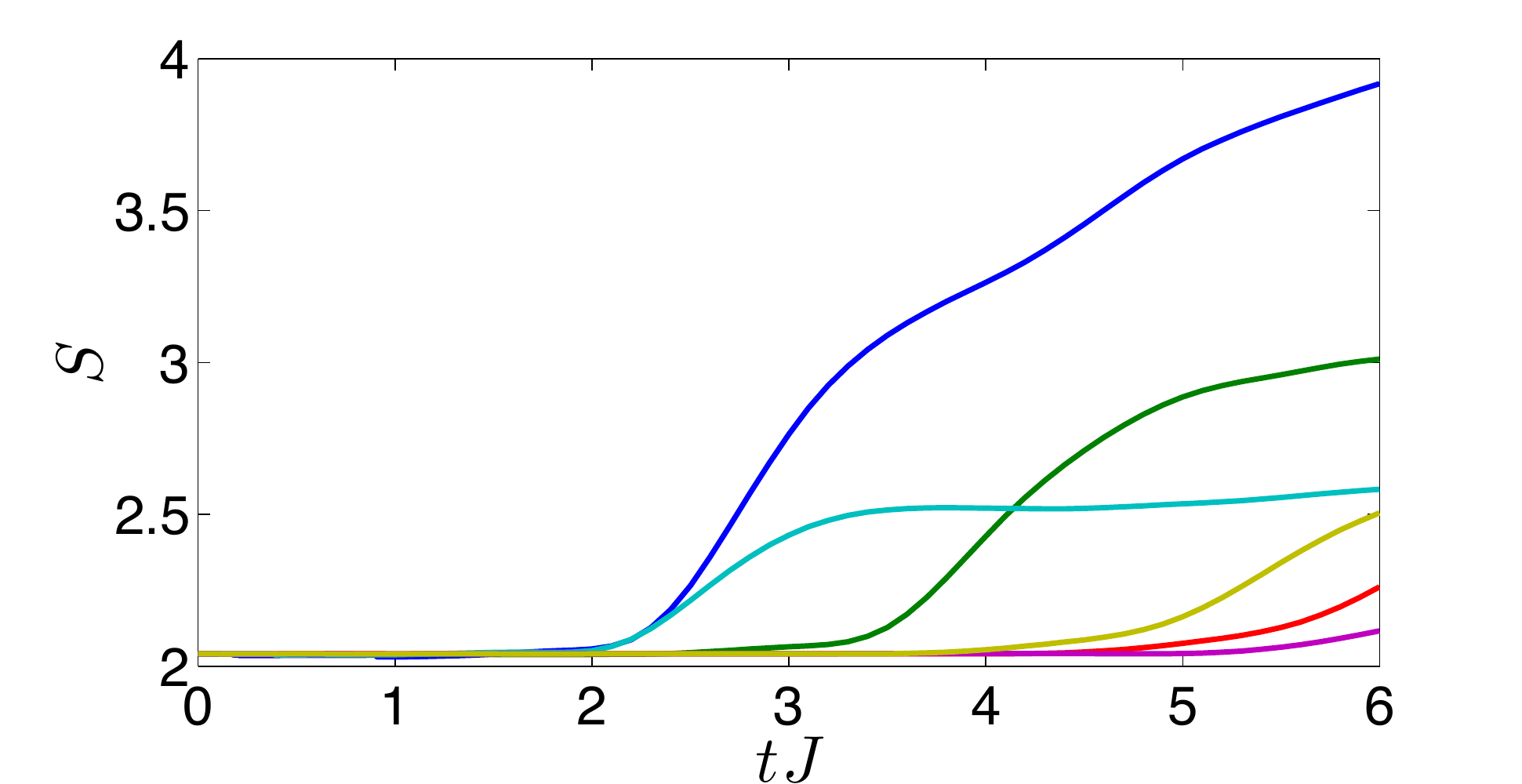}\end{center}
\caption{Six example trajectories showing time-dependent DMRG simulations of the dynamics of 48 bosons on 48 lattice sites in a master equation with $H=H_{BH}$ eq.~\eqref{HBH}, and jump operators $c_m=a_m^\dagger a_m$, and $\gamma=0.02J$. Here we investigate the increase in the difficulty of the simulation as a function of time by plotting the von Neumann entropy across a splitting in the centre of the system (with 24 sites on each side). We see that there are different delays to the increase in $S_{\rm vN}$, which depend on the timing of quantum jumps as well as their location with respect to the point of the bipartite splitting. All of these trajectories are exact calculations (t-DMRG truncation errors $\epsilon_T$ lead to errors in $S_{\rm vN}$ that are smaller than the line thickness), with $D=256$.} \label{fig:spementropy}
\end{figure}

As an example of the entanglement growth after quantum jumps, in Fig.~\ref{fig:spementropy}, we plot example values of the von Neumann entropy in the centre of a system as a function of time for bosons on an optical lattice with light scattering in the form discussed in section \ref{sec:hardcorebosons}. We show six trajectories that are representative of relatively typical behaviour, and we can see how the difficulty of the simulation increases with time. Note that there are different delays to the increase in $S_{\rm vN}$, which depend on both the timing of quantum jumps and also their location with respect to the point of the bipartite splitting. It should be noted that for local jumps, it can be favourable numerically to use different matrix dimensions $D$ for different matrices along the chain, in order to keep a fixed value of $\epsilon_T$. This is especially straight-forward when using a Trotter decomposition eq.~\eqref{trotter} and applying local time evolution operators. In general, jump operators that are local in space, and so provide a \emph{local quench} for the system should lead to a slower entanglement growth than global quenches \cite{Calabrese2007}, which should favour the use of quantum trajectories methods to describe the resulting dynamics.

As with the case of exact diagonalisation methods, matrix product state methods can be substantially optimised by restricting the values that are stored to particular symmetry sectors. This was long used as a means of optimising states in DMRG \cite{Schollwock2005}. Because of this, the added numerical advantages of quantum trajectories over density matrix propagation arising from additional symmetries present in the evolution of individual trajectories apply to t-DMRG methods in the same was as was discussed in section \ref{sec:ed} for exact diagonalisation methods. In Ref.~\cite{Daley2009}, calculations were performed for master equations describing three-body loss using quantum trajectories methods. In that case, the calculation of each trajectory was optimised by making use of a U(1) symmetry corresponding to total particle number conservation, because at each point in the evolution, each trajectory can have a fixed total particle number, even though this can change in time. Note that optimisation to such symmetries also provides a good choice when there is a freedom in choosing the jump operators in the way described in section \ref{sec:physicalinterp}.

As mentioned above, there is also an alternative way to compute dynamics of open quantum systems (as described, e.g., by a master equation for the density matrix), by using matrix product state to encode a density matrix directly \cite{Verstraete2004,Zwolak2004}. We can, e.g., replace states by density matrices and operators by super-operators. The dimension of the local Hilbert space then increases from $d$ to $d^2$. Whether such representations or quantum trajectories methods using stochastically propagated pure states is more efficient depends strongly on the particular problem being solved, and is still an area of ongoing research. In comparing the two methods it is important to make several general remarks.

Firstly, the most clear trade-off for the density matrix approach vs. quantum trajectories is that in the density matrix approach the local dimension of the Hilbert space must be increased (giving rise to a computational cost increase that is usually of the order of $d^3$) whereas in the quantum trajectories case it is important to sample sufficient numbers of trajectories in order to obtain accurate statistics for the quantities that are being computed. Given that the number of trajectories required for computation of global few-body correlation functions with a relative accuracy of a few percent tends to be of the order of several hundred, the density matrix approach could have advantages, especially in the case that the local Hilbert space dimension is small (e.g., $d=2$ for a spin-1/2 chain). This advantage is much less clear for lattice models, especially Bose-Hubbard models with potentially large on-site occupation, or ladder systems. For the Fermi Hubbard model on a two-leg ladder, for example, it is typical to take $d=16$.

Secondly, and perhaps more importantly, the matrix size $D$ required to solve the problem will depend strongly on the details of the problem. For a density matrix representation, the matrix dimension $D$ has to be increased in order to allow for classical correlations to be represented as well as quantum entanglement. This is directly analogous to the mathematics leading to a lack of classical correlations in the Gutzwiller ansatz-type density matrix representation presented in section \ref{sec:gutzwiller}. Often, this will mean that $D$ will have to be much larger for an accurate representation of the state in a density matrix form directly than would be required for states propagated in a quantum trajectories formalism. There is one important caveat for this: If the state becomes substantially mixed, then by propagating pure states, we can end up having more quantum entanglement represented than would be required for a mixed state representation of the same dynamics. One avenue of research for the future is the question of how to optimise the choice of jump operators (as discussed in sec.~{sec:physicalinterp}) in order to produce pure states in a quantum trajectories formulation that would minimise the amount of entanglement generated in time. 

As a general rule, short-time dissipative dynamics that begin with a pure state are straight-forwardly described by integrating t-DMRG with quantum trajectories \cite{Daley2009,Kantian2009,Barmettler2011}. Several examples of the use of quantum trajectories integrated with t-DMRG to describe open many-body systems in an AMO context are described in the next section.

\section{Open many-body AMO systems}
\label{sec:amosys}\label{amosys}

As we discussed above, the study of strongly interacting AMO systems has made it possible to realise open many-body quantum systems in which the usual quantum optics approximations apply. This makes it possible to have microscopically well-controlled models for many-body systems in an analogous form to the single-particle models of quantum optics. In this section, we give several examples, focusing initially on cold quantum gases in optical lattices \cite{Jaksch1998,Jaksch2005,Lewenstein2007,Bloch2008b,Lewenstein2012}, where we begin with a discussion of the corresponding microscopic models before presenting an overview of dissipation due to light scattering and due to particle losses. We then discuss dissipative state preparation in this system, as well as corresponding experiments with other experimental systems, including trapped ions, Bose-Einstein condensates in an optical cavity, and Rydberg atoms.

\subsection{Microscopic description for cold quantum gases and quantum simulation} 
\label{sec:qsim}

A key starting point for a lot of recent progress with ultracold atomic gases is the possibility to realise microscopically well controlled models for strongly interacting systems, and to manipulate the parameters of those models via external fields. This comes about because the atomic physics associated with magnetic and optical trapping potentials is very well understood \cite{Pethick2002,Metcalf1999}, and for the experimentally relevant limit of dilute gases, inter-particle interactions are dominated by two-particle interactions. Moreover, because the collision energies of atoms are very small, the scattering (which typically involves complicated interatomic potentials with many bound states) reduces to a universal regime where the behaviour can be described by a single parameter, the s-wave scattering length $a_{\rm scatt}$ \cite{Dalibard1998}\cite{Castin2001}. Moreover, when $a_{\rm scatt}$ is small compared with typical scattering momentum scales, we can make a Born approximation for the scattering. For cold bosonic atoms, we then write the well-controlled microscopic Hamiltonian in terms of bosonic field operators\footnote{Although atoms consist of protons, neutrons, and electrons, a tightly bound atom at low energy can be assumed to maintain bosonic or fermionic character provided short-distance physics can be eliminated, and the typical separation of atoms is much larger than their size.} $\hat{\psi}(\mathbf{r})$ that obey $\left[ \hat{\psi}(\mathbf{x}), \hat{\psi}^\dagger(\mathbf{y})\right]=\delta(\mathbf{x}-\mathbf{y})$ as:
\begin{equation}
\hat{H}_{\rm atoms}\approx\int d^{3}r\,\hat{\psi}^{\dag}(\mathbf{r})\,\left[-\frac{\hbar^{2}}{2m}\nabla^{2}+V_{0}(\mathbf{r})\right]\,\hat{\psi}(\mathbf{r})+\frac{g}{2}\int d^{3}r\,\,\,\hat{\psi}^{\dag}(\mathbf{r})\,\hat{\psi}^{\dag}(\mathbf{r})\,\,\hat{\psi}(\mathbf{r})\,\hat{\psi}(\mathbf{r}), \label{hatoms}
\end{equation}
where $m$ is the atomic mass, $g=4\pi\hbar^2a_{\rm scatt}/m$, $V_{0}(\mathbf{r})$ is the effective single-particle potential (generated by external magnetic fields and optical traps).

Not only is this model well-understood microscopically, but experiments offer precise time-dependent control, both over $V_{0}$, through external magnetic fields and optical traps based on the AC-Stark Shift \cite{Pethick2002,Metcalf1999}, and over $a_{\rm scatt}$, through optical and magnetic Feshbach resonances \cite{Chin2010}. While the gas described here is in principle weakly interacting, Feshbach resonances can be used to increase the interaction strength towards a strongly interacting regime. Alternatively, even while maintaining values of $a_{\rm scatt} \sim 100 a_0$, where $a_0$ is the Bohr radius, we can achieve a strongly interacting system by loading atoms into an optical lattice potential, formed via the AC-Stark shift with standing waves of laser light. Because the temperature and energy scales can be straight-forwardly made much smaller than the bandgap in an optical lattice potential \cite{Jaksch1998}, this makes it possible to directly realise simple lattice models for atoms in a single band of an optical lattice, such as the Bose-Hubbard model eq.~\eqref{HBH}, or extended versions of this that allow for additional confinement and/or superlattice potentials in the form 
\[
{H_{\rm BH+}}=-J\sum_{\langle j,l\rangle}{a}_{j}^{\dag}{a}_{l}+\frac{U}{2}\sum_{l}{a}_{l}^{\dag}{a}_{l}({a}_{l}^{\dag}{a}_{l}-1)+\sum_{l}\epsilon_{l}{a}_{l}^{\dag}{a}_{l},\]
where as before $\langle j,l\rangle$ denotes a sum over all combinations of neighbouring
sites, and $\epsilon_{i}$ is the local energy offset of each site. In order to obtain the Bose-Hubbard model, we expand the field operator $\hat{\psi}(\mathbf{r})$ in terms of Wannier function modes for the lowest Bloch Band centred at each site location $\mathbf{r_l}$, $w_0(\mathbf{r}-\mathbf{r}_l)$ \cite{Jaksch1998,Kohn1959} as $\hat{\psi}(\mathbf{r})=\sum_l w_0(\mathbf{r}-\mathbf{r}_l) a_l$. We then see that terms involving off-site interactions and tunnelling to more than the nearest neighbour site can be neglected in lattices deeper than about $5 E_R$, where $E_R=\hbar^2 k_L^2/(2m)$, and $k_L$ is the wavenumber of the laser generating a cubic lattice through forming standing waves in three dimensions \cite{Jaksch1998,Jaksch2005}. 

It is possible to realise a large array of lattice models in this way, and to obtain an array of spin models as limiting cases \cite{Lewenstein2007}. Lattice geometries are specified by the optical trapping potentials, and it is also possible to confine atoms to low-dimensional systems by making the lattice sufficiently deep along particular axis directions that the atoms do not tunnel along those directions on typical experimental timescales \cite{Jaksch2005}. Because $J$ depends exponentially on the depth of the lattice, it is relatively straight-forward to make tunnelling timescales very long indeed \cite{Jaksch2005}. A particularly important model that can also be realised in this way is the (fermionic) Hubbard model \cite{Jordens2008,Esslinger2010}, 
\[
{H_{\rm FH}}=-J\sum_{\langle l,j\rangle,\sigma}\hat{a}_{l\sigma}^{\dag}\hat{a}_{j\sigma}+U\sum_{l}\hat{a}_{l\uparrow}^\dag\hat{a}_{l\uparrow}\hat{a}_{l\downarrow}^\dag\hat{a}_{l\downarrow}+\sum_{l,\sigma}\epsilon_{l}\hat{a}_{l\sigma}^\dag\hat{a}_{l\sigma},\]
which is well known as a description for strongly interacting electrons in the solid state. Here, $\hat{a}_{l\sigma}$ are fermionic operators obeying standard
anti-commutator relations, and $\sigma\in\{\uparrow,\downarrow\}$.
This is a simple example of the many two-species models that can be
engineered with atoms in optical lattices. It is similarly possible to create mixtures of bosons and fermions in a lattice, as well as multi-band Hubbard models \cite{Lewenstein2007, Pichler2010}.

This level of control over ultracold atomic gases offers enormous new opportunities for the field of analogue quantum simulation \cite{Cirac2012,Bloch2012}. In particular, by realising lattice models that are analytically and computationally intractable, there is great potential to use these systems as a special-purpose analogue quantum computer to determine the properties of these models. A simple but important example of this is the simulation of the 2D Hubbard model, the ground state of which under certain conditions is still debated. Moreover, the ability to construct an enormous range of models opens possibilities for realising theoretical ideas that have been predicted, but never observed, in a real physical system. In each of these ways, there are many possibilities for cold atoms in optical lattices to contribute strongly to our understanding of many-body quantum physics in the next few decades.

At the same time, the use of these systems as quantum simulators also offers up key challenges. Firstly, there are always imperfections in an experimental implementation that go beyond the Hamiltonian models written down here. Technically, this can include things such as noise on lattice potentials \cite{Pichler2012,Pichler2013}, and also spontaneous emissions, or incoherent scattering of light from the lasers generating the optical lattice potential \cite{Gerbier2010,Pichler2010, Dalibard1985,Gordon1980,Ellinger1994}. It is important to characterise and control the effects that these imperfections have on the many-body states in the lattice, in order to restrict heating and ensure that the most interesting many-body states can be reached \cite{McKay2011}. Secondly, many of the most interesting many-body states specifically require low temperatures for their realisation, especially those that arise in perturbation theory, e.g., as spin superexchange terms $\propto J^2/U$ in strongly interacting regimes where $U\ll J$ \cite{Lewenstein2007}. While current experimental temperatures have reached regimes of hundreds of picokelvin \cite{McKay2011} -- which is impressive on an absolute scale -- these temperatures are often higher than scales of the order of $J^2/U$ in optical lattices \cite{Jordens2010,Fuchs2011,Greif2013}. 

However, the same understanding of the atomic physics that allows for design and engineering of lattice and spin models in these systems also allows us to microscopically model dissipative processes, including both the imperfections that can lead to heating, and also engineered dissipative processes that could provide a new route to cooling and preparation of many-body states. Below we investigate several examples of these processes, beginning with light scattering in an optical lattice and particle loss, and continuing on to engineered dissipation, and state preparation in driven, dissipative systems. 

\subsection{Light scattering}
\label{sec:lightscattering}

Incoherent light scattering from trapped atoms has been studied in several different contexts over the past few decades. This has included the study of heating of single particles in optical dipole traps \cite{Dalibard1985,Gordon1980}, and more recently directly in the context of optical lattice potentials \cite{Gerbier2010,Pichler2010}. Where early studies with single atoms focussed on the rate of increase of energy, this is typically not sufficient to completely characterise how the many-body state changes in an optical lattice. Instead, we have to look more closely at the full out-of-equilibrium dynamics of the many-body system. This is more reminiscent of studies of the measurement of Bose-Einstein condensates in the late 1990s, which sought to understand the origins of interference patterns in experiments produced by multiple trapped condensates in a way that went beyond the assumption of each condensate having a fixed phase \cite{Cirac1996,Wong1996,Jack1996,Yoo1997,Castin1997,Ruostekoski1997,Ruostekoski1998,Dunningham1999,Dalvit2002}. These studies included relative phase measurement directly via light scattering   \cite{Ruostekoski1997,Ruostekoski1999,Saba2005}, as well as decoherence of a Bose-Einstein condensate due to such scattering processes \cite{Ruostekoski1998}. 

For atoms in an optical lattice, it is possible to derive a many-body master equation to describe incoherent scattering of light in the system. This treatment involves the generalisation of the optical Bloch equations from section \ref{sec:twolevel} to many atoms, as was done by Lehmberg \cite{Lehmberg1970,Lehmberg1970a}, and the inclusion of the atomic motion \cite{Ellinger1994}, in order to properly account for the mechanical effects of light on atoms \cite{Kazantsev1989,Cohen-Tannoudji1989,Cohen-Tannoudji1998}.

For a single particle, the resulting master equation for the system density operator $\rho$ was derived in several places \cite{Dalibard1985,Gordon1980,Gerbier2010,Pichler2010}, and takes the form
\begin{eqnarray}
\frac{d}{dt}{\rho}&=&-{\rm i}\left[H_{\rm atom},\rho\right] - \frac{\Gamma}{2} \int d^2\mathbf{u}N(\mathbf{u})\left( c^{\dag}_{\mathbf{u}}c_{\mathbf{u}}\rho+\rho\,c^{\dag}_{\mathbf{u}}c_{\mathbf{u}}-2c_{\mathbf{u}}\rho\,c^{\dag}_{\mathbf{u}}\right),\nonumber
\end{eqnarray}
where the Hamiltonian for the single atom with ground state $|g\rangle$ and excited state $|e\rangle$, as depicted in Fig.~\ref{fig:largefrequency}b, is given by 
\begin{equation}\label{eq:single_article_Hamiltonian}
H_{\rm atom}=\frac{\hat{\mathbf{p}}^{\,2}}{2m}-\Delta \ket{e}\bra{e}-\left(
\frac{\Omega_0(\hat{\mathbf{x}})}{2} \ket{e}\bra{g}
+{\rm H.c.}\right).
\end{equation}
The decay rate is given by $\Gamma$, and the jump operators in the master equation are $c_{\mathbf{u}}=e^{-{\rm i}k_{A}\mathbf{u}\cdot
\hat{\mathbf{x}}} \ket{g}\bra{e}$, corresponding to a decay $\ket{e}\rightarrow\ket{g}$ and a momentum recoil. For the recoil, $k_{A}$ is the wavenumber of a photon at the atomic transition frequency, and $k_A \approx k_L$, where $k_L$ is the wavenumber of the laser driving the classical transition at a spatially-dependent Rabi frequency $\Omega_0(\hat{\mathbf{x}})$ and with detuning $\Delta$. The distribution of scatted photons, $N(\mathbf{u})$, is determined by the distribution of dipole radiation about the dipole $\hat{\mathbf{d}}$ between the states $|g\rangle$ and $|e\rangle$ as 
\begin{equation}
N(\mathbf{u})=\frac{3}{8\pi}\left[1-\left(\hat{\mathbf{d}}\cdot\mathbf{u}\right)^2\right].
\end{equation}
Here, the equations of motion have been written in a frame rotating with the laser frequency, so that the optical frequencies have been eliminated, and we have made the standard set of approximations outlined in section \ref{sec:approxamo}. In the limit where $\Delta\gg \Omega_0$, we can adiabatically eliminate the excited state \cite{Gardiner2005}, giving an effective equation of motion for the system density operator for the ground state $\rho_g$, 
\begin{align}
\frac{d}{dt}\rho_g&=-{\rm i}(H_{\textrm{eff}} \rho_g-\rho_g  H_{\textrm{eff}}^{\dag})+{\mathcal{J}}{\rho_g}.
\end{align}
Here,
\begin{eqnarray} H_{\textrm{eff}}&=&\frac{\hat{\mathbf{p}}^{\,2}}{2m}+\frac{|\Omega_0(\hat{\mathbf{x}})|^2}{4\Delta}-{\rm i}\frac{1}{2}\frac{\Gamma|\Omega_0(\hat{\mathbf{x}})|^2}{4\Delta^2}\nonumber\\
&\equiv&\frac{\hat{\mathbf{p}}^{\,2}}{2m}+V_{\rm opt}(\hat{ \mathbf{x}}) -{\rm i}\frac{\gamma_{\rm eff}(\hat{\mathbf{x}})}{2}, \label{voptgeff}
\end{eqnarray}
showing how the optical potential $V_{\rm opt}(\mathbf{x})$ arises from the AC-Stark shift in this two-level system, as well as how the effective spontaneous emission rate $\gamma_{\rm eff}(\mathbf{x})$ appears. Note that we have also taken the limit $\Delta \gg \Gamma$ to simplify eq.~\eqref{voptgeff}. The recycling term in the master equation is given by
\begin{equation}
{\mathcal{J}}{\rho_g}=\Gamma\int d^2\mathbf{u}\,N(\mathbf{u})\left[
e^{-{\rm i}k_{A}\mathbf{u}\cdot\hat{\mathbf{x}}}\frac{\Omega_0(\hat{\mathbf{x}})}{2\Delta}\right]\rho_g \left[ e^{ik_{A}\mathbf{u}\cdot\hat{\mathbf{x}}}\frac{\Omega_0^{\ast}(\hat{\mathbf{x}})}{2\Delta}\right].
\end{equation}
From this, we clearly see the effect of the spontaneous emission on the motion of the atom in the ground state, based on the effective jump operator $\tilde c_{\mathbf{u}}(\hat{\mathbf{x}})\equiv e^{-{\rm i}k_{A}\mathbf{u}\cdot\hat{\mathbf{x}}}\Omega_0(\hat{\mathbf{x}})/(2\Delta)$, which contains the momentum recoils associated with the absorption of a laser photon from the classical field [$\Omega_0(\mathbf{x})$] followed by spontaneous scattering of a photon in the direction $\mathbf{u}$.

In Ref.~\cite{Pichler2010}, this derivation was generalised to $N$ bosonic atoms in a laser field, where the dynamics of the reduced density operator $\rho$, now for the many-body system, was derived. Because the dissipative dynamics are again dominated by the large optical frequency, all of the usual quantum optics approximations from section \ref{sec:approxamo} can be made here. As in the single-particle case, the excited state is adiabatically eliminated, making it possible to write a master equation in second quantisation using the field operators  $\hat{\psi}(\mathbf{x})$,
\begin{eqnarray}\label{eq:master_equation}
\dot \rho&=&-{\rm i} \left( H_{\textrm{eff}}\rho-\rho H_{\textrm{eff}}^\dag \right) +\mathcal{J}\rho,
\end{eqnarray}
with effective Hamiltonian
\begin{eqnarray}
H_{\rm eff} &=&H_{\rm atoms} + H_{\rm eff, rad},
\end{eqnarray}
where $H_{\rm atoms}$ was given in eq.~\eqref{hatoms} up to the additional note that the effective potential $V_0(\mathbf{r})=V_{\rm opt}(\mathbf{r})$ as given in eq.~\eqref{voptgeff} above, and the effects of radiative coupling to the field are given by the effective Hamiltonian term
\begin{flalign}
H_{\rm eff,rad}= &-{\rm i}\frac{1}{2}\int d^3x \frac{\Gamma| \Omega_0(\mathbf{x})|^2}{4\Delta^2} \hat{\psi}^{\dag}(\mathbf{x})\hat{\psi}(\mathbf{x})\nonumber\\
& -{\rm i}\frac{1}{2}\iint d^3xd^3y \frac{\Gamma\Omega_0(\mathbf{y})\Omega_0^{\ast}(\mathbf{x})}{4\Delta^2}F(k_{A}(\mathbf{x}-\mathbf{y})) \fd{x}{}\fd{y}{}\f{y}{}\f{x}{}\nonumber\\
&+\iint d^3xd^3y \frac{\Gamma\Omega_0(\mathbf{y})\Omega_0^{\ast}(\mathbf{x})}{4\Delta^2}G(k_{A}(\mathbf{x}-\mathbf{y})) \fd{x}{}\fd{y}{} \f{y}{}\f{x}{} 
\end{flalign}
and the recycling term \cite{Pichler2010}
\begin{flalign}
\mathcal{J}\rho&=\iint d^3x d^3y \frac{\Gamma\Omega_0(\mathbf{x})\Omega_0(\mathbf{y})}{4\Delta^2} F(k_{A}(\mathbf{x}-\mathbf{y}))
 \hat{\psi}^{\dag}(\mathbf{x})\hat{\psi}(\mathbf{x})\rho
\hat{\psi}^{\dag}(\mathbf{y})\hat{\psi}(\mathbf{y}),&&\label{eq:dissipation}
\end{flalign}
where the fuctions $F$ and $G$ are defined as 
\begin{flalign}
F(\mathbf{z})&=\int d^2{u}\,N(\mathbf{u})e^{-{\rm i}\mathbf{u}\cdot\mathbf{z}}&&\nonumber\\
&=\frac{3}{2}\left\{\frac{\sin z}{z}\left(1-(\hat{\mathbf{d}}\cdot\hat{\mathbf{z}})^2\right)+\right.
\left.\left(1-3(\hat{\mathbf{d}}\cdot\hat{\mathbf{z}})^2\right)\left(\frac{\cos z}{z^2}-\frac{\sin z}{z^3}\right)\right\},\\
G(\mathbf{z})&=-\frac{1}{z^3}\mathcal{P}\int_{-\infty}^{\infty}\frac{d\zeta}{2\pi}\frac{\zeta^3}{\zeta-z}F(\zeta\mathbf{z}/z)&&\nonumber\\
&=\frac{3}{4}\left\{\left((\hat{\mathbf{d}}\cdot\hat{\mathbf{z}})^2-1\right)\frac{\cos z}{z}+\right.
\left.\left(1-3(\hat{\mathbf{d}}\cdot\hat{\mathbf{z}})^2\right)\left(\frac{\sin z}{z^2}+\frac{\cos z}{z^3}\right)\right\},
\end{flalign}
and $\mathcal{P}$ is used to denote the Cauchy principal value integral.
If the short-range collisional physics Hamiltonian accounting for short range collision physics in the presence of laser fields gives rise to additional dissipative processes, than this can also be accounted for by an additional two-body loss term in the effective Hamiltonian \cite{Pichler2010}.

\begin{figure}[tbh]
\begin{center}
\includegraphics[width=12cm]{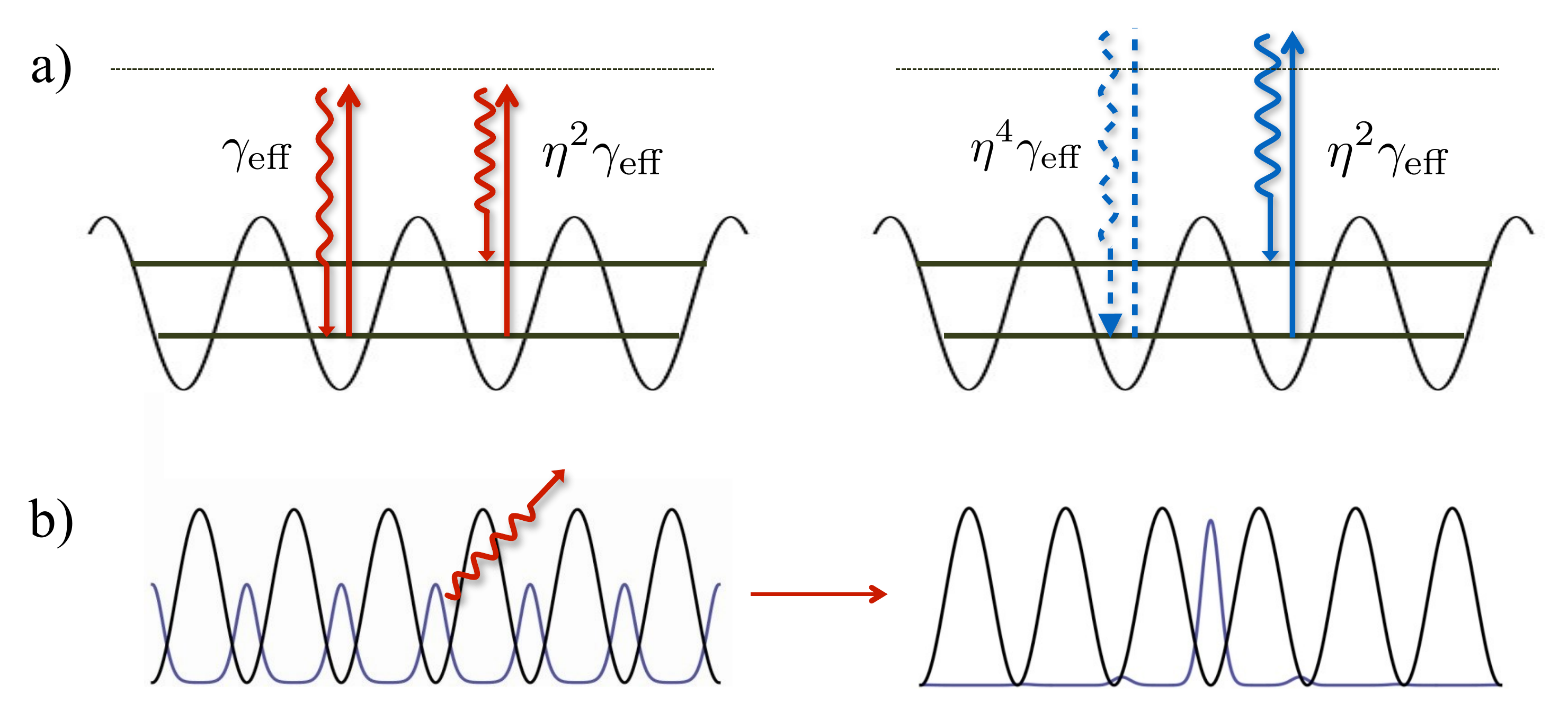}
\end{center} 
\caption{Schematic plot of the effects of spontaneous emissions in a deep optical lattice. For an atom beginning in the lowest band of the lattice, a spontaneous emission (a) can transfer it to the first excited band with a probability $\eta^2$, were $\eta$ is the Lamb-Dicke parameter (see the text for details), and (b) will always lead to localisation of the atom, in the sense that the spontaneous emission provides information to the environment about the location of the atom.}\label{fig:spemschematic}
\end{figure}

In eq.~\eqref{eq:dissipation}, the function $F(k_{A}(\mathbf{x}-\mathbf{y}))$ is localised in $(\mathbf{x}-\mathbf{y})$ on a length scale given by $1/k_{A}$, i.e., $\lambda_A/(2\pi)$. As a result, particles undergoing spontaneous emission can be reinterpreted as a continuous measurement of the position of the atoms up to a position resolution of length scale $\lambda_A/(2\pi)$. Phrased another way, as the atoms scatter light incoherently in random directions, they provide their environment with information about where they are. This scattering does two things then, as depicted in Fig.~\ref{fig:spemschematic}. Firstly, with a probability given by $\eta^2$, where $\eta=2 \pi a_{\rm trap}/\lambda\sim 0.3$ for optical lattices with depth around 10$E_R$, an atom beginning in the lowest band will be transferred to a higher band via the momentum recoil. This depends on the size of the trap wavefunction $a_{\rm trap}$ in each lattice site\footnote{Note that this transfer can also be interpreted as measurement of the position of the atom within the lattice site.}. Then, whether or not the atom is transferred to a higher band, a delocalised atom will be localised within the lattice, in the sense that a coherent superposition of an atom delocalised over different lattice sites will be transferred to a mixed state of the atom on different lattice sites. Note that this happens because the localisation scale is $\lambda_A/(2\pi)$, whereas the lattice spacing for a lattice formed by counter-propagating laser beams is $\lambda_A/2$. This is the opposite limit to what might occur in solid state experiments, where the lattice spacing is often much shorter than typical optical wavelengths.

\begin{figure}[tbh]
\begin{center}
\includegraphics[width=8cm]{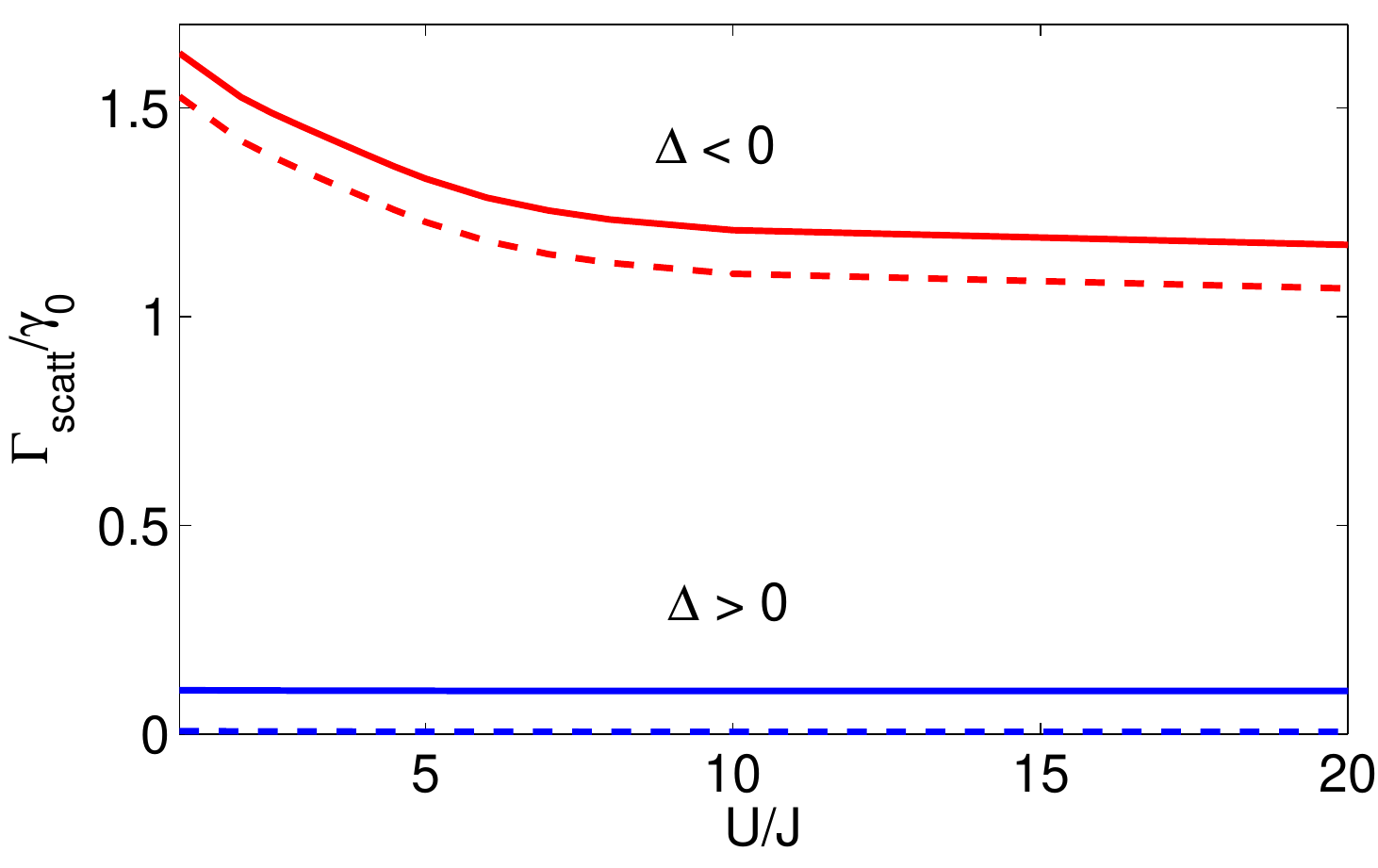}
\end{center} 
\caption{Comparison between the total process rate $\Gamma_{\rm scatt}
/\gamma_0$ (solid lines) and the rate for processes back to the lowest band (dashed lines) for $\Delta<0$ (upper lines) and a $\Delta>0$ (lower lines) in the ground state of a 1D lattice (Using DMRG ground states for Bosons in a 1D lattice with lattice depth in the $x$, $y$ and $z$ directions $V_x=V_y=30 \,E_R, V_z=10 \,E_R$). The rates include scattering from all three lattice generating beams. Reprinted figure from H.~Pichler, A.~J.~Daley, and P.~Zoller, Phys. Rev. A {\bf 82}, 063605 (2010) \cite{Pichler2010}. Copyright 2010 by the American Physical Society.}\label{fig:pichler1}
\end{figure}

The rate with which spontaneous emission events occur is then dependent on the sign of the detuning for the atoms. For red detuning $\Delta<0$, the AC-Stark shift potential derived in eq.~\eqref{voptgeff} shows that atoms are located where the light is brightest, whereas for $\Delta>0$, the atoms are located where the light is darkest. This results in a difference in scattering rates between blue and red light $\sim \eta^2$ \cite{Gordon1980,Dalibard1985}, with the dominant process for blue-detuned fields being the transfer of atoms to the first excited band. This can be seen in the results from Ref.~\cite{Pichler2010} for scattering rates from many bosons in an optical lattice, which are reproduced in Fig.~\ref{fig:pichler1}. There the expected rate of scattering per particle $\Gamma_{\rm scatt}$ is plotted for atoms in red and blue detuned optical lattices, normalised to $\gamma_0$, which is the value of $\gamma_{\rm eff}$ (from eq.~\ref{voptgeff}) at the maximum depth of the lattice. The initial state involves bosons in a 1D optical lattice at unit filling, and we see that in the red detuned case, for small interactions $U/J$, multiple occupancy of particles on lattice sites gives rise to a super-radiant enhancement of the scattering rate.

\begin{figure}[tb]
\begin{center}
\includegraphics[width=12cm]{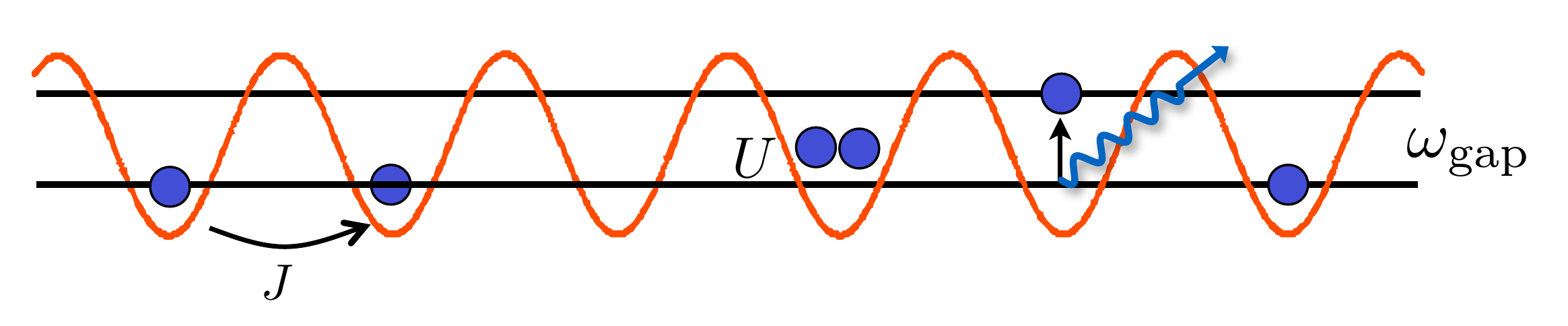}
\end{center} 
\caption{Schematic diagram indicating the energy scales involved in preventing thermalisation of the bandgap energy after a spontaneous emission. Atoms transferred to a higher band cannot thermalise the energy $\omega_{\rm gap}$ on typical experimental timescales, because this energy scale is much larger than the energy scales $J$ associated with tunnelling and $U$ associated with interactions in the lowest band.}\label{fig:spemhigherband}
\end{figure}

Although the scattering rates are different by almost an order of magnitude between red and blue detuning, the rate of energy increase is identical for the two cases, and also equal to $N$ times the single particle result \cite{Pichler2010,Gerbier2010,Dalibard1985}. This might seem to imply that red and blue detunings have the same effect on a many body state, but this is not the case. The reason for this comes back to the question of whether energy added to the system is thermalised in an out of equilibrium many-particle system. Here, the largest contribution towards the increase in energy in the system comes from transfer of particles to higher Bloch bands. But, as depicted in Fig.~\ref{fig:spemhigherband}, the bandgap energy $\omega_{\rm gap}\gg U,J$, so that the energy scales of dynamics in the lowest band are much smaller than the bandgap energy. This means that only a very high-order process in $J/\omega_{\rm gap}$ and $U/\omega_{\rm gap}$ will allow the sharing of the bandgap energy amongst atoms in the lowest band. These processes will not take place on typical experimental timescales, and so the system will not thermalise the energy input from the scattering on the relevant timescales.

\begin{figure}[tb]
\begin{center}
\includegraphics[width=12cm]{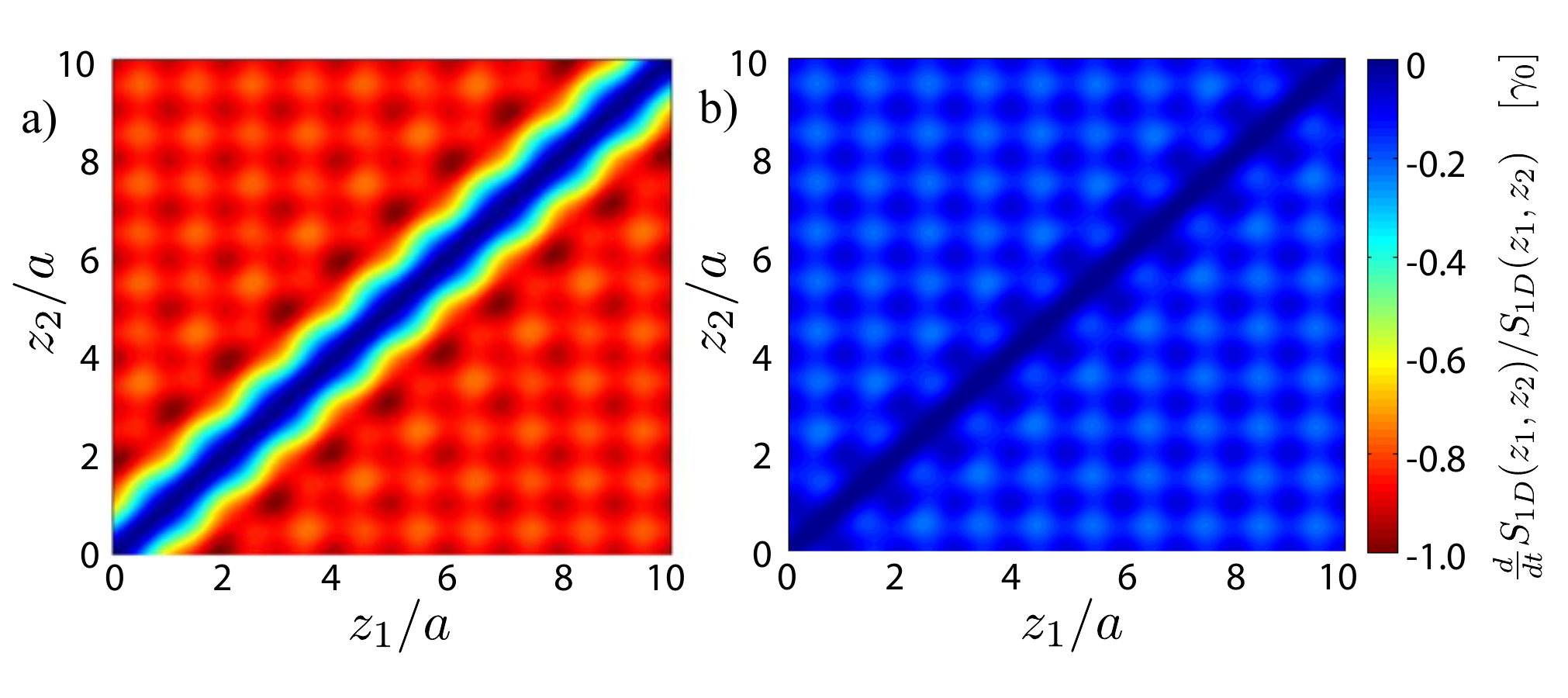}
\end{center}
\caption{Relative rate of change of the (integrated) single particle density matrix $S_{1D}(z_1,z_2)=\iint dxdy\langle\hat{\psi}^{\dag}(x,y,z_1)\hat{\psi}(x,y,z_2)\rangle$ in an effectively 1D lattice with depth in recoil energies given by ($V_x=V_y=30\,E_R$; $V_z=10\,E_R$). In the tightly bound transversal directions the atoms are assumed to be in the lowest band. Scattering from all three lattice generating laser beams is taken into account (with weights corresponding to the lattice depths). (The lattice constant is denoted by $a$.). Reprinted figure from H.~Pichler, A.~J.~Daley, and P.~Zoller, Phys. Rev. A {\bf 82}, 063605 (2010) \cite{Pichler2010}. Copyright 2010 by the American Physical Society. } \label{fig:pichler2}
\end{figure}

For this reason, it is important to study the complete out-of-equilibrium many-body physics of many atoms as described by the master equation. As a first step in this direction, it is possible to calculate the rate of change of the single-particle density matrix $\langle \hat \psi^\dagger(\mathbf{x})\psi(\mathbf{y})\rangle$ at short times using perturbation theory. This correlation function is important, because it characterises the off-diagonal long-range order in a superfluid phase, as well as the expontential localisation of particles in a Mott Insulator. This calculation was performed in Ref.~\cite{Pichler2010}, and the results are shown in Fig.~\ref{fig:pichler2}. This shows that the key effect of spontaneous emissions is to lead to localisation of the atoms, and therefore a direct decrease in the off-diagonal elements. This rate of decrease is proportional to the scattering rate, and therefore is much larger for red-detuned lattices than for blue-detuned lattices. It is also more important in the superfluid state, where the off-diagonal correlations are initial strong, as opposed to the Mott Insulator state, where the atoms are anyway initially exponentially localised. 

\begin{figure}[tb]
  \centering
  \includegraphics[width=12cm]{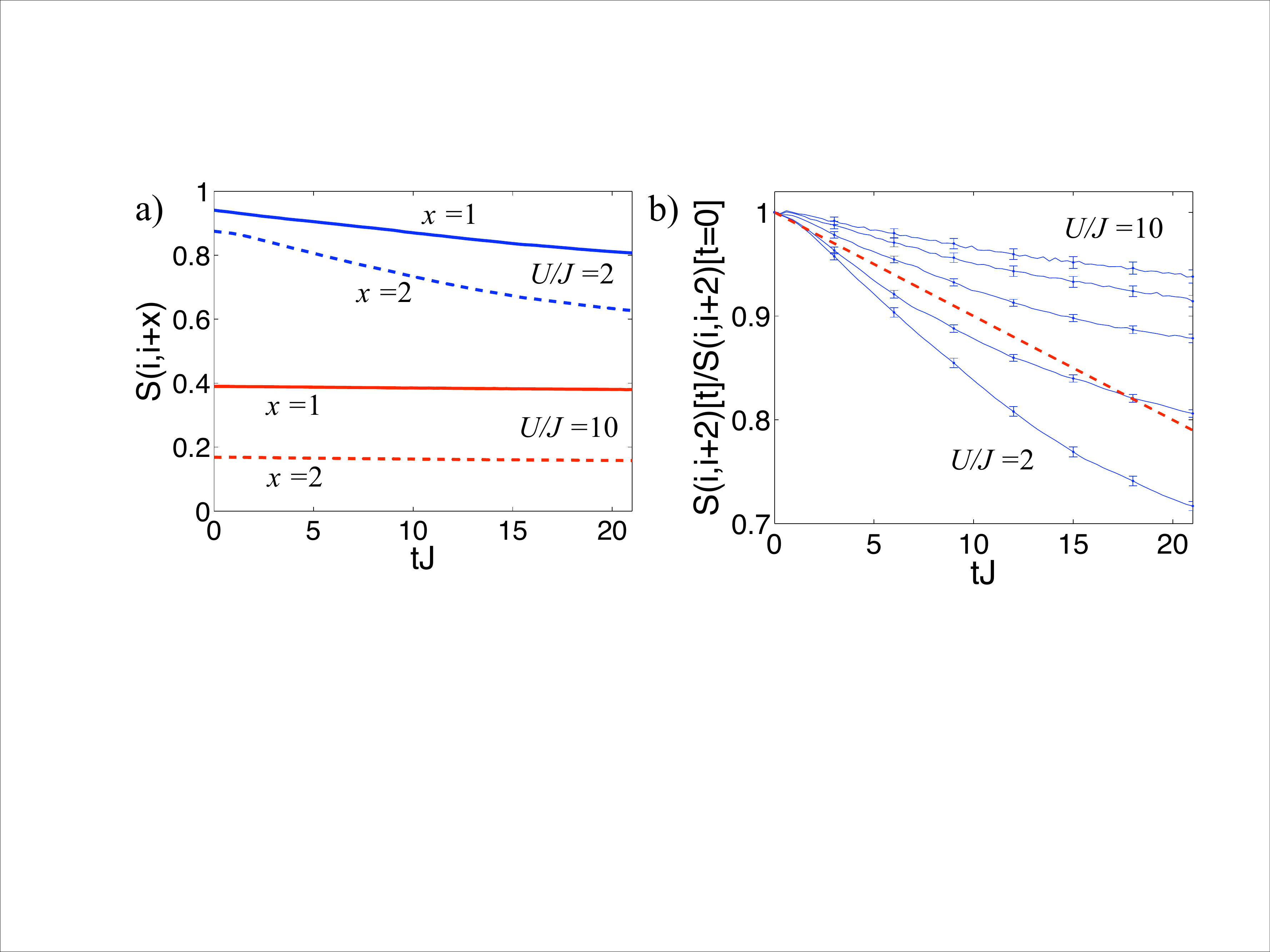}
  \caption{Comparison of the decay of off-diagonal correlations in the single particle density matrix $S(i,j)=\langle a_i^\dag a_j  \rangle$ as a function of time. These are computed from the master equation for the lowest band Eq.~\eqref{eq:QT_master_eq} with $\gamma/J=0.01$ for 24 particles on a 24 site lattice by combining quantum trajectories methods with t-DMRG. (a) Comparison of the decay of off-diagonal elements $\sum_i S(i,i+x)$ when $U/J=2$ (superfluid regime, upper lines) and $U/J=10$ (Mott Insulator regime, lower lines). In each case we show correlations at separation distances $x=1$ (solid lines) and $x=2$ (dashed lines), each of which are averaged over $i$. We see that the correlations for the Mott Insulator state at $U/J=10$ are almost constant, whereas for $U/J=2$, the decay becomes more rapid than at initial times.  (b) Comparison of $S(i,i+2)$ averaged over $i$ and normalized to the value at time $t=0$,  $|S(i,i+2)|[t=0]$ for $U/J=2,4,6,8,10$ (seen here from bottom to top). The dashed line shows the corresponding result from perturbation theory. Each computation was averaged over 1000 trajectories, and error bars are shown in (b). For (a), the statistical errors fit inside the line thickness. Adapted from H. Pichler et al., Phys. Rev. A {\bf 82}, 063605 (2010) \cite{Pichler2010}. Copyright 2010 by the American Physical Society.} \label{fig:pichler3}
\end{figure}

To go beyond this perturbation theory result, in Ref.~\cite{Pichler2010}, a combination of t-DMRG with quantum trajectories is used to solve the master equation for atoms in the lowest band. Under the approximation that atoms do not get transferred to higher bands, and making the approximation that atoms are localised exactly on a single site, eq.~\eqref{eq:master_equation} reduces to
\begin{equation}
\label{eq:QT_master_eq}
\dot{\rho}=-{\rm i}[H_{BH},\rho]-\sum_{i}\frac{\gamma}{2}\left( n_i n_i\rho+\rho n_i n_i-2n_i\rho n_i \right),
\end{equation}
where $n_i=a^\dagger_i a_i$, $\gamma$ is the effective scattering rate on a single site, and $H_{BH}$ is the Bose-Hubbard Hamiltonian for the lowest band from eq.~\eqref{HBH}.
In Fig.~\ref{fig:pichler3} we reproduce the results for the time dependence of the single-particle density matrix in the lowest band, $\langle a_i^\dagger a_j\rangle$. It can be seen that in the Mott Insulator limit, studying the proper interplay between coherent and dissipative dynamics in this system shows that the system is even more robust towards spontaneous emissions than might have been expected from perturbation theory, as local correlations are reestablished by coherent dynamics. The superfluid state, on the other hand, exhibits a faster rate of decay for off-diagonal elements of the single-particle density matrix at medium times than had been predicted using perturbation theory. This appears to result from better thermalisation of the energy being added to the system from spontaneous emission events. 

\begin{figure}[tb]
  \centering
  \includegraphics[width=11cm]{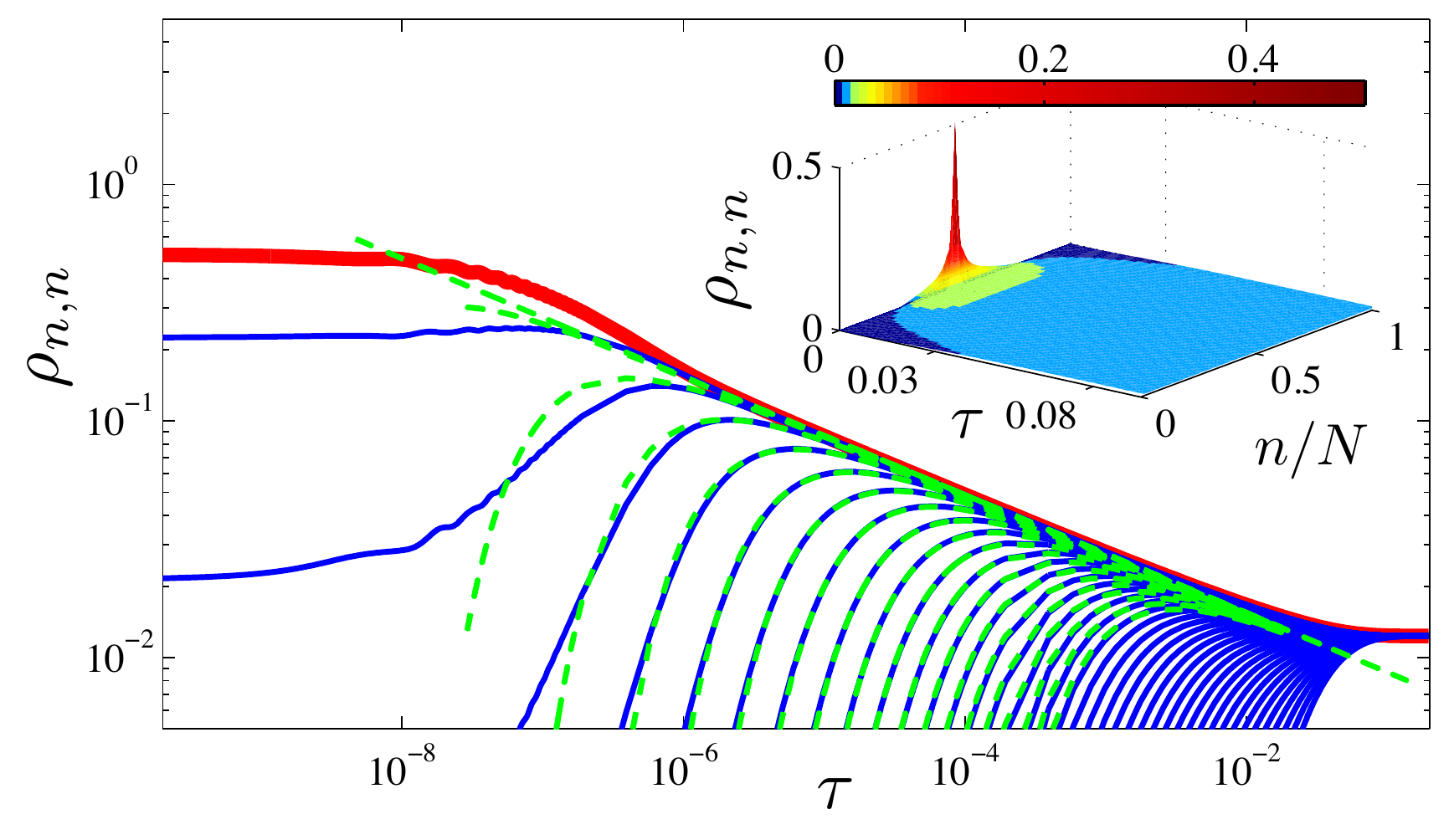}
  \caption{Diagonal terms of the density matrix $\rho_{n,n}$, with $ {\rho}=\sum \rho_{n,m} \ket{n}\bra{m}$ versus rescaled time $\tau$ in 
log-log form: the exact solutions of Eq.~(\ref{eq:QT_master_eq}) for $n=41$(top)$...80$(bottom)
are represented by blue continuous lines; the element $n=40$ (thick red line) is proportional to the probability of finding 
a balanced configuration $P_b$; the corresponding diffusion density $p(x=n/N,\tau)$ Eq.~(5) of Ref.~\cite{Poletti2012} up 
to $n/N=60$ are shown in dashed green lines. Inset: 3D plot of the evolution. 
Parameters: $U/J=20$, $\hbar\gamma/J=1$, $N=80$ and $L=2$. Reprinted figure with permission from D.~Poletti, J.-S.~Bernier, A.~Georges, and C.~Kollath, Phys. Rev. Lett {\bf 109}, 045302 (2012). Copyright 2012 by the American Physical Society. } \label{fig:kollath2}
\end{figure}

\begin{figure}[tb]
  \centering
  \includegraphics[width=9cm]{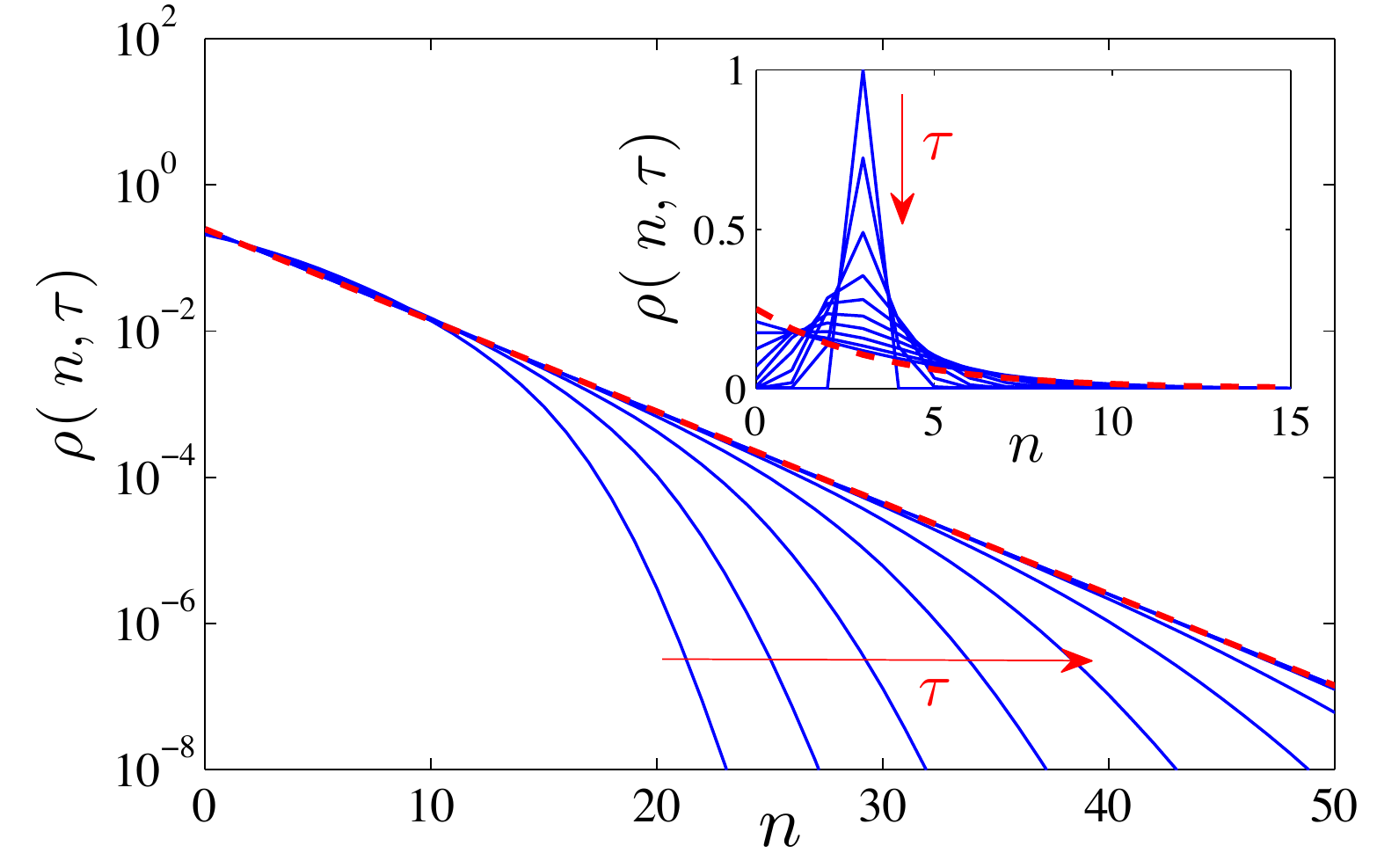}
  \caption{Numerical evolution of the density matrix elements $\rho(n,\tau)$ (solid lines), where the density operator is given by the ansatz $\bigotimes_j\left[\sum_{n}\rho(n,t)\ket{n}\bra{n}\right]$, in a semi-logarithmic plot, versus $n$ for large rescaled times $\tau$ between 0.1 and 50 (not equidistant) in the direction of the (red) arrow. The inset shows the same evolution at shorter times $\tau$ between 0.0002 and 0.1 (not equidistant) in the direction of the (red) arrow in a linear plot. Parameters: $N/L=3$, $U/\hbar\gamma=10$. The (red) dashed lines show the (analytical) asymptotic limit. Reprinted figure with permission from D.~Poletti, P.~Barmettler, A.~Georges, and C.~Kollath,  Phys. Rev. Lett {\bf 111}, 195301 (2013) \cite{Poletti2013}. Copyright 2013 by the American Physical Society. } \label{fig:kollath1}
\end{figure}

The interplay between interactions and dissipation also leads to interesting dynamics in on-site particle occupation numbers for bosons. In particular, dynamics resulting from spontaneous emissions generate significant probability that states with multiple occupancies on lattice sites -- which are disfavoured due to interaction energies -- will appear, and these probabilities continue to grow over time. In Refs.~\cite{Poletti2012,Poletti2013}, Poletti et al.~investigate these dynamics beginning from eq.~(\ref{eq:QT_master_eq}). They observe a number of striking effects, including the tendency of interactions to slow down decoherence, because spontaneous emissions distinguish only between states with different on-site particle numbers. Strong interactions lead to exponential localisation of particles at given lattice sites, with small amplitudes in perturbation theory for particles to be found at neighbouring sites. This results in anomalous diffusion in a particle number basis, even for a two-site model with large occupation \cite{Poletti2012}. For a lattice with many sites, it further gives rise to glass-like behaviour, exhibiting an anomalous diffusive evolution in configuration space at short times and dynamics dominated by rare events at long times \cite{Poletti2013}.

An example of these dynamics for a two-site system is reprinted from Ref.~\cite{Poletti2012} in Fig.~\ref{fig:kollath2}, and shows how the probability distribution for different occupation numbers on two sites diffuses as a function of time. Beginning with equal number of particles on each site, the initial behaviour is an exponential change away from this configuration, which is replaced by power-law decay with an exponent of 1/4, as the dynamics become dominated by processes populating states with high local occupancy. Related behaviour can be seen for a lattice with many sites, as demonstrated by a figure reprinted here from Ref.~\cite{Poletti2013} in Fig.~\ref{fig:kollath1}. The calculation shown involves an approximate solution to the master equation eq.~(\ref{eq:QT_master_eq}), obtained using a separable and translationally invariant ansatz for the density operator as $\bigotimes_j\left[\sum_{n}\rho(n,t)\ket{n}\bra{n}\right]$. Here the sum over $j$ includes all of the lattice sites and that over $n$ includes all of the possible occupations of each site. Again, it is clear that the long-time behaviour is dominated by anomalous diffusion that populates states with high onsite occupation numbers. Such dynamics should be directly observable in current experiments, especially in setups with local site addressing in a quantum gas microscope and control over light scattering rates. Initial experimental studies of dynamics in the presence of light scattering and density measurements were recently performed by Patil et al.~\cite{Patil2014}.

For fermionic species this interplay between interactions and dissipation can similarly lead to important many-body effects. In the case of multiple fermionic species, where the dissipation distinguishes between the two species, this can lead to a build-up of long-range correlations beginning from a fermionic Mott Insulator \cite{Bernier2013}. The same types of effects can also protect states of spin-ordered fermions, especially where spontaneous emission does not distinguish between different spin-species \cite{Sarkar2014}.

\begin{figure}[tb]
  \centering
  \includegraphics[width=10cm]{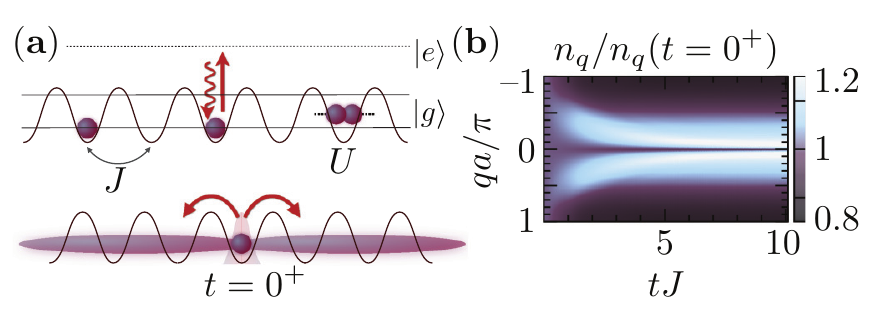}
  \caption{  (a) Absorption and spontaneous emission of a lattice photon
leads effectively to localization of single atoms. Tunnelling and interactions between
atoms then redistribute the energy added to the system. (b) Localization of an atom
in space corresponds initially to a distribution of the atom over the whole Brillouin
  zone (the tails of the quasi-momentum distribution are lifted). Subsequent unitary 
  evolution leads to a broadened quasi-momentum distribution, {\it i.e.}, the $n_{q=0}$ peak and the tails
  decrease, while the small quasi-momentum components increase (t-DMRG
  results, $U=2J$, $N=48$ particles on $L=48$ sites, $d_l=6$,
  $D=512$. Reprinted figure from J.~Schachenmayer, L.~Pollet, M.~Troyer, and A.~J.~Daley, Phys. Rev. A {\bf 89}, 011601(R) (2014) \cite{Schachenmayer2014}. Copyright 2014 by the American Physical Society.} \label{fig:schachenmayer1}
\end{figure}

Another important question related to the interplay between dissipative and coherent dynamics is that of whether spontaneous emission events that leave atoms in the lowest band of an optical lattice lead to energy increases that can be thermalised on typical experimental timescales. Though atoms being placed in higher bands cannot thermalise the bandgap energy, as we have already discussed, there is experimental evidence that suggests as a system heats due to spontaneous emissions that the momentum distributions can be quantitatively compared with thermal momentum distributions calculated via Quantum Monte Carlo methods \cite{Trotzky2010}.

In Ref.~\cite{Schachenmayer2014}, this is studied using quantum trajectory methods combined with t-DMRG, initially considering the thermalisation after a single spontaneous emission event (averaged over the different lattice sites on which the event can occur), and then generalising this to continuous light scattering. As shown in Fig.~\ref{fig:schachenmayer1}, after a spontaneous emission occurs, initially the momentum distribution develops strong tails as a result of a single atom being localised spatially on a lattice site, and therefore completely delocalised over the Brillouin zone in quasimomentum. As a result of interactions with other particles, the quasimomentum distribution then gradually relaxes towards a steady-state value. As a particular way of determining to what extent simple single-particle quantities have thermalised after a particular length of time, the time dependence of the kinetic energy and the quasimomentum distribution is then calculated, and compared with the thermal values from a canonical ensemble in which the temperature is chosen to match the \emph{total energy} in the system.

\begin{figure}[tb]
  \centering
  \includegraphics[width=10 cm]{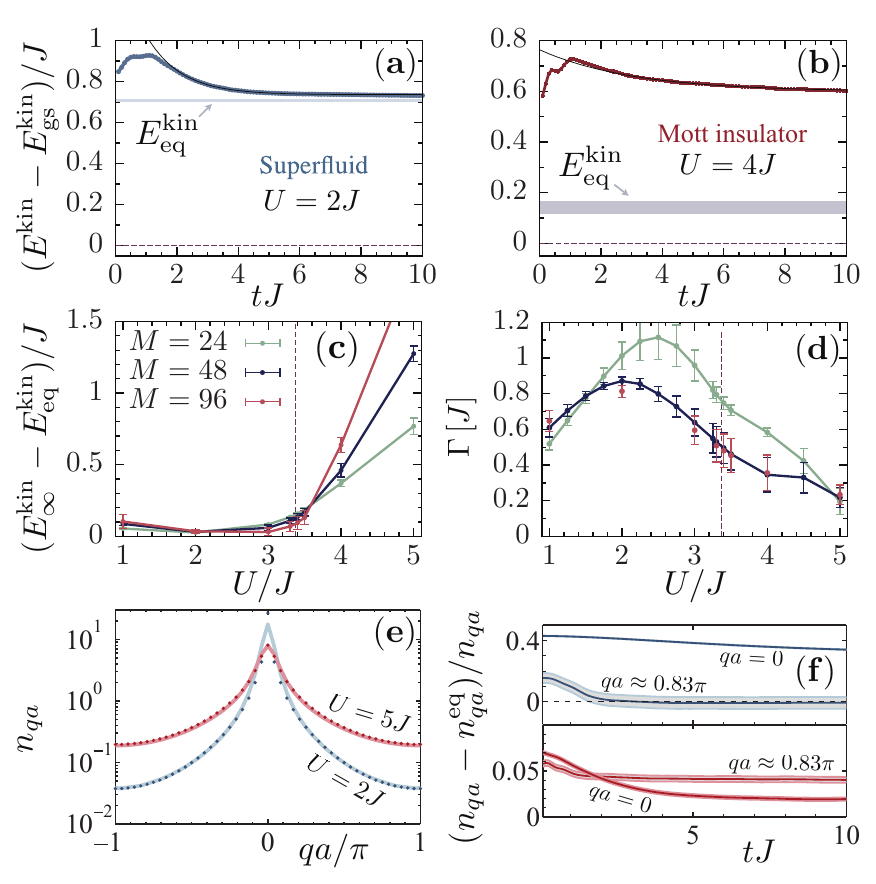}
  \caption{Time evolution after a single spontaneous
emission (averaged over jumps on all possible sites). (a - b) For
a superfluid initial state ($U=2J$), the kinetic energy relaxes
to the equilibrium value obtained from a Quantum Monte Carlo (QMC) calculation
$E_{\rm eq}^{\rm kin}$. For MI states ($U=4J$), the energy
relaxes, but not to $E_{\rm eq}^{\rm kin}$. The zero value of
kinetic energy for this plot is the ground state kinetic energy
$E_{\rm gs}^{\rm kin}$. (c) The difference of the infinite time
value of the kinetic energy (obtained from an extrapolation of an
exponential fit) to the Monte Carlo equilibrium energy. For MI
states with $U/J \gtrsim 3.37$, the difference increases rapidly
for system sizes $M=24,48,96$ sites (d) The decay rate extracted
from the exponential fit as a function of $U$. (e) Comparison of
the time-evolved quasi-momentum distribution at $t=10/J$ (dots)
to the equilibrium distribution from a QMC calculation. (f)
Differences between the two distributions as a function of time
for the $qa=0$ peak and for a large quasi-momentum of $qa=(40/48) \pi$. In
the superfluid ($U=2J$), the components for large momenta relax rapidly to thermal values, for $qa\sim 0$, the relaxation timescale is much longer. In the MI with $U=5J$, the same is true, but for large momenta there is a discrepancy to the thermal value.
(t-DMRG results, $d_l=6$, $D=256,512$; error bars represent
fitting errors and statistical errors from QMC.  Reprinted figure from J.~Schachenmayer, L.~Pollet, M.~Troyer, and A.~J.~Daley, Phys. Rev. A {\bf 89}, 011601(R) (2014) \cite{Schachenmayer2014}. Copyright 2014 by the American Physical Society.} \label{fig:schachenmayer2}
\end{figure}

It is found that both in the Mott Insulator and the superfluid regime, the system relaxes to a steady state value on a short timescale. As can be seen in Fig.~\ref{fig:schachenmayer2}a, in the superfluid regime the value to which the kinetic energy relaxes is equal to the expected thermal value $E_{\rm eq}^{\rm kin}$ in a canonical ensemble with the same total energy. Such a relaxation has also been recently explored for Luttinger liquids heated in a related manner \cite{Buchhold2014}. However, in the Mott Insulator regime (Fig.~\ref{fig:schachenmayer2}b), while the system relaxes rapidly to a steady-state value, this is not the same value as expected from a canonical ensemble with the same total energy. As shown in Fig.~\ref{fig:schachenmayer2}c, such discrepancies begin at the phase transition point, and become larger for larger $U/J$. This sudden transition is in contrast to the variation of the relaxation rate as a function of $U/J$, which is smoothly and slowly varying (Fig.~\ref{fig:schachenmayer2}d). Similar results and timescales are seen for the high-quasimomentum components of the quasimomentum distribution, as shown in Figs.~\ref{fig:schachenmayer2}e,f. For low-quasimomentum values also on the superfluid side the relaxation time is much longer, because these momentum values depend heavily on correlation functions over long distances. In Ref.~\cite{Schachenmayer2014}, it is shown that this behaviour, as a local quench, is strongly dependent on the low-energy spectrum of the Hamiltonian, which leads to the strong changes from the superfluid regime to the Mott Insulator regime. 

Microscopic understanding and control over spontaneous emissions in many-body systems also has other applications, including the potential to observe Pauli-blocking of spontaneous emissions \cite{Sandner2011,Busch1998} for Fermi gases. This can be used as an ingredient in state preparation schemes that produce pairing of fermions dissipatively \cite{Diehl2010,Yi2012}

With this recent progress, it has been possible to understand various basic features of heating and decoherence of many-particle states with cold atoms in optical lattices. In the near future it should be possible to investigate such effects for a much broader range of many-body states and Hamiltonians. This will provide both a better fundamental understanding of decoherence in many-body states, as well as important practical information for controlling heating in experiments. In the near future it will be important to understand better how the dynamical processes that atoms undergo as they are loaded into a lattice or many-body states are prepared interacts with the dissipation to produce heating during this phase of the experiment. It will also be important to go further into other dissipative mechanisms and their effects, including classical noise \cite{Pichler2012,Pichler2013} and particle loss (see the next section).

\subsection{Particle loss}
\label{sec:particleloss}

Another dissipative process with strongly interacting cold quantum gases that can be treated in experimentally relevant regimes using open systems techniques from quantum optics is particle loss. In collisional loss events, the binding energy released in the collision is typically much larger than the depth of the corresponding trap. In this regime, this energy acts as a large scale, exactly in the sense of section \ref{sec:approxamo}. In particular in the case that atoms lost in such events will leave the lattice without additional collisions, this allows Born and Markov approximations to be made, and the derivation of a corresponding master equation. This has been investigated theoretically for two-body \cite{Syassen2008,Garcia-Ripoll2009}, three-body \cite{Daley2009}, and single-body \cite{Barmettler2011} loss processes for cold atoms in optical lattices.

One particularly useful and counter-intuitive result that emerges is that these systems are often found in a regime where losses can be suppressed by a continuous quantum Zeno effect \cite{Itano1990,Misra1977}. This has been observed in experiments for two-body losses of molecules \cite{Syassen2008,Yan2013}, and initial indications have also been seen for three-body loss in caesium atoms \cite{Mark2012}. In this section we introduce the continuous quantum Zeno effect, before reviewing these observations and the consequences for many-body physics in these systems.

\subsubsection{Continuous quantum Zeno effect}
\label{sec:zeno}

The quantum Zeno effect was first discussed by Misra and Sudarshan in Ref.~\cite{Misra1977}, and involves the tendency of repeated measurements on a quantum system to inhibit coherent processes. In an idealised scenario, this can be formulated with projective measurements as follows: if we consider, e.g., a two level system that begins in the state $\ket g$ and is coupled resonantly to state $\ket e$ with effective Rabi frequency $\Omega$, then after a short time $\Delta \tau$, the probability a measurement finding the system to be in state $\ket e$, $P_e \approx \Omega^2 \Delta \tau^2 /4$. Because this result $\propto \Delta \tau^2$, if measurements are repeated in equally spaced intervals up to a time $T$, then as we increase the number of measurements $T/\Delta \tau$ by taking $\Delta \tau\rightarrow 0$, we see that the probability of finding the system in the excited state at time $T$ also goes monotonically to zero. This effect was first observed in trapped ions by Itano et al.~\cite{Itano1990}, and an anti-Zeno effect that arises under more general conditions has also been measured \cite{Fischer2001}. 

The \emph{continuous} quantum Zeno effect is a generalization of these ideas to continuous measurements \cite{Itano1990,Gagen1993}, in the sense that when a system is coupled to its environment, then the environment can be seen to be continuously performing measurements on the system. In essence it amounts to the fact that a strong dissipative process in a quantum system can suppress a related coherent process. Let us consider the specific case of a two-level atom described by the optical Bloch equations from section \ref{sec:twolevel}. In the presence of spontaneous decay of the excited state, we can write the quantum trajectories formulation of the evolution with an effective Hamiltonian given by
\begin{equation}
H_{\rm eff} = - \frac{\Omega}{2} \sigma_x - \Delta \sigma_+\sigma_-  - {\rm i}\frac{\Gamma}{2} \sigma_+\sigma_-.
\end{equation}
If the system begins in the ground state $\ket g$, then for small $\Gamma$ it will undergo damped Rabi oscillations as computed in section \ref{sec:twolevel}. Let us now consider the overdamped regime $\Gamma \gg \Delta, \Omega$. For the general case where $\Omega \ll \Delta$ and/or $\Omega \ll \Gamma$, we can apply perturbation theory to the effective Hamiltonian, taking $H_0=-(\Delta + {\rm i}\Gamma/2)\sigma_+\sigma_-$ and $H_1=- \frac{\Omega}{2} \sigma_x$, and we see that we can write the effective energy shift for the ground state, $E_g$ as:
\begin{equation}
E_g=\frac{\bra g H_1 \ket e \bra e H_1 \ket g}{\Delta + {\rm i}\Gamma/2}= \frac{\Omega^2}{4\Delta^2 + \Gamma^2} (\Delta - {\rm i} \Gamma/2)
\end{equation}
Based on the physical interpretation of quantum trajectory methods, we know that the probability of spontaneous emission occurring after time $t$, $P_s(t)$ is given by $1-\Vert \ket \Psi \Vert$, where $\Psi$ is the corresponding wavefunction evolving under the effective Hamiltonian. In the perturbation theory limit, we therefore have
\begin{equation}
P_s(t)=\left\Vert \exp\left[ -{\rm i} \frac{\Omega^2}{4\Delta^2 + \Gamma^2} (\Delta - {\rm i} \Gamma/2) t \right] \ket g  \right\Vert^2 =\exp\left[- \frac{\Omega^2}{4\Delta^2 + \Gamma^2} \Gamma t\right],
\end{equation}
so that the effective photon scattering rate is given by $\Gamma_{\rm scatt}\approx \Omega^2 \Gamma / (4\Delta^2+\Gamma^2)$. In the limit where $\Gamma \gg \Delta$, we see that this reduces to $\Gamma_{\rm scatt}\approx \Omega^2 /\Gamma$, i.e., as we continue to increase $\Gamma$, the scattering rate decreases.

This is the essence of the continuous quantum Zeno effect - for sufficiently strong dissipative processes, the dissipation will suppress coherent processes (in this case the classical drive with Rabi frequency $\Omega$) that occupy the states coupled to that dissipation. This leads to the counterintuitive situation that increasing the rate of dissipation for specific states actually leads to less actual dissipation in total (at least for a system beginning in a state that doesn't directly undergo dissipative processes -- if we begin in the state $\ket e$, it always decays at a rate $\Gamma$). This can also be rephrased in two other ways - one is that the linewidth of the excited state is so broad that coupling into it is suppressed. The other is that increasing $\Gamma$ increases the effective coupling to the environment, and therefore the rate at which the environment obtains information -- or \emph{measures} the system, analogously to the quantum Zeno effect with projective measurements.

\subsubsection{Two-body loss}

Evidence for the quantum Zeno effect resulting from two-body loss in a many-particle system was first obtained by Syassen et al.~\cite{Syassen2008}. Their setup involved weakly bound Feshbach molecules of Rb$_2$ \cite{Chin2010} held either in 3D optical lattice potentials, or confined to move in 1D tubes, in a regime where collisions between molecules resulted in the transfer of one molecule to a much more deeply bound state, and the separation of the atoms forming the second molecule. The binding energy released in this collision is typically much larger than the depth of the optical trap confining the molecules to begin with, resulting in loss of the particles involved in the collision. Because of the large energy scale involved in the inelastic collision, and because the particles are ejected immediately from the trap, the approximations made in section \ref{sec:approxamo} are justified, and a master equation can be derived to describe this process. Detailed analysis of these master equations are presented in Ref.~\cite{Garcia-Ripoll2009} for the 3D lattice case, and in \cite{Durr2009} for the case of particles in 1D tubes, based on a Lieb-Liniger model. In each case, a master equation can be derived for the density operator $\rho$ of the system, tracing over the states associated with the products of the collision. Assuming an effective zero-range potential to describe two-body loss in an equivalent form to the description of elastic interactions between particles, this can be written in terms of a field operator for bosonic Feshbach molecules in the initial state, $\hat \psi_m (\mathbf{r})$ as \cite{Garcia-Ripoll2009}
\begin{eqnarray}
\frac{d\rho}{dt}&=&-{\rm i} [H_{\rm mol},\rho]\nonumber\\ & & -\frac{L_2}{2} \int d^3 r \left\{ 2 [\hat\psi_m(\mathbf{r})]^2\rho [\hat\psi_m^\dagger(\mathbf{r})]^2 - [\hat\psi_m^\dagger(\mathbf{r})]^2 [\hat\psi_m(\mathbf{r})]^2\rho - \rho [\hat\psi_m^\dagger(\mathbf{r})]^2 [\hat\psi_m(\mathbf{r})]^2  \right\},\nonumber \\ 
\end{eqnarray}
where $L_2$ is the two-particle loss coefficient, and the coherent hamiltonian $H_{\rm mol}$ takes the same form as the atomic hamiltonian eq.~\eqref{hatoms} in section \ref{sec:qsim},
\begin{equation}
{H}_{\rm mol}\approx\int d^{3}r\,\hat{\psi}_m^{\dag}(\mathbf{r})\,\left[-\frac{\hbar^{2}}{2m}\nabla^{2}+V_{0}(\mathbf{r})\right]\,\hat{\psi}_m(\mathbf{r})+\frac{g}{2}\int d^{3}r\,\,[\hat{\psi}_m^{\dag}(\mathbf{r})]^2\,[\hat{\psi}_m(\mathbf{r})]^2.  \label{hmols}
\end{equation}
Here, $m$ is again the particle mass (now the mass of a Feshbach molecule), and $g=4\pi \hbar^2 a_{\rm scatt} / m$, where $a_{\rm scatt}$ is the real part of the scattering length for two Feshbach molecules ($L_2$ may be interpreted as the imaginary part of an effective two-particle scattering length including inelastic processes \cite{Garcia-Ripoll2009}). 

In the case of a 3D optical lattice, this can be rewritten as an equation of motion for molecules in the lowest band of the lattice, analogously to the derivation of the Bose-Hubbard model that was discussed in section \ref{sec:qsim}. The resulting master equation is given by \cite{Garcia-Ripoll2009}
\begin{equation}
\frac{d \rho}{dt} =- {\rm i}\left(H_{\rm eff,m} \rho - \rho H_{\rm eff,m}^\dag \right) +{\Gamma_0} \sum_l \hat m_l^2 \rho (\hat m_l^\dag)^2,
\label{3bodylossmastereq}
\end{equation}
with bosonic annihilation operators for Feshbach molecules in Wannier function modes denoted $m_l$, and the effective Hamiltonian
\begin{equation}
H_{\rm eff,m}=H_{\rm BH,m}-{\rm i} \frac{\Gamma_0}{2} \sum_l (\hat m_l^\dag)^2 \hat m_l^2.
\end{equation}
Here, $H_{\rm BH,m}$ is the Bose-Hubbard Hamiltonian for molecules on the lattice, taking an identical form to $H_{\rm BH}$ from eq.~\eqref{HBH}, but with bosonic molecule operators $m_l$, and $\Gamma_0$ denotes the effective on-site loss rate when two molecules occupy the same site of the lattice. As for the previous discussion of the Bose-Hubbard model in section \ref{sec:qsim}, we require all parameters to be small compared with the band gap energy, and assume that terms involving off-site processes other than nearest neighbour tunnelling are negligibly small (see below for further comments on this). 

In order to interpret more directly the dynamics described by this master equation, we can expand the density operator in terms of the contributions with fixed particle number $N$, $\rho_{N}$.  Beginning with $N_0$ particles in the system, $\rho=\sum_{N=N_0,N_0-2,N_0-4,...} \rho_{N}$, and
\begin{equation}
\dot \rho_{N} = - {\rm i}\left(H_{\rm eff} \rho_N- \rho_N H_{\rm eff}^\dag \right) +\Gamma_0 \sum_i \hat m_i^2 \rho_{N+2} (\hat m_i^\dag)^2. 
\end{equation}
We can then consider the loss as having two effects: it produces states with two less particles, with the corresponding coupling between density matrices given by the recycling term, and for a fixed particle number, it provides an effective two-body on-site interaction term with an imaginary coefficient. This term generates the quantum Zeno effect analogously to the case of a two-level system discussed in section \ref{sec:zeno}.

\begin{figure}[tb]
  \centering
  \includegraphics[width=10cm]{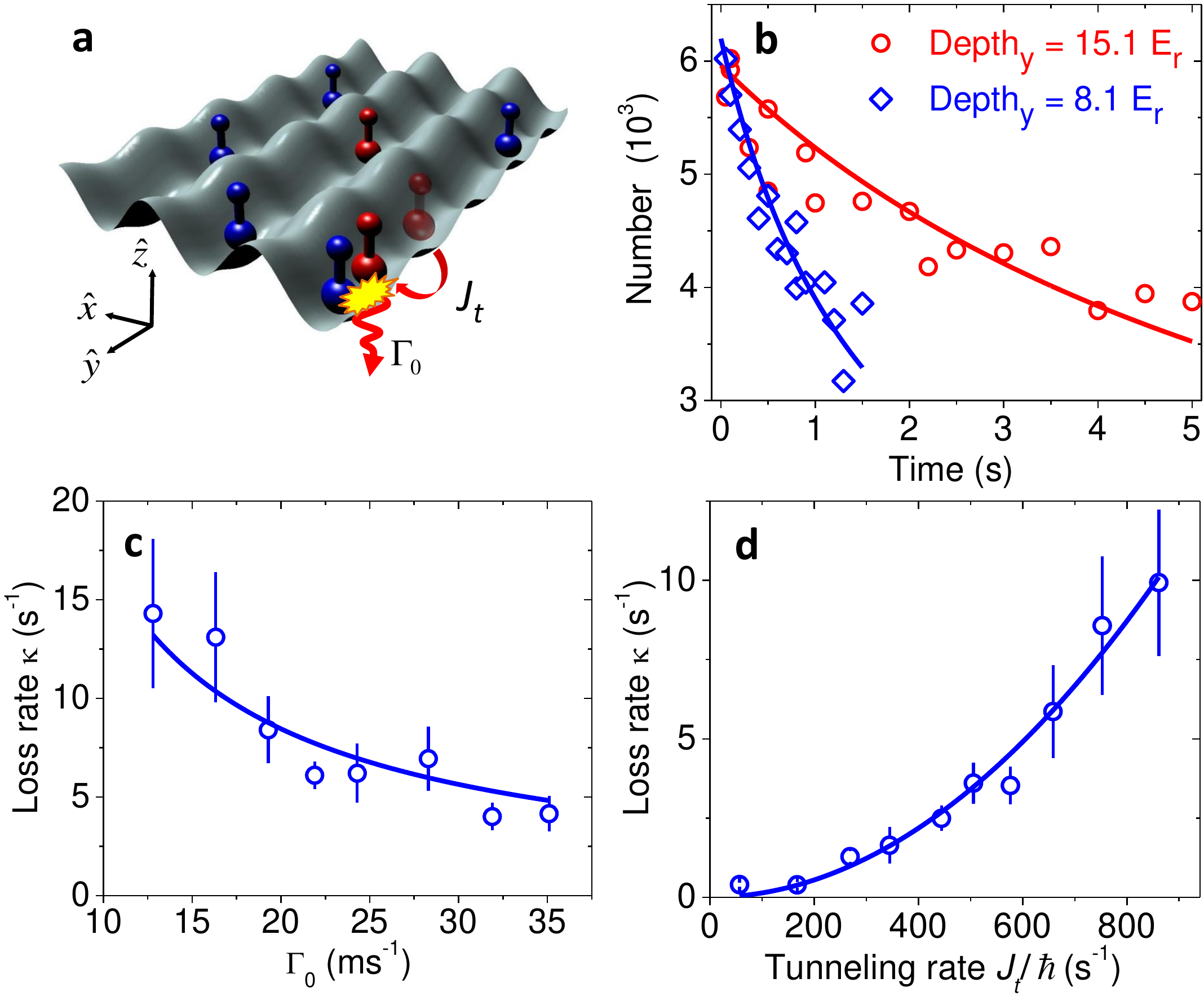}
  \caption{Quantum Zeno effect for polar molecules in a 3D lattice.
\textbf{~a},~The lattice depths along $x$ and $z$ are kept at $40$~$E_R$, while the lattice depth along $y$ is reduced to allow tunneling along the $y$ direction at a rate $J_t/\hbar$. Once two molecules in different spin states tunnel to the same site, they are lost due to chemical reactions at a rate $\Gamma_0$.
\textbf{b},~Number loss of $|\downarrow \rangle$ state molecules versus time is shown for lattice depths along $y$ of $8.1$~$E_R$ and $15.1$~$E_R$.
\textbf{c},~The number loss rate $\kappa$ versus $\Gamma_0$ fits to a $1/\Gamma_0$ dependence, which is consistent with the quantum Zeno effect.
\textbf{d},~The number loss rate $\kappa$ versus $J_t$ fits to a $(J_t)^2$ dependence, as predicted from the quantum Zeno effect. Reprinted by permission from Macmillan Publishers Ltd: Nature, B.~Yan, S.~A.~Moses, B.~Gadway, J.~P.~Covey, K.~R.~A.~Hazzard, A.~M.~Rey, D.~S.~Jin, and J.~Ye, Nature {\bf 501}, 521 (2013), copyright 2013.} \label{fig:ye1}
\end{figure}

\begin{figure}[tb]
  \centering
  \includegraphics[width=\textwidth]{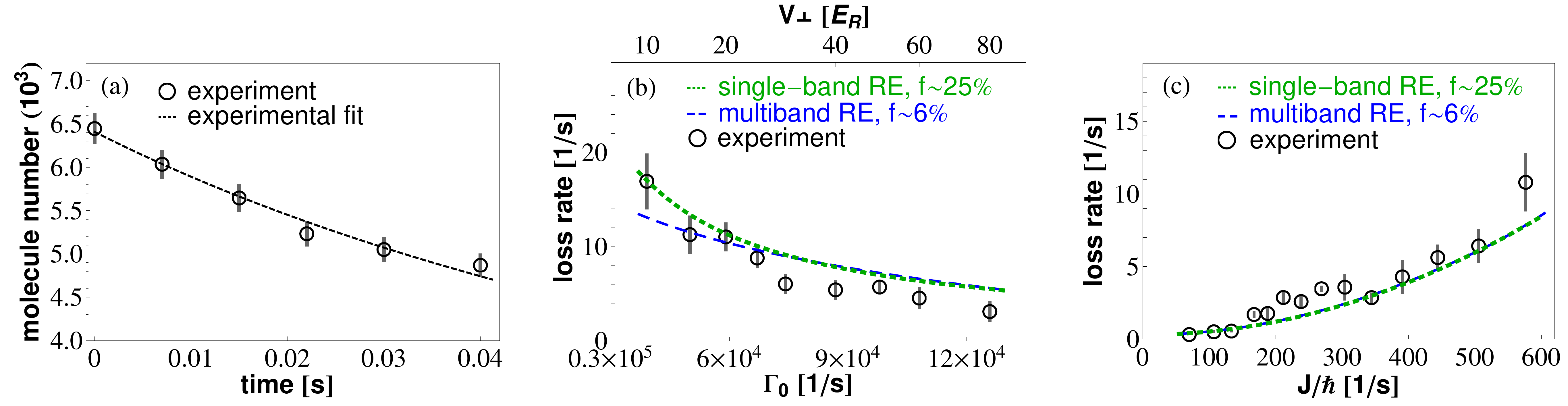}
  \caption{(a) Measured number loss of $\ket{\downarrow}$ molecules for an axial (transverse) lattice depth of $V_y=5\,E_R$ ($V_\perp=25\,E_R$) (circles) and best fit using a rate equation, $dN_{\downarrow}/dt=-\kappa N_{\downarrow}(0)/ [N_{\downarrow}(t)]^2$  (black dashed line). 
(b) Number loss rate, $\kappa$, as a function of $\Gamma_0$  (fixing $J\approx570~$Hz and varying the bare on-site rate via $V_\perp$).  (c) Number loss rate, $\kappa$, versus $J$ for fixed $\Gamma_0\approx87$~kHz (varying $V_y$ and adjusting $V_{\perp}$  accordingly).  $V_y$ ($V_{\perp}$) was varied from 5 to $16\,E_R$ ($20$ to $40\,E_R$). Black circles are experimental measurements (error bars represent one standard error).  Green short-dashed lines show solutions of the rate equation using an effective loss rate $\Gamma_{\rm{eff}}\propto J^2/\Gamma_0$ (single-band approximation). The blue long-dashed line shows the multiband rate equation using $\tilde{\Gamma}_{\rm{eff}}\propto \tilde{J}^2/\tilde{\Gamma}_0$, in which $\tilde J$ and $\tilde \Gamma_0$ are renormalized coefficents based on the effective multi-band theory. The multiband and single-band rate equation results were obtained by fixing the filling fraction to be $ 6\%$, and $25\%$ respectively. Panels (b) and (c) directly manifest the continuous quantum Zeno effect: in (b) the measured loss rate $\kappa$ decreases with increasing on-site $\Gamma_0$; in (c) a fit to the experimental data  supports $\kappa\propto J^2$, with a $\chi^2$ (sum of the squared fitting errors) several times smaller than for a linear fit. Reprinted figure with permission from B.~Zhu et al., Phys. Rev. Lett. {\bf 112}, 070404 (2014). Copyright 2014 by the American Physical Society.} \label{fig:rey1}
\end{figure}

An identical treatment can also be performed to describe two-particle losses in ground-state molecules due to chemical reactions, as was recently done by Yan et al.~\cite{Yan2013} for the case of KRb molecules on a lattice, which can reach to form Rb$_2$ and K$_2$ molecules, again with a release of binding energy that is much larger than the lattice depth \cite{Yan2013,Zhu2014}. In Fig.~\ref{fig:ye1}, we reprint their results demonstrating the suppression of loss due to the quantum Zeno effect. In particular, by allowing tunnelling only along one direction in a 3D optical lattice, they make it possible to independently control the tunnelling $J$ and the on-site loss rate $\Gamma_0$ by controlling the lattice depth along the 1D axis and transverse to it \cite{Yan2013,Zhu2014}. The setup, initially involving KRb molecules in separate lattice sites, is depicted in Fig.~\ref{fig:ye1}a, and as a function of time leads to losses of molecules as depicted in Fig.~\ref{fig:ye1}b. By fitting a loss rate coefficient $\kappa$ based on the time-dependence of the molecule number $N_{\downarrow}(t)$,
\begin{equation}
\frac{dN_{\downarrow}}{dt}=-\kappa \frac{N_{\downarrow}(0)}{ [N_{\downarrow}(t)]^2},
\end{equation}
they can investigate the change in the effective loss rate as $J$ and $\Gamma_0$ are varied. The results shown in Figs.~\ref{fig:ye1}c,d demonstrate very clearly the suppression of loss as $\Gamma_0$ is increased, by showing that $\kappa\propto J^2/\Gamma_0$. The underlying mechanism for this is the quantum Zeno effect in the form discussed in section \ref{sec:zeno}, i.e., a strong dissipative loss process on a given site will suppress coherent tunnelling processes that populate sites with two particles, forming states that can potentially undergo loss.

As was noted above, the form we wrote based on the Bose-Hubbard model for particle loss in the lowest band of an optical lattice requires a series of assumptions, especially neglecting coupling to higher bands and off-site loss processes. In Ref.~\cite{Zhu2014}, a further theoretical analysis of the results of Yan et al.~\cite{Yan2013} was performed, taking into account additional terms involving higher bands and off-site loss processes. In Fig.~\ref{fig:rey1}, we reprint part of their theoretical comparison to experimental data, involving both the simple single-band loss model discussed above, and a more general multi-band model. They find that both models give quantitatively very similar results, and confirm that the quantum Zeno effect suppression of the losses survives these extensions of the model. 

These effects naturally generalise to other systems undergoing two-body loss in which the energy released is larger than the trapping potential, and the resulting product particles are ejected from the system. For example, group-II atoms in metastable excited states would undergo similar loss processes. In Ref.~\cite{Foss-Feig2012}, Foss-Feig et al.~investigated this process for short-range s-wave collision processes. It was shown that this can lead to an interesting effect, because many-body states that are completely anti-symmetric under exchange of fermions in space do not undergo loss. If the Hamiltonian fulfils the requirement that spatially symmetric states can be eigenstates of the Hamiltonian (which is true, e.g., for two-species fermions in a harmonic trapping potential), then such states can be long lived. Then, if we begin with $N$ particles in the system in randomly chosen initial single-particle states, then of the order of $\sqrt{N}$ particles are expected to remain in a steady state. Moreover, because the resulting state is completely anti-symmetric in space, it will be completely symmetric in spin, and could be used for Heisenberg-limited spectroscopy. This results in a result for quantum-enhanced spectroscopy where the precision is not degraded by the loss of atoms, but collisional shifts are substantially reduced, potentially affording substantial improvements in accuracy \cite{Foss-Feig2012}. 

Such collisional loss in group-II atoms has also been proposed as a means to implement dissipative blockade gates for quantum computing \cite{Daley2011,Daley2008b}, in which the strong loss plays the role of an effective interaction for the gate. Two-particle loss was also shown to generate pairing where Rydberg states are designed to be distance-selective \cite{Ates2012}. 

Collective excitation dynamics including two-body loss have already been measured for group-II atoms in optical dipole traps \cite{Martin2013}. In these experiments, high-stability lasers make it possible to probe a system with relatively weak interactions using a Ramsey sequence on the clock transition, in such a way that the interactions become dominant. These experiments are also remarkable in that collective quantum dynamics emerge in the spin degree of freedom, despite relatively high motional temperatures for the atoms.

\subsubsection{Three-body loss}

While the suppression of two-body loss processes for molecules can be very important in stabilising quantum gases in experiments with molecules, and is of substantial fundamental interest, for many atomic species similar effects can be achieved by increasing the elastic two-body interactions (e.g., using Feshbach resonances \cite{Chin2010}). However, the potential to suppress three-body loss processes via a quantum Zeno effect opens very different opportunities, because it implies the possibility to produce effective three-body interactions, which are otherwise difficult to engineer in dilute quantum gases. Three-body loss for atoms involves collisional processes in which two particles form a molecule, and given that the resulting binding energy is large compared with the trap depth, the resulting molecule and atom will be ejected from the system. Such suppression of three-body loss events was first discussed for bosons in optical lattices in Ref.~\cite{Daley2009}, and the first evidence for such a suppression was recently obtained in experiments with Caesium atoms in an optical lattice in Ref.~\cite{Mark2012}. 

\begin{figure}[tbh]
\begin{center}
\includegraphics[width=6cm]{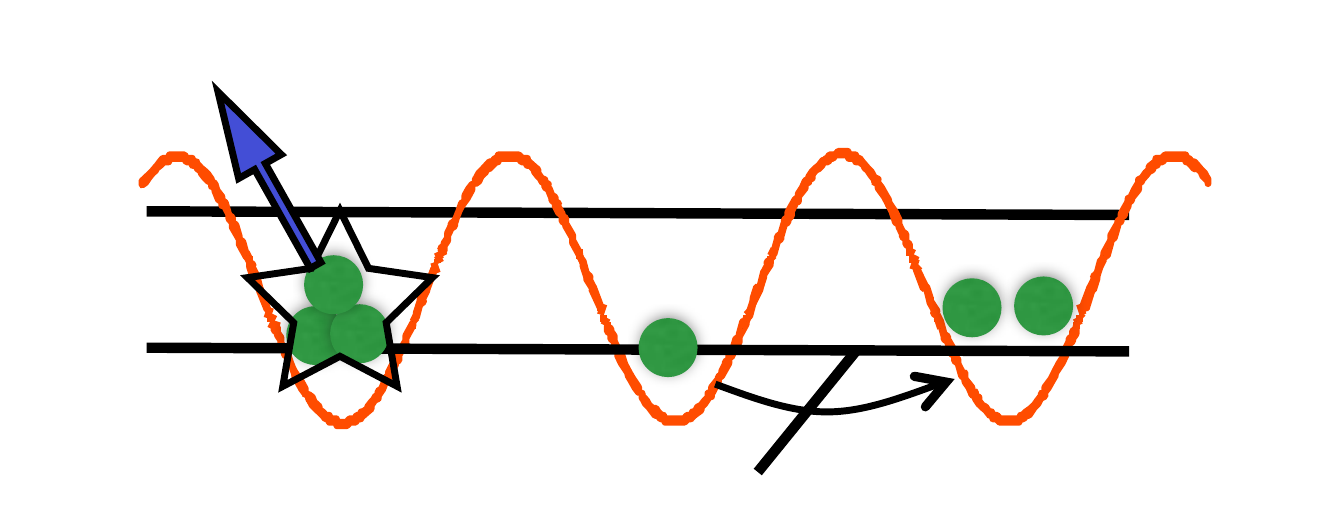}
\end{center} 
\caption{Schematic showing the continuous quantum Zeno effect in the case of three-body loss in an optical lattice. If we have a large on-site rate of loss $\gamma_3$, such that the tunnelling in the lowest Bloch band $J \ll \gamma_3$, then tunnelling processes that populate a single site with three particles (i.e., where a particle on one site tunnels onto an already doubly-occupied site) will be suppressed. In this form, large on-site loss rates can counterintuitively lead to a reduction in the overall rate of loss, and can act as an effective dynamical interaction, which suppresses triple occupation.}\label{fig:3bodyloss}
\end{figure}

The basic process, which is depicted in Fig.~\ref{fig:3bodyloss}, takes the same form as that for two-body loss. In a 3D optical lattice potential, if the on-site loss rate $\gamma_3$ for three atoms is very large, then this strong dissipative process can suppress coherent tunnelling that would otherwise populate triply occupied sites. In this form, both the effective resulting rate of three-body loss, and three-body occupation on a lattice site can be strongly suppressed. As with the case of two-body losses, in the limit that the binding energy released in the collision is much larger than the trap depth, and the products of the collision are immediately ejected from the lattice, the loss process can be described by a master equation. Restricting again to the lowest band of an optical lattice, the equation of motion for the system density operator $\rho$ is given by \cite{Daley2009}
\begin{equation}
\frac{d \rho}{dt} =- {\rm i} \left(H_{\rm eff} \rho - \rho H_{\rm eff}^\dag \right) +\frac{\gamma_3}{6} \sum_i  a_i^3 \rho ( a_i^\dag)^3,
\label{meq3body}
\end{equation}
with the effective Hamiltonian given by
\begin{equation}
H_{\rm eff}=H_{\rm BH}-{\rm i} \frac{\gamma_3}{12} \sum_i ( a_i^\dag)^3 a_i^3,
\end{equation}
where $H_{\rm BH}$ is the Bose-Hubbard hamiltonian [eq.\eqref{HBH}], and $\gamma_3$ is three-body loss rate for three particles occupying a single lattice site. 

\begin{figure}[t]
\par
\begin{center}
\includegraphics[width=8cm]{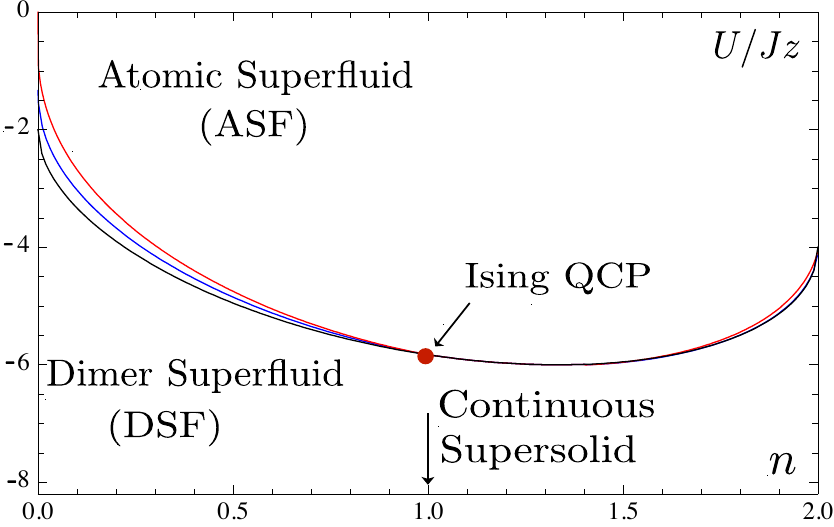}
\end{center}
\par\vspace{-0.5cm}
\caption{Phase diagram for the Bose-Hubbard model with a three-body hard-core constraint, and $U<0$. The black curve
represents the mean-field phase border, while red (light gray) and blue (dark gray) curves include
shifts due to quantum fluctuations in $d=2,3$. 
An Ising quantum critical point
is predicted in the vicinity of unit filling. The continuous supersolid is
reached asymptotically at unit filling. Reprinted figure from S.~Diehl, M.~Baranov, A.~J.~Daley, and P.~Zoller, Phys. Rev. Lett {\bf 104}, 165301 (2010). Copyright 2010 by the American Physical Society.
}
\label{fig:diehl}%
\end{figure}

In Ref.~\cite{Daley2009}, the dynamics is considered beginning in a subspace where no lattice sites are triply occupied. Analogously to the case of the two-level system in section \ref{sec:zeno}, in the limit $\gamma_3\gg J,U$ it is possible to perform perturbation theory on the effective Hamiltonian projected with an operator $P$ onto the subspace of states with at most two atoms per site. In second order-perturbation theory the effective model is given by the projected effective Hamiltonian \cite{Daley2009}
\begin{equation}
H_{\rm eff}^P \approx  P H_{\rm BH} P - {\rm i} \frac{6J^2}{\gamma_3} P\sum_j  c_j^{\dag} c_jP. \label{eq:model}
\end{equation}
Here the operator $ c_j = ( a_j^2/\sqrt{2}) \sum_{k\in {\mathcal N}_j}  a_k$, with ${\mathcal N}_j$ denoting the set of nearest neighbours of site $j$. In this way, the term $PH_{\rm BH}P$ describes dynamics in a Bose-Hubbard Hamiltonian in the presence of the constraint $(a_i^\dag)^3\equiv 0$ \cite{Paredes2007}, and corrections to this description describe three-body recombination, which in the projected effective Hamiltonian begins from two particles present on a site neighbouring a single particle. These three-body loss events occur at a rate $\propto J^2/\gamma_3$, so that we see the continuous quantum Zeno effect again arising naturally in the limit $\gamma_3/J \gg 1$. For timescales that are shorter than the effective loss timescale, we are left with dynamics described by a constrained Bose-Hubbard model.

This constrained Bose-Hubbard model exhibits particularly interesting many-body properties \cite{Daley2009,Diehl2010a,Paredes2007}, beginning with the fact that the three-body constraint stabilises the system for attractive two-body interactions $U<0$. In the absence of the constraint, such attractive interactions would favour build-up of all bosons on one site, but with the maximum on-site occupancy limited to two, the system undergoes a second-order quantum phase transition. Using Gutzwiller mean-field techniques (Sec.~\ref{sec:gutzwiller}), this phase transition can be described as a transition from a usual atomic superfluid with mean-field parameter $\langle a_l \rangle\neq 0 $ (and also $\langle a_l^2\rangle\neq 0$) to a dimer superfluid, where $\langle a_l \rangle=0 $, but $\langle a_l^2\rangle\neq 0$. In Fig.~\ref{fig:diehl}, we reprint the phase diagram from Ref.~\cite{Diehl2010a}, showing the phase boundary at zero temperature in mean-field and including quantum fluctuations, as a function of $U/(Jz)$, where $z$ is the number of nearest neighbour sites, and $n$, the filling fraction in the system. This phase diagram has been investigated in further detail numerically and analytically \cite{Diehl2010b,Diehl2010d,Bonnes2011,Lee2010}, and exhibits an Ising critical point between the atomic superfluid and dimer superfluid phases near unit filling $n=1$. For stronger attractive interactions, bound dimers can form a phase with a coexistence of dimer superfluid and charge density wave order, in a type of continuous supersolid \cite{Diehl2010a,Diehl2010d}. Similar related three-body physics for bosons has been studied in a variety of contexts, including nearest-neighbour interactions that are realisable with polar molecules \cite{Capogrosso-Sansone2009,Buchler2007b, ChenYC2011,Ng2011}, and finite on-site three-body interactions \cite{Chen2008,Mazza2010,Silva-Valencia2011,Safavi-Naini2012,singh2012,Sowinski2012,Daley2013}.

\begin{figure}[tb]
\begin{center}
\includegraphics[width=10cm]{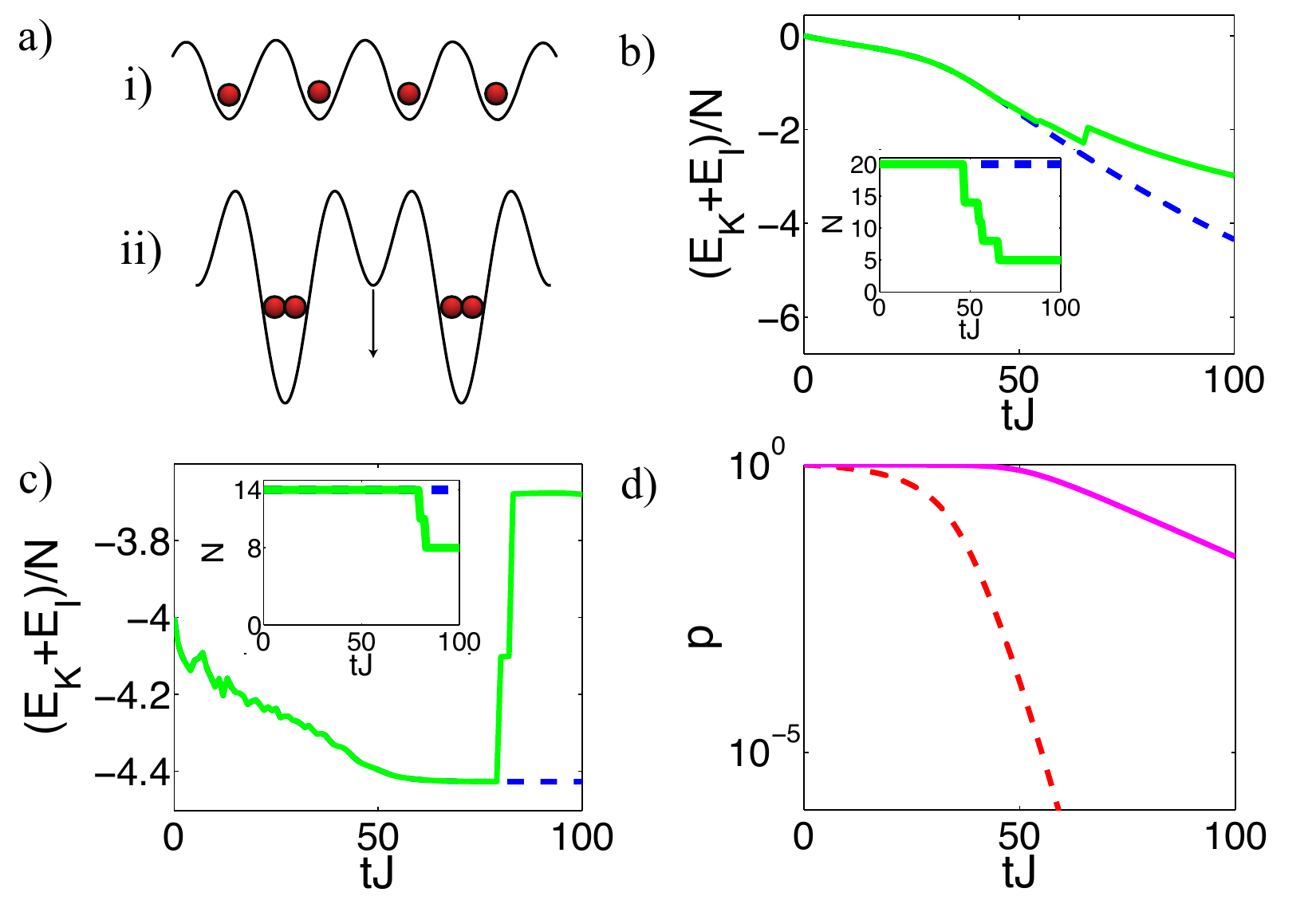}
\end{center} 
\caption{Dynamics of adiabatic ramps into a dimer superfluid regime. (a) We begin with (i) a Mott insulator state (ramping $U/J$), and (ii) a state with pre-prepared dimers in a superlattice (removing the superlattice). (b)-(c) The sum of kinetic ($E_K$) and interaction ($E_I$) energy and (inset) particle number as a function of time for two example trajectories, one with no loss events (dashed lines) and one with several loss events (solid lines). Here, (b) shows a ramp from $U/J=30$ to $U/J=-8$, with $U(t)=\alpha J/(100 + 3tJ)+\gamma$, with $\alpha$ and $\gamma$ ramp parameters, and (c) shows a ramp with a superlattice potential, $\varepsilon_l=V_0 \cos(2\pi l /3)$, where $V_0\approx30 J\exp(-0.1 tJ)$, adjusted so that $V_0(tJ=100)=0$, with fixed $U/J=-8$. In each case, $\gamma_3=250J$. For (b), we use 20 atoms on 20 lattice sites, for (c), 14 atoms on 23 lattice sites. (d) Plot showing the probability that no loss event has occurred after time $t$ for the ramps in (b) (dashed line) and (c) (solid). Reprinted figure from A.~J.~Daley, J.~M.~Taylor, S.~Diehl, M.~Baranov, and P.~Zoller, Phys. Rev. Lett {\bf 102}, 040402 (2009). Copyright 2009 by the American Physical Society.}\label{fig:daley3body}
\end{figure}

In an experiment real losses will also occur, making it important to model the complete dissipative many-body dynamics. For this purpose, in Ref.~\cite{Daley2009}, quantum trajectories techniques were first combined with t-DMRG to solve the master equation \eqref{meq3body}. In this paper, quantum trajectories were used to study the adiabatic preparation of a dimer superfluid state in the presence of three-body loss. Example results are shown in Fig.~\ref{fig:daley3body}, including ramps from an initial Mott Insulator state with a single atom per site (Fig.~\ref{fig:daley3body}b), and ramping in a superlattice as depicted in Fig.~\ref{fig:daley3body}a, using a site-dependent energy offset $\epsilon_l$, as was defined in $H_{\rm BH+}$ in section \ref{sec:qsim}. Figs~\ref{fig:daley3body}b,c show the total energy $\langle H_{\rm BH} \rangle$ and total particle number for example trajectories in a ramp that would adiabatically realise the ground state in the absence of losses. The dashed lines show trajectories without particle losses, whereas the solid lines show examples including losses. By propagating a state under the effective Hamiltonian, it is possible to determine for each ramp type what the probability of having no losses will be, which is depicted as a function of time in Fig.~\ref{fig:daley3body}d. These results clearly demonstrate advantages for the superlattice ramp, because the superlattice holds particles apart for a long time, and by beginning with a finite attractive interaction always includes elastic interactions to help suppress three-body occupation. The ramp from the Mott insulator, in contrast, passes through a regime of small coherent on-site interactions, where the loss is increased. In Ref.~\cite{Daley2009}, it is further shown that dimer superfluid order can survive some loss events, and that preparation of this state via the superlattice ramp should be realistic for experimentally realisable parameters, especially in Caesium. 

These effects are also especially interesting for fermions in an optical lattice, where large three-body loss rates have been measured for three-component Lithium gasses \cite{Ottenstein2008}. There it is found that the three-body constraint induced for short times by the quantum Zeno effect will suppress formation of trimers for attractive interactions \cite{Kantian2009, Privitera2011,Titvinidze2011}, stabilising an atomic colour superfluid state with two-component BCS pairing in a three-component Fermi gas \cite{Rapp2008,Rapp2007}. Quantum Zeno suppression generating effective three-body interations has also been discussed for trapped gasses not in an optical lattice, including weakly interacting Bose-Einstein condensates \cite{Schutzhold2010}, and strongly interacting systems where Pfaffian-like states can be formed via the induced effective interactions \cite{Roncaglia2010}. Such effective interactions can also stabilise p-wave pairing in strongly interacting fermions \cite{Han2009}.

\subsubsection{Single-particle loss}

Single-particle atomic losses can be generated and controlled in a number of forms in an optical lattice, including photon scattering that drives atoms to momentum states with energy greater than the lattice depth. In certain regimes, this can lead to quantum optical analogues, in which the trapped atoms represent few-level systems, and atoms in untrapped states play the role of a radiation field \cite{Vega2008,Navarrete-Benlloch2011}. By tuning the laser field that drives atoms out of the lattice, super-radience and non-Markovian emission of atoms was predicted in Refs.~\cite{Vega2008,Navarrete-Benlloch2011}. 

Local single-particle losses can be generated in experiments either by collisions with a background gas \cite{McKay2011}, or deliberately using individual addressing with an electron beam \cite{Gericke2008} or with light in quantum gas microscopes \cite{Weitenberg2011} that have single-site resolution in an optical lattice \cite{Bakr2009,Sherson2010}. As in the previous cases, the energy transferred to atoms that are removed from the lattice in this manner is much larger than the trap depth, and for these processes, a master equation analogous to eq.~\eqref{3bodylossmastereq}, but for single-particle loss can be derived \cite{Barmettler2011,Witthaut2011,Kordas2013,Vidanovic2014}. Theoretically, this master equation has been solved using quantum trajectory techniques with t-DMRG \cite{Barmettler2011} and a time-dependent Gutzwiller ansatz \cite{Vidanovic2014}. The dissipative dynamics produces complex effects, including the generation of entanglement under appropriate conditions \cite{Barmettler2011,Kordas2013}, and the ability to detect and control supersolid phases \cite{Vidanovic2014}.

\begin{figure}[tb]
  \centering
  \includegraphics[width=12cm]{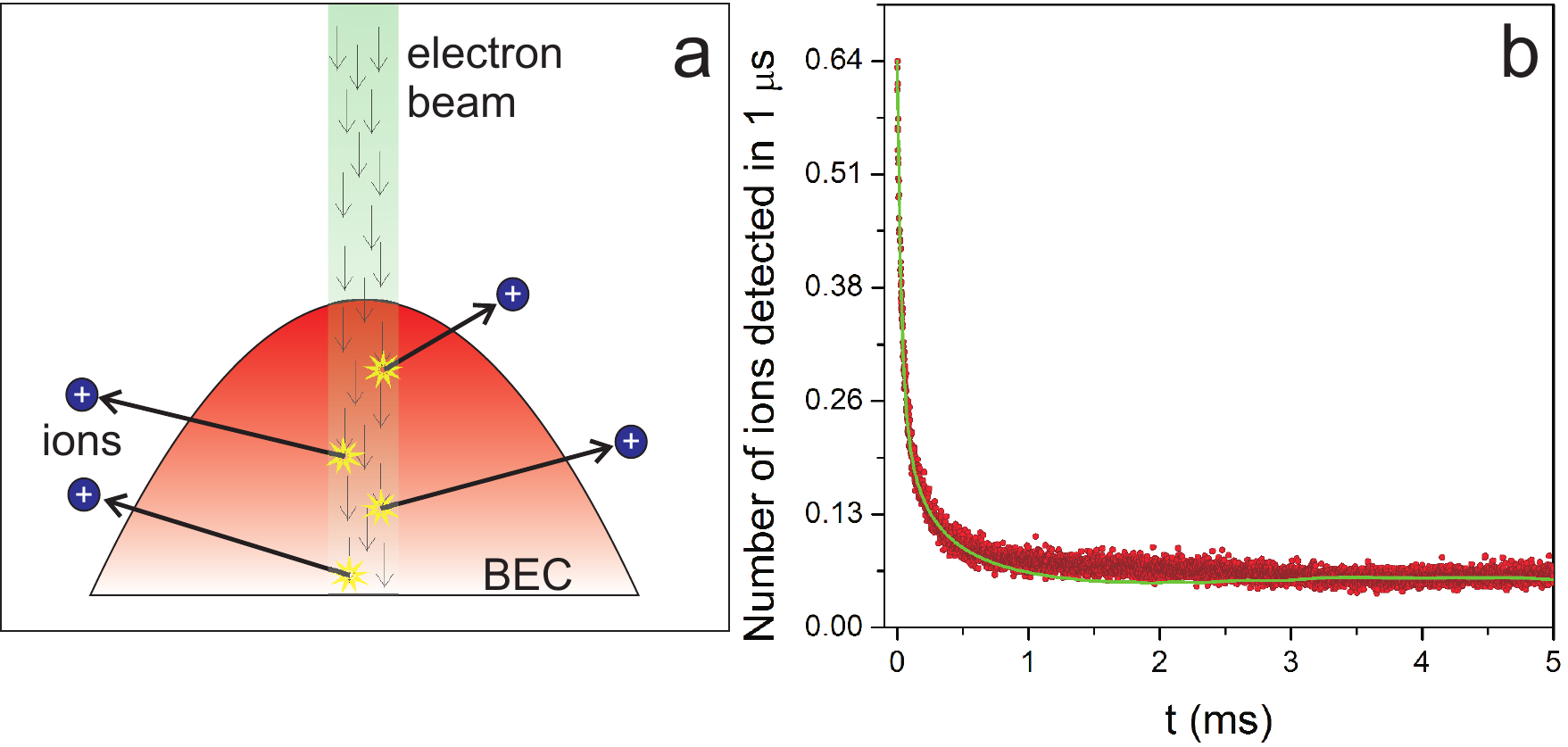}
  \caption{a) The electrons locally collide with the atoms constantly dissipating the BEC. b) Temporal resolved signal from the ion detector. The bin size is 1 $\mu$s. Points are experimental data averaged over 1800 experimental repetitions, while the solid curve is the numerical simulation (see text). After 5 ms we typically collect $\simeq450$ ions. Reprinted figure with permission from G.~Barontini, R.~Labouvie, F.~Stubenrauch, A.~Vogler, V.~Guarrera, and H.~Ott, Phys. Rev. Lett. {\bf 110}, 035302 (2013). Copyright 2013 by the American Physical Society.} \label{fig:ott1}
\end{figure}

\begin{figure}[tb]
  \centering
  \includegraphics[width=12cm]{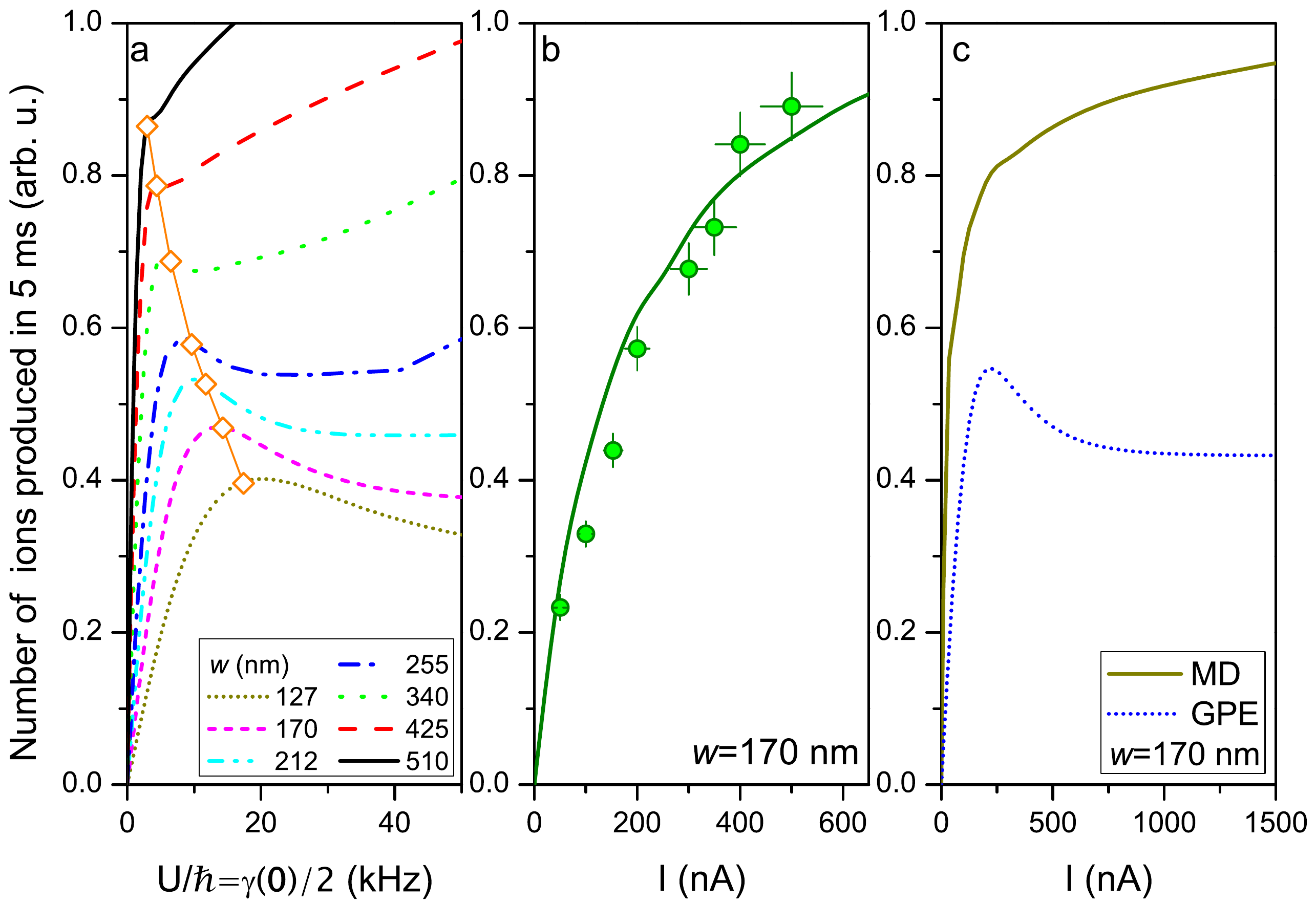}
  \caption{a) Theoretical curves of the number of ions produced in 5 ms as a function of $U/\hbar=\gamma_1(\mathbf{0})$ solving equation (\ref{dGPE}) for different values of the beam width $w$ \cite{Barontini2013}. The values of $U_M$ obtained using the approximate expression given in the text are shown as open diamonds over the corresponding curves. b) Number of ions measured after 5 ms of dissipation for a thermal cloud as a function of the EB current, with beam width $w=170 (7)$ nm. The solid line is the result of the corresponding numerical simulation using the molecular dynamics method. c) Comparison between the theoretical curves of the number of produced ions as a function of $I$ for the BEC and the \emph{corresponding} classical analogue (beam width $w=170 (7)$ nm) \cite{Barontini2013}. Reprinted figure with permission from G.~Barontini, R.~Labouvie, F.~Stubenrauch, A.~Vogler, V.~Guarrera, and H.~Ott, Phys. Rev. Lett. {\bf 110}, 035302 (2013). Copyright 2013 by the American Physical Society.} \label{fig:ott3}
\end{figure}

Recently, the quantum Zeno effect has been observed in experiments with Bose-Einstein condensates subjected to local losses due to an electron beam \cite{Barontini2013,Zezyulin2012}, in which atoms from the cloud are ionised and detected as depicted in Fig.~\ref{fig:ott1}a. In the weakly interacting case, the dynamics under the effective Hamiltonian can be modelled by a time-dependent Gross-Pittaevskii equation with an additional imaginary term \cite{Witthaut2008,Barontini2013},
\begin{equation}
i\hbar\frac{\partial\psi(\textbf{x},t)}{\partial t}=\left(-\frac{\hbar^2\nabla^2}{2m}+V_{ext}(\textbf{x})+g|\psi(\textbf{x},t)|^2-{\rm i}\hbar\frac{\gamma_1(\textbf{x})}{2}\right)\psi(\textbf{x},t).
\label{dGPE}
\end{equation} 
Here $\psi(\textbf{x},t)$ is the condensate wavefunction, obeying the constraint $\int|\psi(\textbf{x},t)|^2d\textbf{x}=N(t)$, $V_{ext}(\textbf{x})$ is an external trapping potential, and $\gamma_1(\textbf{x})$ describes the profile of the single-particle loss in space. In Figs.~\ref{fig:ott1}b and \ref{fig:ott3}, we reprint experimental and theoretical results from Ref.~\cite{Barontini2013}. In Fig.~\ref{fig:ott3}a, the dependence of the number of ions detected as a function of the strength of the electron beam is plotted, and it can be seen that this clearly saturates in time, as a direct result of the quantum Zeno effect. This saturation is observed clearly in Ref.~\cite{Barontini2013}, and is contrasted to the case of a thermal cloud for which the dynamics are classical, and do not exhibit this effect is shown in Figs.~\ref{fig:ott3}b,c. These effects in weakly interacting systems have also been discussed theoretically for the case of a two-mode Bose-Einstein condensate in a double-well trap \cite{Witthaut2008}, and in the context of decoherence due to atom loss in bosonic Josephson junctions \cite{Pawlowski2013,Spehner2014}.

\subsection{State preparation in driven, dissipative many-body systems}

A key aspect of open systems as they appear in quantum optics involves the new tools to control quantum systems that are provided by dissipation. In particular, optical pumping \cite{Brossel1952,Hawkins1953} and laser cooling \cite{Metcalf1999} are dissipative processes that have become a key foundation for the majority of modern atomic physics experiments. The general idea behind these procedures is always to engineer a dissipative process in such a way that an arbitrary initial state of the system $\rho_0$ is driven via dissipation [e.g., as described by a master equation \eqref{mastereq}] into a steady state that is a chosen pure state $\ket{\Psi_s}$, $\rho(t\rightarrow \infty) \rightarrow \ket{\Psi_{s}}\bra{\Psi_s}$. In this process, entropy is removed from the system to the environment. 

The idea of dissipative processes for state preparation and driving in quantum simulators has seen rapid recent progress. In this section, we will give a brief summary of some key directions, beginning with engineered dissipation in the presence of a reservoir gas, and going on to discuss implications of this for dissipative driving towards important many-body states. For several of these topics, further details can be found in the recent review on engineered dissipation and quantum simulation by M\"uller et al., Ref.~\cite{Muller2012}. The drive towards finding new means to cool many-body systems and prepare important many-body states is particularly important in light of current experimental challenges to reach lower temperatures in order to realise certain particularly interesting aspects of the many-body physics predicted for cold atoms in optical lattices \cite{McKay2011}. 

\begin{figure}[tb]
  \centering
  \includegraphics[width=8cm]{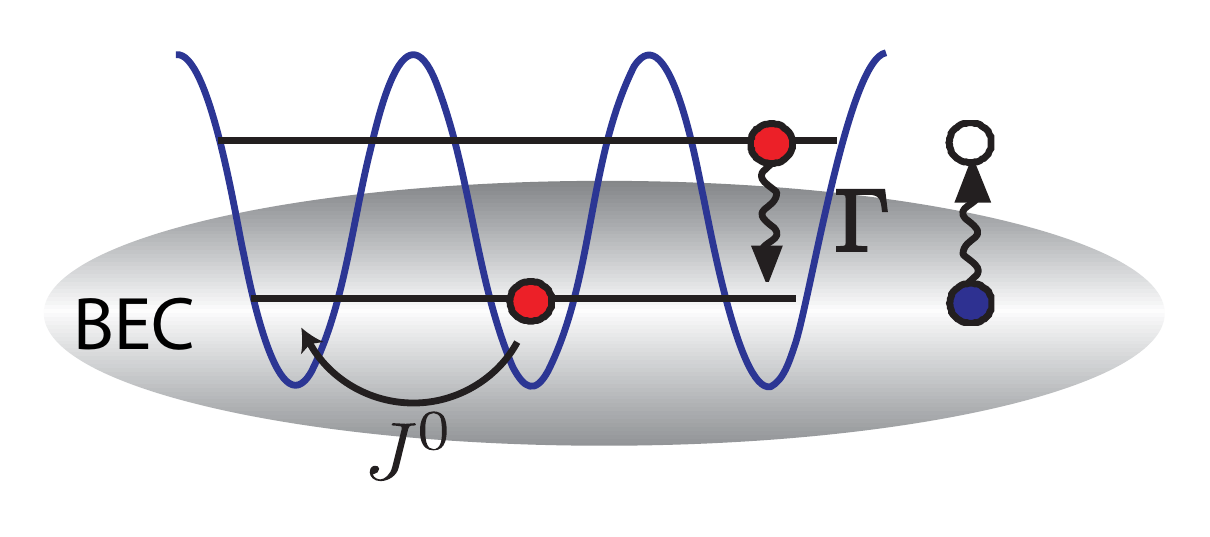}
  \caption{Schematic picture of atoms in an optical lattice immersed in a Bose-Einstein condensate of another species that is not trapped by the lattice. Processes in which an atom in an excited band of the lattice decays to the lowest band are generated by spontaneous emission of Bogoliubov excitations in the condensate. This process is analogous to spontaneous emissions of photons, but takes place on much smaller energy and momentum scales, allowing for tools such as laser cooling and optical pumping to be reworked in a new context. Figure adapted from A.~Griessner, A.~J.~Daley, S.~R.~Clark, D.~Jaksch, and P.~Zoller, New J. Phys {\bf 9}, 44 (2007).} \label{fig:immersion}
\end{figure}

The example of engineered dissipation that is closest to the systems we have discussed up to now is the case in which dissipation is produced and controlled for cold atoms in an optical lattice by immersing the system in a reservoir gas of a second species of atoms \cite{Daley2004a}. This second species is not trapped by the lattice, and is usually chosen to be in a Bose-Einstein condensate, in a broad trapping potential so that the gas is almost uniform. In Refs.~\cite{Daley2004a,Griessner2006,Griessner2007} a master equation is derived for this system, making the assumption that the typical Bloch band separation in the lattice, $\omega_g$ is much larger than the corresponding energy scales in the reservoir gas, including its chemical potential, $\mu_R$. In this scenario, the quantum optics assumptions from section \ref{sec:approxamo} are satisfied because the Bloch band separation $\omega_g$ plays the role of the large energy and frequency scale, and the master equation is well-justified for realistic experimental parameters. 

The resulting dissipative process is drawn schematically in Fig.~\ref{fig:immersion}. Atoms in higher bands of the lattice are sympathetically cooled via density-density interactions with atoms in the reservoir gas, with the atoms in the lattice decaying to the lowest band, while a Bogoliubov excitation is produced in the reservoir gas. As $\omega_g \gg \mu_R$, this excitation is typically a particle branch excitation of high energy \cite{Pitaevskii2003}, which can be allowed to rapidly leave the trap, ensuring that the Markov approximation remains valid \cite{Daley2004a}. Also, because the temperature of the reservoir gas, $T_R$ satisfies $k_B T_R\ll \mu_R \ll \omega_g$, where $k_B$ is the Boltzmann constant, excitations are not present to reheat lattice atoms to higher bands after they have decayed. 

Moreover, while atoms in higher bands couple strongly to the reservoir, decaying on rapid timescales $\sim 1$kHz for typical parameters in an optical lattice and Bose-Einstein condensate of $^{87}$Rb, atoms in the lowest band couple very weakly, preventing re-heating of atoms within in the lowest band (of width $4J^0$). This occurs because of a mismatch in the dispersion relation of a Bose-Einstein condensate and atoms in the lowest band of the lattice, which in analogy to the Landau criterion for superfluidity \cite{Pitaevskii2003,Pethick2002}, prevents energy and momentum conservation when a single atom in the lattice interacts with a single excitation in the reservoir. Even in the case that these matched, the coupling constant would be small \cite{Griessner2006,Griessner2007}, due to the small structure factor of a Bose-Einstein condensate at low energies \cite{Pitaevskii2003,Pethick2002}. Both the decay of atoms in higher bands and the decoupling for atoms in the lowest band have recently been observed in experiments with Rb atoms in state-dependent traps \cite{Chen2014}. Similar effects have also been studied when motional superposition states are prepared in experiments using Li atoms immersed in a Bose-Einstein condensate of sodium \cite{Scelle2013}. In each of these experiments, the observed decay rates were in agreement with the assumptions made in the derivation of the master equation \cite{Daley2009}.

This combination of behaviours for the atoms in the lattice is strongly mathematically and physically analogous to spontaneous emissions into the vacuum modes of a radiation field. It immediately prompts the question as to whether ideas from quantum optics can be re-used in this context, but now with two-level atoms replaced by motional states or Bloch bands in the lattice, and photons being replaced by Bogoliubov excitations in the reservoir gas. By repeating such ideas on the smaller momentum and energy scales that are associated with excitations in a reservoir gas rather than photons, new possibilities for control and cooling to lower temperatures might then be realised.

\begin{figure}[tb]
  \centering
  \includegraphics[width=10cm]{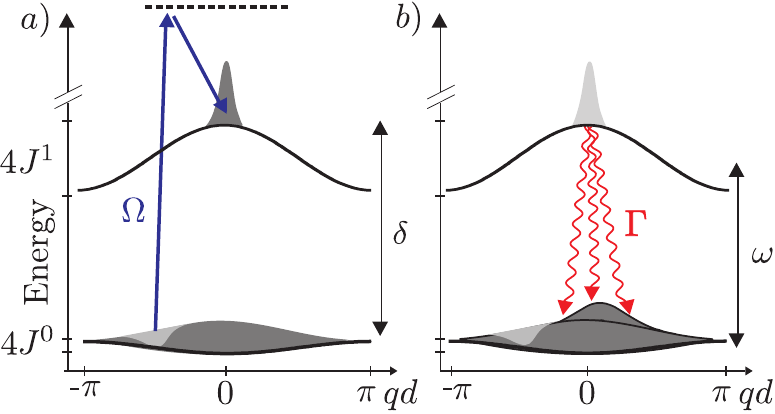}
  \caption{Schematic picture of one excitation and decay step in Raman cooling in an optical lattice. (a) Atoms are transferred from a region with high quasi- momentum $|q| > 0$ in the lowest Bloch band to the first excited band. (b) The collisional interaction with the Bose-Einstein condensate atoms is switched on, the resulting decay of the excited lattice atoms leads to a randomization of the quasi-momentum. Sequences of pulses, each one followed by a decay time $\tau_c$, efficiently excite all atoms outside a narrow region around $q = 0$. Repeating the sequence leads to accumulation of atoms in the dark state region around $q = 0$, i.e., to cooling. Reprinted figure from A.~Griessner, A.~J.~Daley, S.~R..~Clark, D.~Jaksch, and P.~Zoller, New J. Phys {\bf 9}, 44 (2007). Copyright 2007 by IOP Publishing Ltd and Deutsche Physikalische Gesellschaft.} \label{fig:griessner}
\end{figure}

The first example of such a technique is a dark-state laser-cooling scheme based directly on an analogy with Raman dark-state cooling, as proposed by Kasevich and Chu \cite{Kasevich1992}. The key idea behind such schemes such as that of Ref.~\cite{Kasevich1992} and also velocity-selective coherent population trapping \cite{Aspect1988}, is that the desired state is engineered to be a \emph{dark state}, which does not couple to laser pulses that would transfer the atoms to a different internal state. A schematic diagram showing an analogous scheme for cooling non-interacting atoms within the lowest Bloch band of an optical lattice \cite{Griessner2006,Griessner2007} is shown in Fig.~\ref{fig:griessner}. 

The cycle proceeds in two steps: (i) Raman pulses are applied to transfer atoms from the lowest Bloch band of the lattice to an excited Bloch band. Making use of the energy dependence on quasimomentum, together with control over the detuning and the time-dependence of the pulses, these are engineered to transfer only particles with large quasimomentum to the excited band -- in particular, a narrow dark state region near quasimomentum $q=0$ is deliberately not excited. (ii) Coupling to the reservoir gas (which can be left on if it is slower than the Raman pulse, but which can also be switched using Feshbach resonances for appropriate atomic species \cite{Chin2010}) is used to cause the atoms to decay to the lowest band. The quasimomentum will be redistributed in this process, and depending on the relative mass of atoms in the lattice and atoms in the reservoir gas, may be randomly distributed over the whole Brillouin zone \cite{Griessner2007}. 

After repeating this cycle, atoms will begin to collect in the dark state region. Fundamentally, the cooling power of this method is determined only by the selectivity with which the dark state region is not excited, and the narrowness of this region. The final temperature can be much lower than that of the reservoir gas, because exactly as in regular laser-cooling, we are upconverting the energy to be removed from the system in such a way that it falls in an energy regime where there are no thermal excitations in the reservoir. This also clarifies that the decoupling of atoms in the lowest band from the reservoir gas is essential in this scheme, to prevent reheating of atoms already in the dark state.

There are many possible extensions that can be built upon this type of technique. In Ref.~\cite{Griessner2005}, a variant of this method for loading fermions in a state-dependent lattice was discussed. In this proposal, fault-tolerant loading of fermions in an optical lattice was investigated, in which particles are initially coupled to a higher Bloch band, and then cooled to the lowest band by creation of particle-hole pairs. In Refs.~\cite{Diehl2008,Kraus2008}, this was further extended to certain classes of many-body systems. By using additional internal atomic levels, schemes were devised to use a reservoir gas to implement a master equation with jump operators 
\begin{equation}
c_m\equiv c_{\langle l,j\rangle} = (a_l^\dagger + a_j^\dagger)(a_l-a_j), \label{becjump}
\end{equation}
 which results in a coherent driving of the system into a Bose-Einstein condensate on the lattice, i.e., is a designed master equation to produce a steady state $\propto (\sum_l a_l^\dagger)^N |{\rm vac}\rangle$. These ideas were also applied to implement a master equation for which the $\eta$-pairing states \cite{Yang1989} for two-species fermions are the steady state \cite{Diehl2008,Kraus2008}. 

In Refs.~\cite{Diehl2010,Yi2012}, a master equation was found that -- with no direct or induced interactions between particles -- could produce either antiferromagnetic states, or pairing states for two-component fermions. In that case, the dissipative dynamics could be induced with the help of spontaneous emission processes in group-II atoms, which can exhibit Pauli blocking for fermions \cite{Sandner2011}. Quantum trajectory methods were employed together with exact diagonalisation methods (see section \ref{sec:numerics}) in order to demonstrate the time-dependent formation of the corresponding ordered states. Diehl, Bardyn, and their collaborators \cite{Diehl2011,Bardyn2013} also constructed master equations that would drive a system into topologically protected states. Specifically, they investigated atomic quantum wires coupled to a reservoir gas, and showed that with a dissipative drive, dissipative pairing gaps would enforce the isolation of Majorana edge modes \cite{Diehl2011}.

In this context, it is also both interesting and important to understand the competition between dissipation and coherent dynamics. Diehl et al.~\cite{Diehl2010c} treated the effect of interactions in the Hamiltonian on dissipative driving towards a Bose-Einstein condensate based on the jump operators in eq.~\eqref{becjump}, and found that competition between coherent and dissipative dynamics produced a mixed state in which the temperature was effectively controlled by the size of the interactions. Such competition can result further in dissipative phase transitions \cite{Diehl2010c,Sieberer2013a,Honing2012}, and some of the resulting transitions can form new types of universality classes for dissipative phase transitions, which can be studied with functional renormalization group methods \cite{Tauber2013,Sieberer2013}.

\subsection{Connections with other dissipative many-body systems}
\label{sec:othersys}

In addition to cold atoms and molecules in optical traps, there are a wide variety of AMO and solid state systems to which the ideas discussed in this review may be applied, or are already being applied. In this section we briefly summarise these corresponding developments.

A prime example of a system in which dissipative quantum simulation has already been implemented is the case of trapped ions \cite{Blatt2012,Muller2012}, which are also one of the most advanced systems in terms of control over single particles and few-particle coupling for quantum computing \cite{Haffner2008,Monroe2013}. Recent experiments have made use of this extreme control to demonstrate combinations of coherent and dissipative maps in digital quantum simulation \cite{Barreiro2011,Schindler2013}. In these experiments, qubits are prepared and then dynamical elements are applied as quantum gates, demonstrating -- amongst other attributes -- the dissipative preparation of entangled states, and simulations corresponding to many-body spin interactions. These experiments constitute a wonderful starting point for further high-precision studies of dissipative processes in many-particle systems. This is further enhanced by recent developments in realising analogue quantum simulation of spin chains, including transverse Ising and XY models with variable-range interactions, in these systems \cite{Porras2004,Friedenauer2008,Kim2009}. Dynamics in such setups were recently demonstrated in 1D chains \cite{Islam2013}, and also in 2D arrays \cite{Britton2012}. The possibility of combining these advances with control over dissipation is an exciting future direction. Controlled dissipative processes are also useful in a broader quantum computing context, where states can be protected in quantum computations \cite{Verstraete2009,Kliesch2011} or in quantum memories \cite{Pastawski2011} by carefully engineered dissipation. 

Another system where dissipation is inherent in the most fundamental properties involves Bose-Einstein condensates in optical cavities \cite{Ritsch2013,Mekhov2012}. Recent observation of the Dicke phase transition \cite{Baumann2010,Dimer2007} and related optomechanical effects, including the detection of quantised motion and non-classical output light \cite{Brooks2012,Brahms2012} has opened new possibilities for studying many-body dissipative dynamics. This could include the realisation of more complicated collective dissipative spin models \cite{Morrison2008}, and quantum phases associated with the motion of atoms within the cavity \cite{Habibian2013,Torre2013,Buchhold2013}. Atoms in optical lattices within the cavity have also been discussed, including the potential use of cavity fields as a probe to detect detect atom number statistics \cite{Mekhov2007b,Mekhov2007}, or the use of cavity fields for state preparation \cite{Mekhov2009}. Bose-Hubbard dynamics in the presence of quantised light fields have also been discussed in Refs.~\cite{Maschler2008,Niedenzu2010}. Generalisations of this to other photon-atom systems have included studies of strongly interacting photons \cite{Peyronel2012,Reinhard2012,Petrosyan2011,Sevincli2011,Gorshkov2011}, dissipative state preparation and quantum simulation in optomechanical arrays \cite{Tomadin2012,Lechner2013}, and also the potential for photon condensation generated via engineered dissipation in a circuit QED system  \cite{Marcos2012}.

There has similarly been a lot of recent interest in systems of atoms in excited Rydberg states, including possibilities to form crystaline states, and other long-range correlations. In the context of these excitations, several groups have become interested in the effect of decay of the Rydberg levels on the spatial patterns being realised. Their investigations have included the build-up of spatial correlations in the presence of dissipation \cite{Petrosyan2013,Honing2013}, the effects of non-local dissipation \cite{Olmos2013,Ates2012}, and studies of resulting glass-like dynamics and corresponding structures that emerge \cite{Olmos2012,Lesanovsky2013,Lesanovsky2014}.

Also going beyond these AMO systems, there have been several recent treatments of transport systems and dynamics of spin models with open quantum systems \cite{Prosen2014,Karevski2013,Prosen2011,Prosen2010,Znidaric2010,Prosen2008,Cai2013}, including applications of quantum trajectories to solve a master equation for spin dynamics. These studies have also employed matrix product state methods for such calculations \cite{Benenti2009,Prosen2009}. Quantum trajectory methods have also been used to investigate open system dynamics for ions traversing amorphous solids \cite{Minami2003,Seliger2007}. This is based on an approach with a master equation that allows probability flux for particles to be exchanged with the environment \cite{Seliger2005}.

\section{Summary and Outlook}
\label{sec:summary}

In this review we have seen how approaches from quantum optics to treating dynamics in open quantum systems has become important for our understanding of strongly interacting AMO systems. This opens many new opportunities to look at fundamental properties of open many-body systems, especially dynamics far from equilibrium, decoherence of states, and the back-action of continuous measurements. Not only is an understanding of dissipative dynamics necessary to treat processes such as light scattering and loss that are naturally present in quantum simulators, but the possibility to understand and engineer dissipative coupling provides an interesting route towards realisation of lower temperatures and important strongly-interacting many-body states, ranging from paired states of fermions to topological order. 

At the same time, this is also a relatively new subfield with many open questions, ranging from specific questions such as the effect of spontaneous emissions on dynamical processes used to prepare states with cold atoms in optical lattices, to broader questions such as to what extent and under which conditions it is possible to identify universality classes for the steady-state of driven, dissipative many-body systems. While this review has focussed on markovian dissipative systems, which are prevalent in quantum optics and AMO implementations, one important area for future development will be the further development of these ideas in non-markovian regimes. Already there has been a great deal of progress on treating non-markovian effects in open systems \cite{Breuer2007,Breuer2012}, including through the extension of master equation techniques \cite{Breuer2004b,Breuer2007b} and stochastic wavefunction methods \cite{Breuer1999,Breuer2004,Breuer2004c} on expanded Hilbert spaces. In terms of quantum trajectories methods, there have been various extensions to the method to include memory effects of the reservoir for single-particle systems, beginning with work by Imamoglu in Ref.~\cite{Imamoglu1994}, which involved the coupling of a small quantum system to a damped cavity with a finite lifetime, and leading up to recent direct extensions to quantum trajectories methods on the Hilbert space of the reduced system \cite{Piilo2008,Piilo2009}. Recent studies have started combining this with Bose-Einstein condensates, including work where atoms are used as a probe immersed in the condensate \cite{Haikka2013,McEndoo2013}. Combined with strongly interacting systems, such ideas may open new pathways for investigations of fundamental dynamics, or provide practical tools for out-of-equilibrium preparation of many-body states as discussed in section \ref{amosys}. Extensions to non-markovian systems may also bring new opportunities for the study of solid-state systems via these techniques \cite{Breuer2007}.

In each of the ways described in this review, recent advances in AMO experiments and in theoretical tools are opening new opportunities to combine ideas from quantum optics and many-body physics. As discussed at the end of section \ref{sec:othersys}, numerical and analytical techniques with open quantum systems and quantum trajectories methods are also beginning to find applications not only in quantum simulators, but also outside the traditional AMO context. Understanding and even tailoring the dynamics of open systems will in this way lead to better control over a wide variety of systems, and to new possibilities to explore many-body physics in and out of equilibrium.

\section*{Acknowledgements}
I would also like to thank my colleagues and collaborators, Lars Bonnes, Sebastian Diehl, Adrian Kantian, Stephan Langer, Andreas L\"auchli, Hannes Pichler, Saubhik Sarkar, Johannes Schachenmayer, and Ulrich Schollw\"ock for insightful discussions on the application of these methods to many-body systems, and particularly Scott Parkins, who first introduced me to quantum trajectories methods and open quantum systems. I would also like to thank Anton Buyskikh, Sebastian Diehl, Stephan Langer, Saubhik Sarkar, Alexandre Tacla, and Guanglei Xu for providing comments on the manuscript, and Johannes Schachenmayer for providing the data from our project in Ref.~\cite{Schachenmayer2014} for the example in Fig.~\ref{fig:spementropy}. I would especially like to thank Peter Zoller for discussions and collaborations on open quantum systems the last 10 years -- the way of thinking about these systems that I learned from him and that I developed while working with him is strongly reflected in many parts of this review.

Work on dissipative systems in Pittsburgh is supported by AFOSR grant
FA9550-13-1-0093, and computational resources were provided by the Center for Simulation and Modeling at the University of Pittsburgh.

\appendix

\section{Continuous measurement and physical interpretation of trajectories}

\label{sec:continuousmeasurement}

In this appendix, we return to the foundations and interpretation of the quantum trajectory methods. Specifically, we will outline how the quantum trajectories interpretation of
dissipative processes can be derived and understood from the perspective of continuous
measurement as it is discussed in quantum optics. This theory was developed initially in the context of photodetection, and the inference of a \emph{conditional evolution} of a pure state wavefunction by considering quantum jumps (see, e.g., \cite{Mollow1975,Gisin1984,Barchielli1986,Blatt1986,Zoller1987}).

 We consider a generic
system interacting with a bath as in section \ref{sec:generalqo}. In this section, we will follow an
exposition similar to that given by Gardiner and Zoller \cite{Gardiner2005},
generalising their approach to multiple interaction channels. In the quantum optics context, these channels can be, e.g., different polarization states of a photon. For our many-body systems, they will often correspond to coupling to spatially separated parts of the reservoir - e.g., in the case of master equations where particles scatter particles from different lattice sites or are lost from particular sites, each channel can reflect processes in a particular lattice site provided the correlation length associated with excitations in the reservoir is small compared with the lattice spacing. 

\subsection{Interaction Hamiltonian}
\label{sec:inthamcm}
We begin by rewriting the interaction Hamiltonian between the system
and the bath in a time-dependent form, transforming 
\begin{equation}
H_{\mathrm{int}}=-{\rm i}\hbar\sum_{l}\int_{0}^{\infty}d\omega\,\kappa_{l}(\omega)\left[x_{l}^{+}b_{l}(\omega)-x_{l}^{-}b_{l}^{\dagger}(\omega)\right]
\end{equation}
 into a rotating frame, i.e., an interaction picture in which the time-dependence of the operators incorporates the time-dependence of the combined Hamiltonian $H_{{\rm sys}}+H_{{\rm env}}$.
The transformed interaction Hamiltonian $\tilde{H}{}_{\mathrm{int}}(t)$
is given by

\begin{eqnarray}
\tilde{H}{}_{\mathrm{int}}(t) & = & {\rm i}\hbar\sum_{l}\int_{\omega_{{\rm sys}}-\xi}^{\omega_{{\rm sys}}+\xi}d\omega\kappa_{l}(\omega)\left[b_{l}^{\dagger}(\omega)e^{i(\omega-\omega_{l})t}x_{l}^{-}-x_{l}^{+}b_{l}(\omega)e^{-{\rm i}(\omega-\omega_{l})t}\right] \label{inthamcmeq}\\
 & \equiv & {\rm i}\hbar\sum_{l}\sqrt{\gamma_{l}}\left(b_{l}^{\dagger}(t)x_{l}^{-}-x_{l}^{+}b_{l}(t)\right),
\end{eqnarray}
where we have used the fact that $[H_{{\rm sys}},x_{l}^{\pm}]\approx \pm\hbar\omega_{l}x_{l}^{\pm}$,
and we have denoted the interval of frequencies over which the system
couples to the environment to be bounded in each case by some frequency
scale $\xi$, based on a rotating wave approximation \cite{Gardiner2005}.
Based on the frequency independence of the interaction constant (which
amounts to the first part of the Markov approximation), we replace
\[
\kappa_{l}(\omega_{l})\rightarrow\sqrt{\frac{\gamma_{l}}{2\pi}},
\]
and we define 
\begin{equation}
b_{l}(t)=\frac{1}{\sqrt{2\pi}}\int_{\omega_{l}-\xi}^{\omega_{l}+\xi}\, b_{l}(\omega)e^{-{\rm i}(\omega-\omega_{l})t}d\omega.
\end{equation}

\subsection{Separation of timescales}

We note that the commutator of these operators, 
\begin{equation}
\left[b_{l}(t),b_{l'}^{\dagger}(t')\right]=\delta_{ll'}\left[\frac{1}{2\pi}\int_{-\xi}^{+\xi}d\omega e^{-{\rm i}\omega(t-t')}\right]=\delta_{ll'}\delta_{\xi}(t-t'),
\end{equation}
where $\delta_{\xi}(t-t')$ denotes a slowly-varying Dirac delta function,
which can be treated as a regular delta function provided that the
timescale $|t-t'|\gg1/\xi$. This will be a key condition for the
interpretation of the dynamics, and essentially will require that
the timescales we consider are always much longer than $1/\xi$. If
we have remaining terms in the system Hamiltonian within this transformed
picture, then we will denote them as $H_{0}$. For two-level atoms,
for example, $H_{0}$ would contain coupling elements at frequencies
$\Omega_{0}$ and detunings $\Delta$. Adding these scales in, we
now have a clear hierarchy of frequency/energy scales in this problem,
\[
\Omega_{0},\Delta,\gamma_{l}\ll\xi\ll\omega_{l}.
\]
Note that this hierarchy implies all of the standard three approximations
- the second inequality implies the rotating wave approximation, and
the first inequality implies the Born approximation. We have already
made the first Markov approximation, and we will later see how these
inequalities imply the second part of the Markov approximation. 

Using these definitions, we now write a state vector for the complete
system of the system and the bath, which obeys the Schr\"{o}dinger equation
\[
\frac{d}{dt}|\Psi(t)\rangle=-\frac{{\rm i}}{\hbar}H_{\mathrm{tot}}|\Psi(t)\rangle
\]
and evolves from an initial condition where the system and the bath
are in a product state, i.e., $\ket{\Psi(0)}=\ket{\psi_{\mathrm{sys}}}\otimes\ket{\mathrm{vac}}$,
where $\ket{\mathrm{vac}}$ denotes the vacuum state in the environment
(note that this can be generalised straight-forwardly to other conditions,
especially thermal states. We thus obtain the equation of
motion:
\begin{equation}
\frac{d}{dt}\ket{\Psi(t)}=\left\{ -\frac{{\rm i}}{\hbar}H_{0}+\sum_{l}\left[\sqrt{\gamma_{l}}b_{l}^{\dagger}(t)x_{l}^{-}-\sqrt{\gamma_{l}}x_{l}^{+}b_{l}(t)\right]\right\} \,\ket{\Psi(t)}
\end{equation}
We then consider integration of the Schr\"odinger equation over time steps
$\Delta t$ so that 
\[
\Omega_{0},\Delta,\gamma_{l}\ll1/(\Delta t)\ll\xi .
\]
This choice of time step will allow the use of the Born approximation,
and the computation of the dynamics in perturbation theory for each
time step, while allowing the use of the rotating wave approximation.
As mentioned in section \ref{sec:physicalinterp}, this choice of time step is necessary
in order to allow the usual approximations in AMO systems to be valid,
and as some authors have pointed out \cite{Plenio1998}, making measurements
faster than this rate would result in a quantum Zeno effect for all
of the dynamics due to the system-bath interaction, leading to non-sensical
results. 

\subsection{Perturbation expansion of the total state}

We then write the perturbation expansion up to second order for the
time evolved state over a short time $\Delta t$:

\begin{align}
|\Psi(\Delta t)\rangle & =\left\{ 1-\frac{{\rm i}}{\hbar}H_{0}\Delta t+\sum_{l}\left[\sqrt{\gamma_{l}}\, x_{l}^{-}\int_{0}^{\Delta t}b_{l}^{\dagger}(t)\, dt-\sqrt{\gamma}x_{l}^{+}\int_{0}^{\Delta t}b(t)\, dt\right]\right.\nonumber \\
 & +\left. \left(-\frac{{\rm i}}{\hbar}\right)^{2}\sum_{l}\gamma_{l}x_{l}^{+}x_{l}^{-}\int_{0}^{\Delta t}dt\,\int_{0}^{t_{2}}dt^{\prime}\, b_{l}(t)b_{l}^{\dagger}(t^{\prime})\right\} |\Psi(0)\rangle,
\end{align}
where we note that all of the second order terms for $l\neq l'$ vanish
because of the action of the operator $b_{l}(t)$ on the vacuum state
for the environment contained in $|\Psi(0)\rangle$. We must keep
the remaining term, because for $\Delta t\gg1/\xi$,
\begin{align}
\int_{0}^{\Delta t}dt\,\int_{0}^{t}dt'\, b_{l}(t)\, b_{l}^{\dagger}(t')|\text{\textrm{vac}}\rangle & =\int_{0}^{\Delta t}dt\,\int_{0}^{t}dt'\,\left[b_{l}(t)\,,b_{l}^{\dagger}(t')\right]|\text{\textrm{vac}}\rangle\nonumber\\
 & =\int_{0}^{\Delta t}dt\,\int_{0}^{t}dt^{\prime}\,\delta_{\xi}(t-t')|\text{\textrm{vac}}\rangle\nonumber\\
 & =\frac{1}{2}\Delta t|\text{\textrm{vac}}\rangle.
\end{align}
 This second order term is then the origin
of our effective Hamiltonian, as we can combine the system Hamiltonian
with the original second order term to give 
\[
H_{{\rm eff}}=H_{0}-\frac{1}{2}\sum_{l}\gamma_{l}x_{l}^{+}x_{l}^{-}
\]
in the expression 
\begin{align}
|\Psi(\Delta t)\rangle & =\left\{ 1-\frac{i}{\hbar}H_{\mathrm{eff}}\,\Delta t+\sqrt{\gamma}\, x_{l}^{-}\,\Delta B_{l}^{\dag}(0)\right\} |\Psi(0)\rangle,
\end{align}
where the increment operators are defined \cite{Gardiner2005} as
\[
\Delta B_{l}(t):=\int_{t}^{t+\Delta t}b_{l}(t')\, dt'.
\]

\subsection{Increment operators}

These increment operators obey the commutation relations 

\begin{eqnarray}
\left[\Delta B_{l}(t),\Delta B_{l'}^{\dag}(t^{\prime})\right] & = & \delta_{ll'}\int_{t}^{t+\Delta t}\int_{t'}^{t'+\Delta t}\left[b_{l}(s),b_{l'}^{\dagger}(s')\right]\, ds\, ds' \nonumber\\
 & = & \delta_{ll'}\int_{t}^{t+\Delta t}\int_{t'}^{t'+\Delta t}\delta_{\xi}(s-s')\, ds\, ds'\nonumber\\
 & = & \left\{ \begin{array}{ll}
\delta_{ll'}\Delta t & t=t^{\prime} \\
0 & t\neq t^{\prime}
\end{array}\right..
\end{eqnarray}
As a result, we can think of each of these operators $\Delta B_{l}^{\dagger}$
as a creation operator for an excitation in the environment, produced
in mode $l$ in the time interval $(t,t+\Delta t]$. Because of the
commutation relations, excitations in non-overlapping time intervals
are orthogonal, by which we see that action of the Markov approximation,
in so far as excitations in the reservoir at different times (that
differ by more than $1/\xi$) are independent. In order to connect
to photon counting statistics, number operators can be constructed
for the modes $l$ at different times $t$, with 
\begin{equation}
n_{{\rm env,}l}(t)=\frac{\Delta B_{l}^{\dag}(t)}{\sqrt{\Delta t}}\frac{\Delta B_{l}(t)}{\sqrt{\Delta t}},
\end{equation}
and measurement of these operators amounts to getting time-dependent
excitation counts in each mode of the environment (naturally course-grained
over the timescale $\Delta t$.

In this expression, the different channels $l$ can correspond to different polarisation channels for photodetectors, or, e.g., in the case of light scattering for many atoms in an optical lattice, could correspond to photons scattered in different parts of the system separated by a larger distance than $\lambda/(2\pi)$. Detection of this would correspond to polarisation-resolved or spatially-resolved detection of photons.

\subsection{Effect on the system and measurement of the environment}

In order to interpret the above result, and to see more clearly the
interpretation in terms of quantum trajectories that appears, we can
rewrite the state after the first step as 

\begin{align}
|\Psi(\Delta t)\rangle & =\left[1-\frac{{\rm i}}{\hbar}H_{\text{\textrm{eff}}}\,\Delta t+\sum_{l}\sqrt{\gamma_{l}}x_{l}^{-}\,\Delta B_{l}^{\dag}(0)\right]|\Psi(0)\rangle\\
 & =\left(1-\frac{{\rm i}}{\hbar}H_{\text{\textrm{eff}}}\,\Delta t\right)|\psi(0)\rangle\otimes|\mathrm{vac}\rangle+\sum_{l}\sqrt{\gamma_{l}\Delta t}x_{l}^{-}|\psi(0)\rangle\otimes\Delta B_{l}^{\dag}(0)|\mathrm{vac}\rangle.
\end{align}
we can interpret this state as a superposition of states with either
no excitations in the environment, or states with one excitation in
one of the modes $l$ produced in the first time interval. 

If we now make a measurement of the state of the environment, we will
see one of two things occurring. Either, we will observe an excitation
in mode $l,$ $\Delta B_{l}^{\dag}(0)|\mathrm{vac}\rangle$, in which
case we obtain the state
\[
|\psi(\Delta t)\rangle=\frac{x_{l}^{-}|\psi(0)\rangle}{||x_{l}^{-}|\psi(0)\rangle||}.
\]
The probability of obtaining this state is given by $\gamma_{l}\Delta t||x_{l}^{-}|\psi(0)\rangle||^{2}$
for each possible channel $l$. Otherwise, we will observe a vacuum
state, in which case the state we obtain is 
\begin{equation}
|\psi(\Delta t)\rangle=\frac{\left(1-\frac{\rmi}{\hbar}H_{\text{\textrm{eff}}}\,\Delta t\right)|\psi(0)\rangle}{\left\Vert \left(1-\frac{\rmi}{\hbar}H_{\text{\textrm{eff}}}\,\Delta t\right)|\psi(0)\rangle\right\Vert },
\end{equation}
which we obtain with probability $\left\Vert \left(1-\frac{\rmi}{\hbar}H_{\text{\textrm{eff}}}\,\Delta t\right)|\psi(0)\rangle\right\Vert ^{2}$.
As we noted in section \ref{sec:physicalinterp}, we always obtain information
about the state, whether or not an excitation is detected.

If on the other hand we do not measure the state of the environment
after the step, then we can trace over the environment, and obtain a reduced
density operator for the system

\begin{align}
\rho(\Delta t) & =\mathrm{tr}_{{\rm env}}|\Psi(\Delta t)\rangle\langle\Psi(\Delta t)| \nonumber\\
 & =\left(1-\frac{{\rm i}}{\hbar}H_{\text{\textrm{eff}}}\,\Delta t\right)\rho(0)\left(1-\frac{{\rm i}}{\hbar}H_{\text{\textrm{eff}}}\,\Delta t\right)^{\dag}+\sum_{l}\gamma_{l}x_{l}^{-}\rho(0)x_{l}^{+}\Delta t,
\end{align}
from which we can infer the evolution under the master equation. 

Further integration steps have similar actions, because the increment
operators $\Delta B_{l}(t)$ from different time steps commute \cite{Gardiner2005}. In
this way, the integration can be pieced together in each step, and
the procedure outlined in section \ref{sec:higherorder} directly recovered. 

\subsection{Further interpretation}
Through the application of continuous measurement theory \cite{Gardiner2005}, we therefore see how a quantum trajectories interpretation can be built up in a system where we have the ability to measure certain modes of the environment, and also how the individual approximations in the quantum optical system give rise to the simplified form of the master equation. This derivation serves to illustrate the key aspects of the physical interpretation of quantum trajectories described in section \ref{sec:physicalinterp}. It also makes it clear how we can think of continuous measurement of the environment (where the measured operators must necessarily be Hermitian), but nonetheless obtain non-Hermitian jump operators for the system.

If it is possible to measure different modes of the environment, then we can determine exactly which type of quantum jumps occurred, and ascribe a clear physical meaning to the individual trajectories. This was first done in this form in the context of photon counting \cite{Dum1992}. When we cannot ascribe a direct physical interpretation, then the average over the trajectories reproduces the corresponding master equation, and quantum trajectories still performs well as a means to simulate this. As also mentioned in \ref{sec:physicalinterp}, there are some situations where it may not be necessary to measure the environment directly, if the existence of particular jumps, or non-existence of particular types of jumps can be inferred from the state of the system. This type of postselection can be very useful, as it can be applied to measurements such as in the case of particle loss in many-body systems (section \ref{sec:particleloss}). This would allow us, e.g., to distinguish cases where loss events had occurred from the dynamics under the effective Hamiltonian $H_{\rm eff}$ in the absence of loss, postselecting on those times where the initial atom number remained conserved. This type of measurement postselection has even been implemented in a quantum gas microscope experiment, in the determination of string order parameters in the Bose-Hubbard model, to realise effective initial states of much lower temperature than would otherwise be achievable \cite{Endres2011}.

\bibliographystyle{tADP}
   \def\eprint#1{arXiv:#1}
   \def\Eprint#1{arXiv:#1}
\bibliography{traj_rev}

\begin{thebibliography}{362}
\providecommand{\natexlab}[1]{#1}

\bibitem[1]{Breuer2007}
H.P. Breuer {\itshape The theory of open quantum systems},    Oxford University
  Press, Oxford England New York, 2007.

\bibitem[2]{Gardiner2005}
C.W. Gardiner and P. Zoller {\itshape Quantum Noise},    Springer, Berlin,
  2005.

\bibitem[3]{Carmichael1993}
H.J. Carmichael {\itshape An Open Systems Approach to Quantum Optics},
  Springer, Berlin, 1993.

\bibitem[4]{Carmichael1999}
H.J. Carmichael {\itshape Statistical Methods in Quantum Optics 1: Master
  Equations and Fokker-Planck Equations},  Physics and Astronomy Online Library
   Springer, 1999.

\bibitem[5]{Weiss2012}
U. Weiss {\itshape Quantum dissipative systems},    World Scientific, Singapore
  Hackensack, N.J, 2012.

\bibitem[6]{Muller2012}
M. M{\"u}ller, S. Diehl, G. Pupillo, and P. Zoller, {\itshape Engineered Open
  Systems and Quantum Simulations with Atoms and Ions}, Advances In Atomic,
  Molecular, and Optical Physics 61 (2012), pp. 1--80.

\bibitem[7]{Nielsen2000}
M. Nielsen and I. Chuang {\itshape Quantum Computation and Quantum
  Information},  Cambridge Series on Information and the Natural Sciences
  Cambridge University Press, 2000.

\bibitem[8]{Monroe2013}
C. Monroe and J. Kim, {\itshape Scaling the Ion Trap Quantum Processor},
  Science 339 (2013), pp. 1164--1169.

\bibitem[9]{Awschalom2013}
D.D. Awschalom, L.C. Bassett, A.S. Dzurak, E.L. Hu, and J.R. Petta, {\itshape
  Quantum Spintronics: Engineering and Manipulating Atom-Like Spins in
  Semiconductors}, Science 339 (2013), pp. 1174--1179.

\bibitem[10]{Devoret2013}
M.H. Devoret and R.J. Schoelkopf, {\itshape Superconducting Circuits for
  Quantum Information: An Outlook}, Science 339 (2013), pp. 1169--1174.

\bibitem[11]{Kimble2008}
H.J. Kimble, {\itshape The quantum internet}, Nature 453 (2008), pp.
  1023--1030.

\bibitem[12]{Deleglise2008}
S. Deleglise, I. Dotsenko, C. Sayrin, J. Bernu, M. Brune, J.M. Raimond, and S.
  Haroche, {\itshape Reconstruction of non-classical cavity field states with
  snapshots of their decoherence}, Nature 455 (2008), pp. 510--514.

\bibitem[13]{Gleyzes2007}
S. Gleyzes, S. Kuhr, C. Guerlin, J. Bernu, S. Deleglise, U. Busk~Hoff, M.
  Brune, J.M. Raimond, and S. Haroche, {\itshape Quantum jumps of light
  recording the birth and death of a photon in a cavity}, Nature 446 (2007),
  pp. 297--300.

\bibitem[14]{Guerlin2007}
C. Guerlin, J. Bernu, S. Deleglise, C. Sayrin, S. Gleyzes, S. Kuhr, M. Brune,
  J.M. Raimond, and S. Haroche, {\itshape Progressive field-state collapse and
  quantum non-demolition photon counting}, Nature 448 (2007), pp. 889--893.

\bibitem[15]{Nogues1999}
G. Nogues, A. Rauschenbeutel, S. Osnaghi, M. Brune, J.M. Raimond, and S.
  Haroche, {\itshape Seeing a single photon without destroying it}, Nature 400
  (1999), pp. 239--242.

\bibitem[16]{Bloch2008}
I. Bloch, {\itshape Quantum coherence and entanglement with ultracold atoms in
  optical lattices}, Nature 453 (2008), pp. 1016--1022.

\bibitem[17]{Haffner2008}
H. H{\"a}ffner, C.F. Roos, and R. Blatt, {\itshape Quantum computing with
  trapped ions}, Physics Reports 469 (2008), pp. 155--203.

\bibitem[18]{Blatt2008}
R. Blatt and D. Wineland, {\itshape Entangled states of trapped atomic ions},
  Nature 453 (2008), pp. 1008--1015.

\bibitem[19]{Monz2011}
T. Monz, P. Schindler, J.T. Barreiro, M. Chwalla, D. Nigg, W.A. Coish, M.
  Harlander, W. H\"ansel, M. Hennrich, and R. Blatt, {\itshape 14-Qubit
  Entanglement: Creation and Coherence}, Phys. Rev. Lett. 106 (2011), p.
  130506.

\bibitem[20]{Leibfried2003}
D. Leibfried, R. Blatt, C. Monroe, and D. Wineland, {\itshape Quantum dynamics
  of single trapped ions}, Rev. Mod. Phys. 75 (2003), pp. 281--324.

\bibitem[21]{Myatt2000}
C.J. Myatt, B.E. King, Q.A. Turchette, C.A. Sackett, D. Kielpinski, W.M. Itano,
  C. Monroe, and D.J. Wineland, {\itshape Decoherence of quantum superpositions
  through coupling to engineered reservoirs}, Nature 403 (2000), pp. 269--273.

\bibitem[22]{Carr2009}
L.D. Carr, D. DeMille, R.V. Krems, and J. Ye, {\itshape Cold and ultracold
  molecules: science, technology and applications}, New Journal of Physics 11
  (2009), p. 055049.

\bibitem[23]{Yan2013}
B. Yan, S.A. Moses, B. Gadway, J.P. Covey, K.R.A. Hazzard, A.M. Rey, D.S. Jin,
  and J. Ye, {\itshape Observation of dipolar spin-exchange interactions with
  lattice-confined polar molecules}, Nature 501 (2013), pp. 521--525.

\bibitem[24]{Hanson2008}
R. Hanson and D.D. Awschalom, {\itshape Coherent manipulation of single spins
  in semiconductors}, Nature 453 (2008), pp. 1043--1049.

\bibitem[25]{Clarke2008}
J. Clarke and F.K. Wilhelm, {\itshape Superconducting quantum bits}, Nature 453
  (2008), pp. 1031--1042.

\bibitem[26]{Reed2012}
M.D. Reed, L. DiCarlo, S.E. Nigg, L. Sun, L. Frunzio, S.M. Girvin, and R.J.
  Schoelkopf, {\itshape Realization of three-qubit quantum error correction
  with superconducting circuits}, Nature 482 (2012), pp. 382--385.

\bibitem[27]{Caldeira1981}
A.O. Caldeira and A.J. Leggett, {\itshape Influence of Dissipation on Quantum
  Tunneling in Macroscopic Systems}, Phys. Rev. Lett. 46 (1981), pp. 211--214.

\bibitem[28]{Leggett1987}
A.J. Leggett, S. Chakravarty, A.T. Dorsey, M.P.A. Fisher, A. Garg, and W.
  Zwerger, {\itshape Dynamics of the dissipative two-state system}, Rev. Mod.
  Phys. 59 (1987), pp. 1--85.

\bibitem[29]{Hu1995}
B.L. Hu and A. Matacz, {\itshape Back reaction in semiclassical gravity: The
  Einstein-Langevin equation}, Phys. Rev. D 51 (1995), pp. 1577--1586.

\bibitem[30]{Lello2013}
L. Lello and D. Boyanovsky, {\itshape Searching for sterile neutrinos from
  $\pi$ and $\mathbf{K}$ decays}, Phys. Rev. D 87 (2013), p. 073017.

\bibitem[31]{Lello2013a}
L. Lello, D. Boyanovsky, and R. Holman, {\itshape Entanglement entropy in
  particle decay},  2013 (2013), p.~1.

\bibitem[32]{Brossel1952}
J. Brossel, A. Kastler, and J. Winter, {\itshape Gr{\'e}ation optique d'une
  in{\'e}galit{\'e} de population entre les sous-niveaux Zeeman de l'{\'e}tat
  fondamental des atomes}, J. Phys. Radium 13 (1952), p. 668.

\bibitem[33]{Hawkins1953}
W.B. Hawkins and R.H. Dicke, {\itshape The Polarization of Sodium Atoms}, Phys.
  Rev. 91 (1953), pp. 1008--1009.

\bibitem[34]{Wineland1978}
D.J. Wineland, R.E. Drullinger, and F.L. Walls, {\itshape Radiation-Pressure
  Cooling of Bound Resonant Absorbers}, Phys. Rev. Lett. 40 (1978), pp.
  1639--1642.

\bibitem[35]{Neuhauser1978}
W. Neuhauser, M. Hohenstatt, P. Toschek, and H. Dehmelt, {\itshape
  Optical-Sideband Cooling of Visible Atom Cloud Confined in Parabolic Well},
  Phys. Rev. Lett. 41 (1978), pp. 233--236.

\bibitem[36]{Metcalf1999}
H. Metcalf and P. Stratenvan~der  {\itshape Laser Cooling and Trapping},
  Graduate Texts in Contemporary Physics  Springer New York, 1999.

\bibitem[37]{Anderson1995a}
M.H. Anderson, J.R. Ensher, M.R. Matthews, C.E. Wieman, and E.A. Cornell,
  {\itshape Observation of {B}ose-{E}instein Condensation in a Dilute Atomic
  Vapor}, Science 269 (1995), p. 198.

\bibitem[38]{Davis1995b}
K.B. Davis, M.O. Mewes, M.R. Andrews, N.J. Drutenvan , D.S. Durfee, D.M. Kurn,
  and W. Ketterle, {\itshape {B}ose-{E}instein Condensation in a Gas of Sodium
  Atoms}, Phys. Rev. Lett. 75 (1995), p. 3969.

\bibitem[39]{Bradley1995a}
C.C. Bradley, C.A. Sackett, J.J. Tollett, and R.G. Hulet, {\itshape Evidence of
  {B}ose-{E}instein Condensation in an Atomic Gas with Attractive
  Interactions}, Phys. Rev. Lett. 75 (1995), p. 1687 {\it{ibid.} \bf{79}}, 1170
  (1997).

\bibitem[40]{DeMarco1999a}
B. DeMarco and D.S. Jin, {\itshape Onset of {F}ermi Degeneracy in a Trapped
  Atomic Gas}, Science 285 (1999), p. 1703.

\bibitem[41]{Truscott2001a}
A.G. Truscott, K.E. Strecker, W.I. McAlexander, G.B. Partridge, and R.G. Hulet,
  {\itshape Observation of {F}ermi Pressure in a Gas of Trapped Atoms}, Science
  291 (2001), p. 2570.

\bibitem[42]{Cirac1995}
J.I. Cirac and P. Zoller, {\itshape Quantum Computations with Cold Trapped
  Ions}, Phys. Rev. Lett. 74 (1995), pp. 4091--4094.

\bibitem[43]{Boykin2002}
P.O. Boykin, T. Mor, V. Roychowdhury, F. Vatan, and R. Vrijen, {\itshape
  Algorithmic cooling and scalable NMR quantum computers}, Proceedings of the
  National Academy of Sciences 99 (2002), pp. 3388--3393.

\bibitem[44]{Baugh2005}
J. Baugh, O. Moussa, C.A. Ryan, A. Nayak, and R. Laflamme, {\itshape
  Experimental implementation of heat-bath algorithmic cooling using
  solid-state nuclear magnetic resonance}, Nature 438 (2005), pp. 470--473.

\bibitem[45]{Chan2011}
J. Chan, T.P.M. Alegre, A.H. Safavi-Naeini, J.T. Hill, A. Krause, S.
  Groblacher, M. Aspelmeyer, and O. Painter, {\itshape Laser cooling of a
  nanomechanical oscillator into its quantum ground state}, Nature 478 (2011),
  pp. 89--92.

\bibitem[46]{Dehmelt1982}
H. Dehmelt, IEEE Trans. Instrum. Meas. 31 (1982), p.~83.

\bibitem[47]{Nagourney1986}
W. Nagourney, J. Sandberg, and H. Dehmelt, {\itshape Shelved optical electron
  amplifier: Observation of quantum jumps}, Phys. Rev. Lett. 56 (1986), pp.
  2797--2799.

\bibitem[48]{Bergquist1986}
J.C. Bergquist, R.G. Hulet, W.M. Itano, and D.J. Wineland, {\itshape
  Observation of Quantum Jumps in a Single Atom}, Phys. Rev. Lett. 57 (1986),
  pp. 1699--1702.

\bibitem[49]{Sauter1986}
T. Sauter, W. Neuhauser, R. Blatt, and P.E. Toschek, {\itshape Observation of
  Quantum Jumps}, Phys. Rev. Lett. 57 (1986), pp. 1696--1698.

\bibitem[50]{Hegerfeldt1992b}
G.C. Hegerfeldt and M.B. Plenio, {\itshape Macroscopic dark periods without a
  metastable state}, Phys. Rev. A 46 (1992), pp. 373--379.

\bibitem[51]{Hegerfeldt1995}
---{}---{}---, {\itshape Spectral structures induced by electron shelving},
  Phys. Rev. A 52 (1995), pp. 3333--3343.

\bibitem[52]{Wiseman2010}
H. Wiseman and G. Milburn {\itshape Quantum Measurement and Control},
  Cambridge University Press, 2010.

\bibitem[53]{Handel2005}
R. Handelvan , J.K. Stockton, and H. Mabuchi, {\itshape Modelling and feedback
  control design for quantum state preparation}, Journal of Optics B: Quantum
  and Semiclassical Optics 7 (2005), p. S179.

\bibitem[54]{Wiseman1993}
H.M. Wiseman and G.J. Milburn, {\itshape Quantum theory of optical feedback via
  homodyne detection}, Phys. Rev. Lett. 70 (1993), pp. 548--551.

\bibitem[55]{Foster2000}
G.T. Foster, L.A. Orozco, H.M. Castro-Beltran, and H.J. Carmichael, {\itshape
  Quantum State Reduction and Conditional Time Evolution of Wave-Particle
  Correlations in Cavity QED}, Phys. Rev. Lett. 85 (2000), pp. 3149--3152.

\bibitem[56]{Bushev2006}
P. Bushev, D. Rotter, A. Wilson, F. Dubin, C. Becher, J. Eschner, R. Blatt, V.
  Steixner, P. Rabl, and P. Zoller, {\itshape Feedback Cooling of a Single
  Trapped Ion}, Phys. Rev. Lett. 96 (2006), p. 043003.

\bibitem[57]{Haroche2013}
S. Haroche and J. Raimond {\itshape Exploring the Quantum: Atoms, Cavities, and
  Photons},  Oxford Graduate Texts  OUP Oxford, 2013.

\bibitem[58]{Raimond2001}
J.M. Raimond, M. Brune, and S. Haroche, {\itshape Manipulating quantum
  entanglement with atoms and photons in a cavity}, Rev. Mod. Phys. 73 (2001),
  pp. 565--582.

\bibitem[59]{Gardiner1985}
C.W. Gardiner and M.J. Collett, {\itshape Input and output in damped quantum
  systems: Quantum stochastic differential equations and the master equation},
  Phys. Rev. A 31 (1985), pp. 3761--3774.

\bibitem[60]{Yurke1984}
B. Yurke and J.S. Denker, {\itshape Quantum network theory}, Phys. Rev. A 29
  (1984), pp. 1419--1437.

\bibitem[61]{Caves1982}
C.M. Caves, {\itshape Quantum limits on noise in linear amplifiers}, Phys. Rev.
  D 26 (1982), pp. 1817--1839.

\bibitem[62]{Ritsch1988}
H. Ritsch and P. Zoller, {\itshape Atomic Transitions in Finite-Bandwidth
  Squeezed Light}, Phys. Rev. Lett. 61 (1988), pp. 1097--1100.

\bibitem[63]{Parkins1988}
A.S. Parkins and C.W. Gardiner, {\itshape Effect of finite-bandwidth squeezing
  on inhibition of atomic-phase decays}, Phys. Rev. A 37 (1988), pp.
  3867--3878.

\bibitem[64]{Lax1963}
M. Lax, {\itshape Formal Theory of Quantum Fluctuations from a Driven State},
  Phys. Rev. 129 (1963), pp. 2342--2348.

\bibitem[65]{Collett1984}
M.J. Collett and C.W. Gardiner, {\itshape Squeezing of intracavity and
  traveling-wave light fields produced in parametric amplification}, Phys. Rev.
  A 30 (1984), pp. 1386--1391.

\bibitem[66]{Plenio1998}
M.B. Plenio and P.L. Knight, {\itshape The quantum-jump approach to dissipative
  dynamics in quantum optics}, Rev. Mod. Phys. 70 (1998), pp. 101--144.

\bibitem[67]{Dalibard1992}
J. Dalibard, Y. Castin, and K. M\o{}lmer, {\itshape Wave-function approach to
  dissipative processes in quantum optics}, Phys. Rev. Lett. 68 (1992), pp.
  580--583.

\bibitem[68]{Molmer1993}
K. M{\o}lmer, Y. Castin, and J. Dalibard, {\itshape Monte Carlo wave-function
  method in quantum optics}, J. Opt. Soc. Am. B 10 (1993), pp. 524--538.

\bibitem[69]{Dum1992}
R. Dum, P. Zoller, and H. Ritsch, {\itshape Monte Carlo simulation of the
  atomic master equation for spontaneous emission}, Phys. Rev. A 45 (1992), pp.
  4879--4887.

\bibitem[70]{Cirac2012}
J.I. Cirac and P. Zoller, {\itshape Goals and opportunities in quantum
  simulation}, Nat Phys 8 (2012), pp. 264--266.

\bibitem[71]{Bloch2012}
I. Bloch, J. Dalibard, and S. Nascimbene, {\itshape Quantum simulations with
  ultracold quantum gases}, Nat Phys 8 (2012), pp. 267--276.

\bibitem[72]{Blatt2012}
R. Blatt and C.F. Roos, {\itshape Quantum simulations with trapped ions}, Nat
  Phys 8 (2012), pp. 277--284.

\bibitem[73]{Aspuru-Guzik2012}
A. Aspuru-Guzik and P. Walther, {\itshape Photonic quantum simulators}, Nat
  Phys 8 (2012), pp. 285--291.

\bibitem[74]{Bloch2008b}
I. Bloch, J. Dalibard, and W. Zwerger, {\itshape Many-body physics with
  ultracold gases}, Rev. Mod. Phys. 80 (2008), pp. 885--964.

\bibitem[75]{Lewenstein2007}
M. Lewenstein, A. Sanpera, V. Ahufinger, B. Damski, A. Sen(De), and U. Sen,
  {\itshape Ultracold atomic gases in optical lattices: mimicking condensed
  matter physics and beyond}, Advances in Physics 56 (2007), pp. 243--379.

\bibitem[76]{Lewenstein2012}
M. Lewenstein, A. Sanpera, and V. Ahufinger {\itshape Ultracold Atoms in
  Optical Lattices: Simulating quantum many-body systems},    OUP Oxford, 2012.

\bibitem[77]{McKay2011}
D.C. McKay and B. DeMarco, {\itshape Cooling in strongly correlated optical
  lattices: prospects and challenges}, Reports on Progress in Physics 74
  (2011), p. 054401.

\bibitem[78]{Pichler2010}
H. Pichler, A.J. Daley, and P. Zoller, {\itshape Nonequilibrium dynamics of
  bosonic atoms in optical lattices: Decoherence of many-body states due to
  spontaneous emission}, Phys. Rev. A 82 (2010), p. 063605.

\bibitem[79]{Poletti2012}
D. Poletti, J.S. Bernier, A. Georges, and C. Kollath, {\itshape
  Interaction-Induced Impeding of Decoherence and Anomalous Diffusion}, Phys.
  Rev. Lett. 109 (2012), p. 045302.

\bibitem[80]{Poletti2013}
D. Poletti, P. Barmettler, A. Georges, and C. Kollath, {\itshape Emergence of
  Glasslike Dynamics for Dissipative and Strongly Interacting Bosons}, Phys.
  Rev. Lett. 111 (2013), p. 195301.

\bibitem[81]{Pichler2012}
H. Pichler, J. Schachenmayer, J. Simon, P. Zoller, and A.J. Daley, {\itshape
  Noise- and disorder-resilient optical lattices}, Phys. Rev. A 86 (2012), p.
  051605.

\bibitem[82]{Pichler2013}
H. Pichler, J. Schachenmayer, A.J. Daley, and P. Zoller, {\itshape Heating
  dynamics of bosonic atoms in a noisy optical lattice}, Phys. Rev. A 87
  (2013), p. 033606.

\bibitem[83]{Schachenmayer2014}
J. Schachenmayer, L. Pollet, M. Troyer, and A.J. Daley, {\itshape Spontaneous
  emission and thermalization of cold bosons in optical lattices}, Phys. Rev. A
  89 (2014), p. 011601.

\bibitem[84]{Griessner2006}
A. Griessner, A.J. Daley, S.R. Clark, D. Jaksch, and P. Zoller, {\itshape
  Dark-State Cooling of Atoms by Superfluid Immersion}, Phys. Rev. Lett. 97
  (2006), p. 220403.

\bibitem[85]{Griessner2007}
---{}---{}---, {\itshape Dissipative dynamics of atomic Hubbard models coupled
  to a phonon bath: dark state cooling of atoms within a Bloch band of an
  optical lattice}, New Journal of Physics 9 (2007), p.~44.

\bibitem[86]{Chen2014}
D. Chen, C. Meldgin, and B. DeMarco, {\itshape Bath-induced band decay of a
  Hubbard lattice gas}, arXiv:1401.5096  (2014).

\bibitem[87]{Griessner2005}
A. Griessner, A.J. Daley, D. Jaksch, and P. Zoller, {\itshape Fault-tolerant
  dissipative preparation of atomic quantum registers with fermions}, Phys.
  Rev. A 72 (2005), p. 032332.

\bibitem[88]{Daley2004a}
A.J. Daley, P.O. Fedichev, and P. Zoller, {\itshape Single-atom cooling by
  superfluid immersion: A nondestructive method for qubits}, Phys. Rev. A 69
  (2004), p. 022306.

\bibitem[89]{Diehl2008}
S. Diehl, A. Micheli, A. Kantian, B. Kraus, H.P. Buchler, and P. Zoller,
  {\itshape Quantum states and phases in driven open quantum systems with cold
  atoms}, Nat Phys 4 (2008), pp. 878--883.

\bibitem[90]{Kraus2008}
B. Kraus, H.P. B\"uchler, S. Diehl, A. Kantian, A. Micheli, and P. Zoller,
  {\itshape Preparation of entangled states by quantum Markov processes}, Phys.
  Rev. A 78 (2008), p. 042307.

\bibitem[91]{Diehl2010}
S. Diehl, W. Yi, A.J. Daley, and P. Zoller, {\itshape Dissipation-Induced
  $d$-Wave Pairing of Fermionic Atoms in an Optical Lattice}, Phys. Rev. Lett.
  105 (2010), p. 227001.

\bibitem[92]{Yi2012}
W. Yi, S. Diehl, A.J. Daley, and P. Zoller, {\itshape Driven-dissipative
  many-body pairing states for cold fermionic atoms in an optical lattice}, New
  Journal of Physics 14 (2012), p. 055002.

\bibitem[93]{Han2009}
Y.J. Han, Y.H. Chan, W. Yi, A.J. Daley, S. Diehl, P. Zoller, and L.M. Duan,
  {\itshape Stabilization of the $p$-Wave Superfluid State in an Optical
  Lattice}, Phys. Rev. Lett. 103 (2009), p. 070404.

\bibitem[94]{Sandner2011}
R.M. Sandner, M. M\"uller, A.J. Daley, and P. Zoller, {\itshape Spatial Pauli
  blocking of spontaneous emission in optical lattices}, Phys. Rev. A 84
  (2011), p. 043825.

\bibitem[95]{Ates2012}
C. Ates, B. Olmos, W. Li, and I. Lesanovsky, {\itshape Dissipative Binding of
  Lattice Bosons through Distance-Selective Pair Loss}, Phys. Rev. Lett. 109
  (2012), p. 233003.

\bibitem[96]{Diehl2011}
S. Diehl, E. Rico, M.A. Baranov, and P. Zoller, {\itshape Topology by
  dissipation in atomic quantum wires}, Nat Phys 7 (2011), pp. 971--977.

\bibitem[97]{Bardyn2013}
C.E. Bardyn, M.A. Baranov, C.V. Kraus, E. Rico, A. {\.I}mamo{\u g}lu, P.
  Zoller, and S. Diehl, {\itshape Topology by dissipation}, New Journal of
  Physics 15 (2013), p. 085001.

\bibitem[98]{Syassen2008}
N. Syassen, D.M. Bauer, M. Lettner, T. Volz, D. Dietze, J.J.
  Garc{\'\i}a-Ripoll, J.I. Cirac, G. Rempe, and S. D{\"u}rr, {\itshape Strong
  Dissipation Inhibits Losses and Induces Correlations in Cold Molecular
  Gases}, Science 320 (2008), pp. 1329--1331.

\bibitem[99]{Garcia-Ripoll2009}
J.J. Garc{\'\i}a-Ripoll, S. D{\"u}rr, N. Syassen, D.M. Bauer, M. Lettner, G.
  Rempe, and J.I. Cirac, {\itshape Dissipation-induced hard-core boson gas in
  an optical lattice}, New Journal of Physics 11 (2009), p. 013053.

\bibitem[100]{Zhu2014}
B. Zhu et~al., {\itshape Suppressing the Loss of Ultracold Molecules Via the
  Continuous Quantum Zeno Effect}, Phys. Rev. Lett. 112 (2014), p. 070404.

\bibitem[101]{Daley2009}
A.J. Daley, J.M. Taylor, S. Diehl, M. Baranov, and P. Zoller, {\itshape Atomic
  Three-Body Loss as a Dynamical Three-Body Interaction}, Phys. Rev. Lett. 102
  (2009), p. 040402.

\bibitem[102]{Mark2012}
M.J. Mark, E. Haller, K. Lauber, J.G. Danzl, A. Janisch, H.P. B\"uchler, A.J.
  Daley, and H.C. N\"agerl, {\itshape Preparation and Spectroscopy of a
  Metastable Mott-Insulator State with Attractive Interactions}, Phys. Rev.
  Lett. 108 (2012), p. 215302.

\bibitem[103]{Schutzhold2010}
R. Sch\"utzhold and G. Gnanapragasam, {\itshape Quantum Zeno suppression of
  three-body losses in Bose-Einstein condensates}, Phys. Rev. A 82 (2010), p.
  022120.

\bibitem[104]{ChenYC2011}
Y.C. Chen, K.K. Ng, and M.F. Yang, {\itshape Quantum phase transitions in the
  attractive extended Bose-Hubbard model with a three-body constraint}, Phys.
  Rev. B 84 (2011), p. 092503.

\bibitem[105]{Privitera2011}
A. Privitera, I. Titvinidze, S.Y. Chang, S. Diehl, A.J. Daley, and W.
  Hofstetter, {\itshape Loss-induced phase separation and pairing for
  three-species atomic lattice fermions}, Phys. Rev. A 84 (2011), p. 021601.

\bibitem[106]{Bonnes2011}
L. Bonnes and S. Wessel, {\itshape Pair Superfluidity of Three-Body Constrained
  Bosons in Two Dimensions}, Phys. Rev. Lett. 106 (2011), p. 185302.

\bibitem[107]{Titvinidze2011}
I. Titvinidze, A. Privitera, S.Y. Chang, S. Diehl, M.A. Baranov, A. Daley, and
  W. Hofstetter, {\itshape Magnetism and domain formation in SU(3)-symmetric
  multi-species Fermi mixtures}, New Journal of Physics 13 (2011), p. 035013.

\bibitem[108]{Diehl2010b}
S. Diehl, M. Baranov, A.J. Daley, and P. Zoller, {\itshape Quantum field theory
  for the three-body constrained lattice Bose gas. II. Application to the
  many-body problem}, Phys. Rev. B 82 (2010), p. 064510.

\bibitem[109]{Diehl2010d}
---{}---{}---, {\itshape Quantum field theory for the three-body constrained
  lattice Bose gas. I. Formal developments}, Phys. Rev. B 82 (2010), p. 064509.

\bibitem[110]{Lee2010}
Y.W. Lee and M.F. Yang, {\itshape Superfluid-insulator transitions in
  attractive Bose-Hubbard model with three-body constraint}, Phys. Rev. A 81
  (2010), p. 061604.

\bibitem[111]{Diehl2010a}
S. Diehl, M. Baranov, A.J. Daley, and P. Zoller, {\itshape Observability of
  Quantum Criticality and a Continuous Supersolid in Atomic Gases}, Phys. Rev.
  Lett. 104 (2010), p. 165301.

\bibitem[112]{Roncaglia2010}
M. Roncaglia, M. Rizzi, and J.I. Cirac, {\itshape Pfaffian State Generation by
  Strong Three-Body Dissipation}, Phys. Rev. Lett. 104 (2010), p. 096803.

\bibitem[113]{Kantian2009}
A. Kantian, M. Dalmonte, S. Diehl, W. Hofstetter, P. Zoller, and A.J. Daley,
  {\itshape Atomic Color Superfluid via Three-Body Loss}, Phys. Rev. Lett. 103
  (2009), p. 240401.

\bibitem[114]{Capogrosso-Sansone2009}
B. Capogrosso-Sansone, S. Wessel, H.P. B\"uchler, P. Zoller, and G. Pupillo,
  {\itshape Phase diagram of one-dimensional hard-core bosons with three-body
  interactions}, Phys. Rev. B 79 (2009), p. 020503.

\bibitem[115]{Ottenstein2008}
T.B. Ottenstein, T. Lompe, M. Kohnen, A.N. Wenz, and S. Jochim, {\itshape
  Collisional Stability of a Three-Component Degenerate Fermi Gas}, Phys. Rev.
  Lett. 101 (2008), p. 203202.

\bibitem[116]{Chen2008}
B.L. Chen, X.B. Huang, S.P. Kou, and Y. Zhang, {\itshape Mott-Hubbard
  transition of bosons in optical lattices with three-body interactions}, Phys.
  Rev. A 78 (2008), p. 043603.

\bibitem[117]{Rapp2008}
A. Rapp, W. Hofstetter, and G. Zar\'and, {\itshape Trionic phase of ultracold
  fermions in an optical lattice: A variational study}, Phys. Rev. B 77 (2008),
  p. 144520.

\bibitem[118]{Rapp2007}
A. Rapp, G. Zar\'and, C. Honerkamp, and W. Hofstetter, {\itshape Color
  Superfluidity and ``Baryon'' Formation in Ultracold Fermions}, Phys. Rev.
  Lett. 98 (2007), p. 160405.

\bibitem[119]{Barontini2013}
G. Barontini, R. Labouvie, F. Stubenrauch, A. Vogler, V. Guarrera, and H. Ott,
  {\itshape Controlling the Dynamics of an Open Many-Body Quantum System with
  Localized Dissipation}, Phys. Rev. Lett. 110 (2013), p. 035302.

\bibitem[120]{Zezyulin2012}
D.A. Zezyulin, V.V. Konotop, G. Barontini, and H. Ott, {\itshape Macroscopic
  Zeno Effect and Stationary Flows in Nonlinear Waveguides with Localized
  Dissipation}, Phys. Rev. Lett. 109 (2012), p. 020405.

\bibitem[121]{Barmettler2011}
P. Barmettler and C. Kollath, {\itshape Controllable manipulation and detection
  of local densities and bipartite entanglement in a quantum gas by a
  dissipative defect}, Phys. Rev. A 84 (2011), p. 041606.

\bibitem[122]{Verstraete2009}
F. Verstraete, M.M. Wolf, and J. Ignacio~Cirac, {\itshape Quantum computation
  and quantum-state engineering driven by dissipation}, Nat Phys 5 (2009), pp.
  633--636.

\bibitem[123]{Kliesch2011}
M. Kliesch, T. Barthel, C. Gogolin, M. Kastoryano, and J. Eisert, {\itshape
  Dissipative Quantum Church-Turing Theorem}, Phys. Rev. Lett. 107 (2011), p.
  120501.

\bibitem[124]{Pastawski2011}
F. Pastawski, L. Clemente, and J.I. Cirac, {\itshape Quantum memories based on
  engineered dissipation}, Phys. Rev. A 83 (2011), p. 012304.

\bibitem[125]{Cho2011}
J. Cho, S. Bose, and M.S. Kim, {\itshape Optical Pumping into Many-Body
  Entanglement}, Phys. Rev. Lett. 106 (2011), p. 020504.

\bibitem[126]{McCutcheon2009}
D.P.S. McCutcheon, A. Nazir, S. Bose, and A.J. Fisher, {\itshape Long-lived
  spin entanglement induced by a spatially correlated thermal bath}, Phys. Rev.
  A 80 (2009), p. 022337.

\bibitem[127]{Vacanti2009}
G. Vacanti and A. Beige, {\itshape Cooling atoms into entangled states}, New
  Journal of Physics 11 (2009), p. 083008.

\bibitem[128]{Daley2011}
A. Daley, {\itshape Quantum computing and quantum simulation with group-II
  atoms}, Quantum Information Processing 10 (2011), pp. 865--884
  10.1007/s11128-011-0293-3.

\bibitem[129]{Daley2008b}
A.J. Daley, M.M. Boyd, J. Ye, and P. Zoller, {\itshape Quantum Computing with
  Alkaline-Earth-Metal Atoms}, Phys. Rev. Lett. 101 (2008), p. 170504.

\bibitem[130]{Foss-Feig2012}
M. Foss-Feig, A.J. Daley, J.K. Thompson, and A.M. Rey, {\itshape Steady-State
  Many-Body Entanglement of Hot Reactive Fermions}, Phys. Rev. Lett. 109
  (2012), p. 230501.

\bibitem[131]{Caballar2014}
R.C.F. Caballar, S. Diehl, H. M\"akel\"a, M. Oberthaler, and G. Watanabe,
  {\itshape Dissipative preparation of phase- and number-squeezed states with
  ultracold atoms}, Phys. Rev. A 89 (2014), p. 013620.

\bibitem[132]{Stannigel2013}
K. Stannigel, P. Hauke, D. Marcos, M. Hafezi, S. Diehl, M. Dalmonte, and P.
  Zoller, {\itshape Constrained dynamics via the Zeno effect in quantum
  simulation: Implementing non-Abelian lattice gauge theories with cold atoms},
  arXiv:1308.0528  (2013).

\bibitem[133]{Diehl2010c}
S. Diehl, A. Tomadin, A. Micheli, R. Fazio, and P. Zoller, {\itshape Dynamical
  Phase Transitions and Instabilities in Open Atomic Many-Body Systems}, Phys.
  Rev. Lett. 105 (2010), p. 015702.

\bibitem[134]{Sieberer2013a}
L.M. Sieberer, S.D. Huber, E. Altman, and S. Diehl, {\itshape Dynamical
  Critical Phenomena in Driven-Dissipative Systems}, Phys. Rev. Lett. 110
  (2013), p. 195301.

\bibitem[135]{Honing2012}
M. H\"oning, M. Moos, and M. Fleischhauer, {\itshape Critical exponents of
  steady-state phase transitions in fermionic lattice models}, Phys. Rev. A 86
  (2012), p. 013606.

\bibitem[136]{Olmos2012}
B. Olmos, I. Lesanovsky, and J.P. Garrahan, {\itshape Facilitated Spin Models
  of Dissipative Quantum Glasses}, Phys. Rev. Lett. 109 (2012), p. 020403.

\bibitem[137]{Lesanovsky2013}
I. Lesanovsky and J.P. Garrahan, {\itshape Kinetic Constraints, Hierarchical
  Relaxation, and Onset of Glassiness in Strongly Interacting and Dissipative
  Rydberg Gases}, Phys. Rev. Lett. 111 (2013), p. 215305.

\bibitem[138]{Lesanovsky2014}
---{}---{}---, {\itshape Out-of-equilibrium structures in strongly interacting
  Rydberg gases with dissipation}, arXiv:1402.2126  (2014).

\bibitem[139]{Petrosyan2013}
D. Petrosyan, M. H\"oning, and M. Fleischhauer, {\itshape Spatial correlations
  of Rydberg excitations in optically driven atomic ensembles}, Phys. Rev. A 87
  (2013), p. 053414.

\bibitem[140]{Honing2013}
M. H\"oning, D. Muth, D. Petrosyan, and M. Fleischhauer, {\itshape Steady-state
  crystallization of Rydberg excitations in an optically driven lattice gas},
  Phys. Rev. A 87 (2013), p. 023401.

\bibitem[141]{Olmos2013}
B. Olmos, D. Yu, and I. Lesanovsky, {\itshape Steady state properties of a
  driven atomic ensemble with non-local dissipation}, arXiv:1308.3967  (2013).

\bibitem[142]{Martin2013}
M.J. Martin, M. Bishof, M.D. Swallows, X. Zhang, C. Benko, J. Stechervon , A.V.
  Gorshkov, A.M. Rey, and J. Ye, {\itshape A Quantum Many-Body Spin System in
  an Optical Lattice Clock}, Science 341 (2013), pp. 632--636.

\bibitem[143]{Tauber2013}
U.C. Tauber and S. Diehl, {\itshape Perturbative Field-Theoretical
  Renormalization Group Approach to Driven-Dissipative Bose-Einstein
  Criticality}, arXiv:1312.5182  (2013).

\bibitem[144]{Sieberer2013}
L.M. Sieberer, S.D. Huber, E. Altman, and S. Diehl, {\itshape Non-equilibrium
  Functional Renormalization for Driven-Dissipative Bose-Einstein
  Condensation}, arXiv:1309.7027  (2013).

\bibitem[145]{Ritsch2013}
H. Ritsch, P. Domokos, F. Brennecke, and T. Esslinger, {\itshape Cold atoms in
  cavity-generated dynamical optical potentials}, Rev. Mod. Phys. 85 (2013),
  pp. 553--601.

\bibitem[146]{Baumann2010}
K. Baumann, C. Guerlin, F. Brennecke, and T. Esslinger, {\itshape Dicke quantum
  phase transition with a superfluid gas in an optical cavity}, Nature 464
  (2010), pp. 1301--1306.

\bibitem[147]{Dimer2007}
F. Dimer, B. Estienne, A.S. Parkins, and H.J. Carmichael, {\itshape Proposed
  realization of the Dicke-model quantum phase transition in an optical cavity
  QED system}, Phys. Rev. A 75 (2007), p. 013804.

\bibitem[148]{Brooks2012}
D.W.C. Brooks, T. Botter, S. Schreppler, T.P. Purdy, N. Brahms, and D.M.
  Stamper-Kurn, {\itshape Non-classical light generated by quantum-noise-driven
  cavity optomechanics}, Nature 488 (2012), pp. 476--480.

\bibitem[149]{Habibian2013}
H. Habibian, A. Winter, S. Paganelli, H. Rieger, and G. Morigi, {\itshape
  Quantum phases of incommensurate optical lattices due to cavity backaction},
  Phys. Rev. A 88 (2013), p. 043618.

\bibitem[150]{Brahms2012}
N. Brahms, T. Botter, S. Schreppler, D.W.C. Brooks, and D.M. Stamper-Kurn,
  {\itshape Optical Detection of the Quantization of Collective Atomic Motion},
  Phys. Rev. Lett. 108 (2012), p. 133601.

\bibitem[151]{Torre2013}
E.G.D. Torre, S. Diehl, M.D. Lukin, S. Sachdev, and P. Strack, {\itshape
  Keldysh approach for nonequilibrium phase transitions in quantum optics:
  Beyond the Dicke model in optical cavities}, Phys. Rev. A 87 (2013), p.
  023831.

\bibitem[152]{Buchhold2013}
M. Buchhold, P. Strack, S. Sachdev, and S. Diehl, {\itshape Dicke-model quantum
  spin and photon glass in optical cavities: Nonequilibrium theory and
  experimental signatures}, Phys. Rev. A 87 (2013), p. 063622.

\bibitem[153]{Peyronel2012}
T. Peyronel, O. Firstenberg, Q.Y. Liang, S. Hofferberth, A.V. Gorshkov, T.
  Pohl, M.D. Lukin, and V. Vuletic, {\itshape Quantum nonlinear optics with
  single photons enabled by strongly interacting atoms}, Nature 488 (2012), pp.
  57--60.

\bibitem[154]{Reinhard2012}
A. Reinhard, T. Volz, M. Winger, A. Badolato, K.J. Hennessy, E.L. Hu, and A.
  Imamoglu, {\itshape Strongly correlated photons on a chip}, Nat Photon 6
  (2012), pp. 93--96.

\bibitem[155]{Petrosyan2011}
D. Petrosyan, J. Otterbach, and M. Fleischhauer, {\itshape Electromagnetically
  Induced Transparency with Rydberg Atoms}, Phys. Rev. Lett. 107 (2011), p.
  213601.

\bibitem[156]{Sevincli2011}
S. Sevin\c{c}li, N. Henkel, C. Ates, and T. Pohl, {\itshape Nonlocal Nonlinear
  Optics in Cold Rydberg Gases}, Phys. Rev. Lett. 107 (2011), p. 153001.

\bibitem[157]{Gorshkov2011}
A.V. Gorshkov, J. Otterbach, M. Fleischhauer, T. Pohl, and M.D. Lukin,
  {\itshape Photon-Photon Interactions via Rydberg Blockade}, Phys. Rev. Lett.
  107 (2011), p. 133602.

\bibitem[158]{Barreiro2011}
J.T. Barreiro, M. Muller, P. Schindler, D. Nigg, T. Monz, M. Chwalla, M.
  Hennrich, C.F. Roos, P. Zoller, and R. Blatt, {\itshape An open-system
  quantum simulator with trapped ions}, Nature 470 (2011), pp. 486--491.

\bibitem[159]{Schindler2013}
P. Schindler, M. Muller, D. Nigg, J.T. Barreiro, E.A. Martinez, M. Hennrich, T.
  Monz, S. Diehl, P. Zoller, and R. Blatt, {\itshape Quantum simulation of
  dynamical maps with trapped ions}, Nat Phys 9 (2013), pp. 361--367.

\bibitem[160]{Tomadin2012}
A. Tomadin, S. Diehl, M.D. Lukin, P. Rabl, and P. Zoller, {\itshape Reservoir
  engineering and dynamical phase transitions in optomechanical arrays}, Phys.
  Rev. A 86 (2012), p. 033821.

\bibitem[161]{Lechner2013}
W. Lechner, S.J.M. Habraken, N. Kiesel, M. Aspelmeyer, and P. Zoller, {\itshape
  Cavity Optomechanics of Levitated Nanodumbbells: Nonequilibrium Phases and
  Self-Assembly}, Phys. Rev. Lett. 110 (2013), p. 143604.

\bibitem[162]{Lindblad1976}
G. Lindblad, {\itshape On the generators of quantum dynamical semigroups},
  Communications in Mathematical Physics 48 (1976), pp. 119--130.

\bibitem[163]{Dum1992a}
R. Dum, A.S. Parkins, P. Zoller, and C.W. Gardiner, {\itshape Monte Carlo
  simulation of master equations in quantum optics for vacuum, thermal, and
  squeezed reservoirs}, Phys. Rev. A 46 (1992), pp. 4382--4396.

\bibitem[164]{Hegerfeldt1992}
G.C. Hegerfeldt and T.S. Wilser, {\itshape Classical and Quantum Systems}, in
  {\itshape Proceedings of the Second International Wigner Symposium}  World
  Scientific, 1992.

\bibitem[165]{Marte1993}
P. Marte, R. Dum, R. Ta\"{\i}eb, P.D. Lett, and P. Zoller, {\itshape Quantum
  wave function simulation of the resonance fluorescence spectrum from
  one-dimensional optical molasses}, Phys. Rev. Lett. 71 (1993), pp.
  1335--1338.

\bibitem[166]{Castin1995}
Y. Castin and K. M\o{}lmer, {\itshape Monte Carlo Wave-Function Analysis of 3D
  Optical Molasses}, Phys. Rev. Lett. 74 (1995), pp. 3772--3775.

\bibitem[167]{Marte1993b}
P. Marte, R. Dum, R. Ta\"{\i}eb, and P. Zoller, {\itshape Resonance
  fluorescence from quantized one-dimensional molasses}, Phys. Rev. A 47
  (1993), pp. 1378--1390.

\bibitem[168]{Bardou1994}
F. Bardou, J.P. Bouchaud, O. Emile, A. Aspect, and C. Cohen-Tannoudji,
  {\itshape Subrecoil laser cooling and L{\'e}vy flights}, Phys. Rev. Lett. 72
  (1994), pp. 203--206.

\bibitem[169]{Carmichael1993b}
H.J. Carmichael, {\itshape Quantum trajectory theory for cascaded open
  systems}, Phys. Rev. Lett. 70 (1993), pp. 2273--2276.

\bibitem[170]{Gardiner1993}
C.W. Gardiner, {\itshape Driving a quantum system with the output field from
  another driven quantum system}, Phys. Rev. Lett. 70 (1993), pp. 2269--2272.

\bibitem[171]{Kochan1994}
P. Kochan and H.J. Carmichael, {\itshape Photon-statistics dependence of
  single-atom absorption}, Phys. Rev. A 50 (1994), pp. 1700--1709.

\bibitem[172]{Itano1990}
W.M. Itano, D.J. Heinzen, J.J. Bollinger, and D.J. Wineland, {\itshape Quantum
  Zeno effect}, Phys. Rev. A 41 (1990), pp. 2295--2300.

\bibitem[173]{Power1996}
W.L. Power and P.L. Knight, {\itshape Stochastic simulations of the quantum
  Zeno effect}, Phys. Rev. A 53 (1996), pp. 1052--1059.

\bibitem[174]{Beige1996}
A. Beige and G.C. Hegerfeldt, {\itshape Projection postulate and atomic quantum
  Zeno effect}, Phys. Rev. A 53 (1996), pp. 53--65.

\bibitem[175]{Garraway1994}
B.M. Garraway and P.L. Knight, {\itshape Comparison of quantum-state diffusion
  and quantum-jump simulations of two-photon processes in a dissipative
  environment}, Phys. Rev. A 49 (1994), pp. 1266--1274.

\bibitem[176]{Moya-Cessa1993}
H. Moya-Cessa, V. Bu\v{z}ek, M.S. Kim, and P.L. Knight, {\itshape Intrinsic
  decoherence in the atom-field interaction}, Phys. Rev. A 48 (1993), pp.
  3900--3905.

\bibitem[177]{Imamoglu1993}
A. Imamoglu, {\itshape Quantum-nondemolition measurements using dissipative
  atom-field coupling: Monte Carlo wave-function approach}, Phys. Rev. A 48
  (1993), pp. 770--781.

\bibitem[178]{Ammann1998}
H. Ammann, R. Gray, I. Shvarchuck, and N. Christensen, {\itshape Quantum
  Delta-Kicked Rotor: Experimental Observation of Decoherence}, Phys. Rev.
  Lett. 80 (1998), pp. 4111--4115.

\bibitem[179]{Doherty2000}
A.C. Doherty, K.M.D. Vant, G.H. Ball, N. Christensen, and R. Leonhardt,
  {\itshape Momentum distributions for the quantum δ-kicked rotor with
  decoherence}, Journal of Optics B: Quantum and Semiclassical Optics 2 (2000),
  p. 605.

\bibitem[180]{Daley2002}
A.J. Daley, A.S. Parkins, R. Leonhardt, and S.M. Tan, {\itshape Diffusion
  resonances in action space for an atom optics kicked rotor with decoherence},
  Phys. Rev. E 65 (2002), p. 035201.

\bibitem[181]{Wimberger2003}
S. Wimberger, I. Guarneri, and S. Fishman, {\itshape Quantum resonances and
  decoherence for δ-kicked atoms}, Nonlinearity 16 (2003), p. 1381.

\bibitem[182]{Gisin1984}
N. Gisin, {\itshape Quantum Measurements and Stochastic Processes}, Phys. Rev.
  Lett. 52 (1984), pp. 1657--1660.

\bibitem[183]{Gisin1989}
---{}---{}---, {\itshape Stochastic quantum dynamics and relativity}, Helv.
  Phys., Acta 62 (1989), p. 363.

\bibitem[184]{Gisin1992}
N. Gisin and I.C. Percival, {\itshape The quantum-state diffusion model applied
  to open systems}, Journal of Physics A: Mathematical and General 25 (1992),
  p. 5677.

\bibitem[185]{Goetsch1994}
P. Goetsch and R. Graham, {\itshape Linear stochastic wave equations for
  continuously measured quantum systems}, Phys. Rev. A 50 (1994), pp.
  5242--5255.

\bibitem[186]{Garraway1994b}
B.M. Garraway and P.L. Knight, {\itshape Evolution of quantum superpositions in
  open environments: Quantum trajectories, jumps, and localization in phase
  space}, Phys. Rev. A 50 (1994), pp. 2548--2563.

\bibitem[187]{Wiseman1993b}
H.M. Wiseman and G.J. Milburn, {\itshape Interpretation of quantum jump and
  diffusion processes illustrated on the Bloch sphere}, Phys. Rev. A 47 (1993),
  pp. 1652--1666.

\bibitem[188]{Cohen-Tannoudji1998}
C. Cohen-Tannoudji, J. Dupont-Roc, and G. Grynberg {\itshape Atom-Photon
  Interactions: Basic Processes and Applications},  A Wiley-Interscience
  publication  Wiley, 1998.

\bibitem[189]{Mahan2000}
G. Mahan {\itshape Many-Particle Physics},  Physics of Solids and Liquids
  Springer, 2000.

\bibitem[190]{Lax1968}
M. Lax, {\itshape Quantum Noise. XI. Multitime Correspondence between Quantum
  and Classical Stochastic Processes}, Phys. Rev. 172 (1968), pp. 350--361.

\bibitem[191]{Press1995}
W.H. Press, S.A. Teukolsky, W.T. Vetterling, and B.P. Flannery {\itshape
  Numerical Recipes in C},    Cambridge University Press, Cambridge, 1995.

\bibitem[192]{Frauchiger2013}
D. Frauchiger, R. Renner, and M. Troyer, {\itshape True randomness from
  realistic quantum devices}, arXiv:1311.4547  (2013).

\bibitem[193]{Holland1998}
M. Holland, {\itshape Unraveling Quantum Dissipation in the Frequency Domain},
  Phys. Rev. Lett. 81 (1998), pp. 5117--5120.

\bibitem[194]{Breslin1995}
J.K. Breslin, G.J. Milburn, and H.M. Wiseman, {\itshape Optimal Quantum
  Trajectories for Continuous Measurement}, Phys. Rev. Lett. 74 (1995), pp.
  4827--4830.

\bibitem[195]{Wiseman2000}
H.M. Wiseman and Z. Brady, {\itshape Robust unravelings for resonance
  fluorescence}, Phys. Rev. A 62 (2000), p. 023805.

\bibitem[196]{Atkins2005}
D.J. Atkins, Z. Brady, K. Jacobs, and H.M. Wiseman, {\itshape Classical
  robustness of quantum unravellings}, EPL (Europhysics Letters) 69 (2005), p.
  163.

\bibitem[197]{Wiseman2005}
H.M. Wiseman and A.C. Doherty, {\itshape Optimal Unravellings for Feedback
  Control in Linear Quantum Systems}, Phys. Rev. Lett. 94 (2005), p. 070405.

\bibitem[198]{Wiseman2012}
H.M. Wiseman and J.M. Gambetta, {\itshape Are Dynamical Quantum Jumps Detector
  Dependent?}, Phys. Rev. Lett. 108 (2012), p. 220402.

\bibitem[199]{Steinbach1995}
J. Steinbach, B.M. Garraway, and P.L. Knight, {\itshape High-order unraveling
  of master equations for dissipative evolution}, Phys. Rev. A 51 (1995), pp.
  3302--3308.

\bibitem[200]{Jaksch1998}
D. Jaksch, C. Bruder, J.I. Cirac, C.W. Gardiner, and P. Zoller, {\itshape Cold
  Bosonic Atoms in Optical Lattices}, Phys. Rev. Lett. 81 (1998), pp.
  3108--3111.

\bibitem[201]{Jaksch2005}
D. Jaksch and P. Zoller, {\itshape The cold atom Hubbard toolbox}, Annals of
  Physics 315 (2005), pp. 52--79.

\bibitem[202]{Sachdev2011}
S. Sachdev {\itshape Quantum Phase Transitions},  Troisi{\`e}me Cycle de la
  Physique  Cambridge University Press, 2011.

\bibitem[203]{Lehmberg1970}
R.H. Lehmberg, {\itshape Radiation from an $N$-Atom System. I. General
  Formalism}, Phys. Rev. A 2 (1970), pp. 883--888.

\bibitem[204]{Lehmberg1970a}
---{}---{}---, {\itshape Radiation from an $N$-Atom System. II. Spontaneous
  Emission from a Pair of Atoms}, Phys. Rev. A 2 (1970), pp. 889--896.

\bibitem[205]{Lux2013}
J. Lux, J. M{\"u}ller, A. Mitra, and A. Rosch, {\itshape Hydrodynamic long-time
  tails after a quantum quench}, arXiv:1311.7644  (2013).

\bibitem[206]{Bauer2011}
B. Bauer et~al., {\itshape The ALPS project release 2.0: open source software
  for strongly correlated systems}, Journal of Statistical Mechanics: Theory
  and Experiment 2011 (2011), p. P05001.

\bibitem[207]{Snoek2007}
M. Snoek and W. Hofstetter, {\itshape Two-dimensional dynamics of ultracold
  atoms in optical lattices}, Phys. Rev. A 76 (2007), p. 051603.

\bibitem[208]{Wernsdorfer2010}
J. Wernsdorfer, M. Snoek, and W. Hofstetter, {\itshape Lattice-ramp-induced
  dynamics in an interacting Bose-Bose mixture}, Phys. Rev. A 81 (2010), p.
  043620.

\bibitem[209]{Zakrzewski2005}
J. Zakrzewski, {\itshape Mean-field dynamics of the superfluid-insulator phase
  transition in a gas of ultracold atoms}, Phys. Rev. A 71 (2005), p. 043601.

\bibitem[210]{Rokhsar1991}
D.S. Rokhsar and B.G. Kotliar, {\itshape Gutzwiller projection for bosons},
  Phys. Rev. B 44 (1991), pp. 10328--10332.

\bibitem[211]{Byczuk2008}
K. Byczuk and D. Vollhardt, {\itshape Correlated bosons on a lattice: Dynamical
  mean-field theory for Bose-Einstein condensed and normal phases}, Phys. Rev.
  B 77 (2008), p. 235106.

\bibitem[212]{Hubener2009}
A. Hubener, M. Snoek, and W. Hofstetter, {\itshape Magnetic phases of
  two-component ultracold bosons in an optical lattice}, Phys. Rev. B 80
  (2009), p. 245109.

\bibitem[213]{Anders2011}
P. Anders, E. Gull, L. Pollet, M. Troyer, and P. Werner, {\itshape Dynamical
  mean-field theory for bosons}, New Journal of Physics 13 (2011), p. 075013.

\bibitem[214]{Pollet2013}
L. Pollet, {\itshape A review of Monte Carlo simulations for the Bose-Hubbard
  model with diagonal disorder}, arXiv:1307.5430  (2013).

\bibitem[215]{Schollwock2011}
U. Schollw{\"o}ck, {\itshape The density-matrix renormalization group in the
  age of matrix product states}, Annals of Physics 326 (2011), pp. 96--192.

\bibitem[216]{Schollwock2005}
U. Schollw\"ock, {\itshape The density-matrix renormalization group}, Rev. Mod.
  Phys. 77 (2005), pp. 259--315.

\bibitem[217]{White1992}
S.R. White, {\itshape Density matrix formulation for quantum renormalization
  groups}, Phys. Rev. Lett. 69 (1992), pp. 2863--2866.

\bibitem[218]{Vidal2003}
G. Vidal, {\itshape Efficient Classical Simulation of Slightly Entangled
  Quantum Computations}, Phys. Rev. Lett. 91 (2003), p. 147902.

\bibitem[219]{Vidal2004}
---{}---{}---, {\itshape Efficient Simulation of One-Dimensional Quantum
  Many-Body Systems}, Phys. Rev. Lett. 93 (2004), p. 040502.

\bibitem[220]{White2004}
S.R. White and A.E. Feiguin, {\itshape Real-Time Evolution Using the Density
  Matrix Renormalization Group}, Phys. Rev. Lett. 93 (2004), p. 076401.

\bibitem[221]{Daley2004}
A.J. Daley, C. Kollath, U. Schollw\"ock, and G. Vidal, {\itshape Time-dependent
  density-matrix renormalization-group using adaptive effective Hilbert
  spaces}, Journal of Statistical Mechanics: Theory and Experiment  (2004), p.
  P04005.

\bibitem[222]{McCulloch2007}
I.P. McCulloch, {\itshape From density-matrix renormalization group to matrix
  product states}, Journal of Statistical Mechanics: Theory and Experiment 2007
  (2007), p. P10014.

\bibitem[223]{Verstraete2008}
F. Verstraete, V. Murg, and J.I. Cirac, {\itshape Matrix product states,
  projected entangled pair states, and variational renormalization group
  methods for quantum spin systems}, Advances in Physics 57 (2008), pp. 143 --
  224.

\bibitem[224]{Crosswhite2008}
G.M. Crosswhite, A.C. Doherty, and G. Vidal, {\itshape Applying matrix product
  operators to model systems with long-range interactions}, Phys. Rev. B 78
  (2008), p. 035116.

\bibitem[225]{Pirvu2010}
B. Pirvu, V. Murg, J.I. Cirac, and F. Verstraete, {\itshape Matrix product
  operator representations}, New Journal of Physics 12 (2010), p. 025012.

\bibitem[226]{Verstraete2004}
F. Verstraete, J.J. Garc\'\i{}a-Ripoll, and J.I. Cirac, {\itshape Matrix
  Product Density Operators: Simulation of Finite-Temperature and Dissipative
  Systems}, Phys. Rev. Lett. 93 (2004), p. 207204.

\bibitem[227]{Zwolak2004}
M. Zwolak and G. Vidal, {\itshape Mixed-State Dynamics in One-Dimensional
  Quantum Lattice Systems: A Time-Dependent Superoperator Renormalization
  Algorithm}, Phys. Rev. Lett. 93 (2004), p. 207205.

\bibitem[228]{Feiguin2005}
A.E. Feiguin and S.R. White, {\itshape Finite-temperature density matrix
  renormalization using an enlarged Hilbert space}, Phys. Rev. B 72 (2005), p.
  220401.

\bibitem[229]{Stoudenmire2010}
E.M. Stoudenmire and S.R. White, {\itshape Minimally entangled typical thermal
  state algorithms}, New Journal of Physics 12 (2010), p. 055026.

\bibitem[230]{Verstraete2004b}
F. Verstraete and J.I. Cirac, {\itshape Renormalization algorithms for Quantum
  Many-Body systems in two and higher dimensions}, arXiv:cond-mat/0407066v1
  (2004).

\bibitem[231]{Vidal2008}
G. Vidal, {\itshape Class of Quantum Many-Body States That Can Be Efficiently
  Simulated}, Phys. Rev. Lett. 101 (2008), p. 110501.

\bibitem[232]{Eisert2010}
J. Eisert, M. Cramer, and M.B. Plenio, {\itshape Colloquium: Area laws for the
  entanglement entropy}, Rev. Mod. Phys. 82 (2010), pp. 277--306.

\bibitem[233]{Amico2008}
L. Amico, R. Fazio, A. Osterloh, and V. Vedral, {\itshape Entanglement in
  many-body systems}, Rev. Mod. Phys. 80 (2008), pp. 517--576.

\bibitem[234]{Calabrese2009}
P. Calabrese and J. Cardy, {\itshape Entanglement entropy and conformal field
  theory}, Journal of Physics A: Mathematical and Theoretical 42 (2009), p.
  504005.

\bibitem[235]{Calabrese2009b}
P. Calabrese, J. Cardy, and B. Doyon, {\itshape Entanglement entropy in
  extended quantum systems}, Journal of Physics A: Mathematical and Theoretical
  42 (2009), p. 500301.

\bibitem[236]{Peschel2009}
I. Peschel and V. Eisler, {\itshape Reduced density matrices and entanglement
  entropy in free lattice models}, Journal of Physics A: Mathematical and
  Theoretical 42 (2009), p. 504003.

\bibitem[237]{Cardy2010}
J. Cardy and P. Calabrese, {\itshape Unusual corrections to scaling in
  entanglement entropy}, Journal of Statistical Mechanics: Theory and
  Experiment 2010 (2010), p. P04023.

\bibitem[238]{Schuch2008}
N. Schuch, M.M. Wolf, F. Verstraete, and J.I. Cirac, {\itshape Entropy Scaling
  and Simulability by Matrix Product States}, Phys. Rev. Lett. 100 (2008), p.
  030504.

\bibitem[239]{Schuch2008b}
N. Schuch, M.M. Wolf, K.G.H. Vollbrecht, and J.I. Cirac, {\itshape On entropy
  growth and the hardness of simulating time evolution}, New Journal of Physics
  10 (2008), p. 033032.

\bibitem[240]{Verstraete2006}
F. Verstraete and J.I. Cirac, {\itshape Matrix product states represent ground
  states faithfully}, Phys. Rev. B 73 (2006), p. 094423.

\bibitem[241]{Sornborger1999}
A.T. Sornborger and E.D. Stewart, {\itshape Higher-order methods for
  simulations on quantum computers}, Phys. Rev. A 60 (1999), pp. 1956--1965.

\bibitem[242]{Garcia-Ripoll2006}
J.J. Garc{\'\i}a-Ripoll, {\itshape Time evolution of Matrix Product States},
  New Journal of Physics 8 (2006), p. 305.

\bibitem[243]{Kantian2010}
A. Kantian, A.J. Daley, and P. Zoller, {\itshape $\eta$ Condensate of Fermionic
  Atom Pairs via Adiabatic State Preparation}, Phys. Rev. Lett. 104 (2010), p.
  240406.

\bibitem[244]{Gobert2005}
D. Gobert, C. Kollath, U. Schollw\"ock, and G. Sch\"utz, {\itshape Real-time
  dynamics in spin-1/2 chains with adaptive time-dependent density matrix
  renormalization group}, Phys. Rev. E 71 (2005), p. 036102.

\bibitem[245]{Prosen2007}
T. Prosen and M. \v{Z}nidari\v{c}, {\itshape Is the efficiency of classical
  simulations of quantum dynamics related to integrability?}, Phys. Rev. E 75
  (2007), p. 015202.

\bibitem[246]{Trotzky2012}
S. Trotzky, Y.A. Chen, A. Flesch, I.P. McCulloch, U. Schollwock, J. Eisert, and
  I. Bloch, {\itshape Probing the relaxation towards equilibrium in an isolated
  strongly correlated one-dimensional Bose gas}, Nat Phys 8 (2012), pp.
  325--330.

\bibitem[247]{Calabrese2005}
P. Calabrese and J. Cardy, {\itshape Evolution of entanglement entropy in
  one-dimensional systems}, Journal of Statistical Mechanics: Theory and
  Experiment 2005 (2005), p. P04010.

\bibitem[248]{Lauchli2008}
A.M. L{\"a}uchli and C. Kollath, {\itshape Spreading of correlations and
  entanglement after a quench in the one-dimensional Bose--Hubbard model},
  Journal of Statistical Mechanics: Theory and Experiment 2008 (2008), p.
  P05018.

\bibitem[249]{Unanyan2010}
R.G. Unanyan, D. Muth, and M. Fleischhauer, {\itshape Short-time versus
  long-time dynamics of entanglement in quantum lattice models}, Phys. Rev. A
  81 (2010), p. 022119.

\bibitem[250]{Lieb1972}
E. Lieb and D. Robinson, {\itshape The finite group velocity of quantum spin
  systems},  28 (1972), pp. 251--257.

\bibitem[251]{Daley2012}
A.J. Daley, H. Pichler, J. Schachenmayer, and P. Zoller, {\itshape Measuring
  Entanglement Growth in Quench Dynamics of Bosons in an Optical Lattice},
  Phys. Rev. Lett. 109 (2012), p. 020505.

\bibitem[252]{Cardy2011}
J. Cardy, {\itshape Measuring Entanglement Using Quantum Quenches}, Phys. Rev.
  Lett. 106 (2011), p. 150404.

\bibitem[253]{Moura-Alves2004}
C. Moura~Alves and D. Jaksch, {\itshape Multipartite Entanglement Detection in
  Bosons}, Phys. Rev. Lett. 93 (2004), p. 110501.

\bibitem[254]{Palmer2005}
R.N. Palmer, C. Moura~Alves, and D. Jaksch, {\itshape Detection and
  characterization of multipartite entanglement in optical lattices}, Phys.
  Rev. A 72 (2005), p. 042335.

\bibitem[255]{Guhne2009}
O. G{\"u}hne and G. T{\'o}th, {\itshape Entanglement detection}, Physics
  Reports 474 (2009), pp. 1--75.

\bibitem[256]{Abanin2012}
D.A. Abanin and E. Demler, {\itshape Measuring Entanglement Entropy of a
  Generic Many-Body System with a Quantum Switch}, Phys. Rev. Lett. 109 (2012),
  p. 020504.

\bibitem[257]{Schachenmayer2013}
J. Schachenmayer, B.P. Lanyon, C.F. Roos, and A.J. Daley, {\itshape
  Entanglement Growth in Quench Dynamics with Variable Range Interactions},
  Phys. Rev. X 3 (2013), p. 031015.

\bibitem[258]{Calabrese2007}
P. Calabrese and J. Cardy, {\itshape Entanglement and correlation functions
  following a local quench: a conformal field theory approach}, Journal of
  Statistical Mechanics: Theory and Experiment 2007 (2007), p. P10004.

\bibitem[259]{Pethick2002}
C. Pethick and H. Smith {\itshape Bose-Einstein Condensation in Dilute Gases},
    Cambridge University Press, 2002.

\bibitem[260]{Dalibard1998}
J. Dalibard, {\itshape Collisional dynamics of ultra-cold atomic gases, in
  Proceedings of the International School of Physics Enrico Fermi, Course CXL:
  Bose -- Einstein condensation in gases, Varenna}, M.~Inguscio, S.~Stringari
  and C.~Wieman,  eds.,  , 1998.

\bibitem[261]{Castin2001}
Y. Castin, {\itshape Bose-Einstein Condensates in Atomic Gases: Simple
  Theoretical Results}, in  {\itshape Coherent atomic matter waves}, R.~Kaiser,
  C.~Westbrook and F.~David,  eds.,    Springer Berlin Heidelberg, 2001, pp.
  1--136.

\bibitem[262]{Chin2010}
C. Chin, R. Grimm, P. Julienne, and E. Tiesinga, {\itshape Feshbach resonances
  in ultracold gases}, Reviews of Modern Physics 82 (2010), pp. 1225--1286.

\bibitem[263]{Kohn1959}
W. Kohn, {\itshape Analytic Properties of Bloch Waves and Wannier Functions},
  Phys. Rev. 115 (1959), pp. 809--821.

\bibitem[264]{Jordens2008}
R. Jordens, N. Strohmaier, K. Gunter, H. Moritz, and T. Esslinger, {\itshape A
  Mott insulator of fermionic atoms in an optical lattice}, Nature 455 (2008),
  pp. 204--207.

\bibitem[265]{Esslinger2010}
T. Esslinger, {\itshape Fermi-Hubbard Physics with Atoms in an Optical
  Lattice}, Annual Review of Condensed Matter Physics 1 (2010), pp. 129--152.

\bibitem[266]{Gerbier2010}
F. Gerbier and Y. Castin, {\itshape Heating rates for an atom in a far-detuned
  optical lattice}, Phys. Rev. A 82 (2010), p. 013615.

\bibitem[267]{Dalibard1985}
J. Dalibard and C. Cohen-Tannoudji, {\itshape Atomic motion in laser light:
  connection between semiclassical and quantum descriptions}, Journal of
  Physics B: Atomic and Molecular Physics 18 (1985), p. 1661.

\bibitem[268]{Gordon1980}
J.P. Gordon and A. Ashkin, {\itshape Motion of atoms in a radiation trap},
  Phys. Rev. A 21 (1980), pp. 1606--1617.

\bibitem[269]{Ellinger1994}
K. Ellinger, J. Cooper, and P. Zoller, {\itshape Light-pressure force in
  \textit{N} -atom systems}, Phys. Rev. A 49 (1994), pp. 3909--3933.

\bibitem[270]{Jordens2010}
R. J\"{o}rdens et~al., {\itshape Quantitative Determination of Temperature in
  the Approach to Magnetic Order of Ultracold Fermions in an Optical Lattice},
  Phys. Rev. Lett. 104 (2010), p. 180401.

\bibitem[271]{Fuchs2011}
S. Fuchs, E. Gull, L. Pollet, E. Burovski, E. Kozik, T. Pruschke, and M.
  Troyer, {\itshape Thermodynamics of the 3D Hubbard Model on Approaching the
  N{\'e}el Transition}, Phys. Rev. Lett. 106 (2011), p. 030401.

\bibitem[272]{Greif2013}
D. Greif, T. Uehlinger, G. Jotzu, L. Tarruell, and T. Esslinger, {\itshape
  Short-range quantum magnetism of ultracold fermions in an optical lattice},
  Science 340 (2013), pp. 1307--1310.

\bibitem[273]{Cirac1996}
J.I. Cirac, C.W. Gardiner, M. Naraschewski, and P. Zoller, {\itshape Continuous
  observation of interference fringes from Bose condensates}, Phys. Rev. A 54
  (1996), pp. R3714--R3717.

\bibitem[274]{Wong1996}
T. Wong, M.J. Collett, and D.F. Walls, {\itshape Interference of two
  Bose-Einstein condensates with collisions}, Phys. Rev. A 54 (1996), pp.
  R3718--R3721.

\bibitem[275]{Jack1996}
M.W. Jack, M.J. Collett, and D.F. Walls, {\itshape Coherent quantum tunneling
  between two Bose-Einstein condensates}, Phys. Rev. A 54 (1996), pp.
  R4625--R4628.

\bibitem[276]{Yoo1997}
S.M. Yoo, J. Ruostekoski, and J. Javanainen, {\itshape Interference of two
  Bose--Einstein condensates}, Journal of Modern Optics 44 (1997), pp.
  1763--1774.

\bibitem[277]{Castin1997}
Y. Castin and J. Dalibard, {\itshape Relative phase of two Bose-Einstein
  condensates}, Phys. Rev. A 55 (1997), pp. 4330--4337.

\bibitem[278]{Ruostekoski1997}
J. Ruostekoski and D.F. Walls, {\itshape Nondestructive optical measurement of
  relative phase between two Bose-Einstein condensates}, Phys. Rev. A 56
  (1997), pp. 2996--3006.

\bibitem[279]{Ruostekoski1998}
J. Ruostekoski, M.J. Collett, R. Graham, and D.F. Walls, {\itshape Macroscopic
  superpositions of Bose-Einstein condensates}, Phys. Rev. A 57 (1998), pp.
  511--517.

\bibitem[280]{Dunningham1999}
J.A. Dunningham and K. Burnett, {\itshape Phase Standard for Bose-Einstein
  Condensates}, Phys. Rev. Lett. 82 (1999), pp. 3729--3733.

\bibitem[281]{Dalvit2002}
D.A.R. Dalvit, J. Dziarmaga, and R. Onofrio, {\itshape Continuous quantum
  measurement of a Bose-Einstein condensate: A stochastic Gross-Pitaevskii
  equation}, Phys. Rev. A 65 (2002), p. 053604.

\bibitem[282]{Ruostekoski1999}
J. Ruostekoski and D.F. Walls, {\itshape Measurement scheme for relative phase
  diffusion between two Bose-Einstein condensates}, Phys. Rev. A 59 (1999), pp.
  R2571--R2574.

\bibitem[283]{Saba2005}
M. Saba, T.A. Pasquini, C. Sanner, Y. Shin, W. Ketterle, and D.E. Pritchard,
  {\itshape Light Scattering to Determine the Relative Phase of Two
  Bose-Einstein Condensates}, Science 307 (2005), pp. 1945--1948.

\bibitem[284]{Kazantsev1989}
A. Kazantsev, G. Surdutovich, and V. I͡Akovlev {\itshape Mechanical Action of
  Light on Atoms},    World Scientific, 1989.

\bibitem[285]{Cohen-Tannoudji1989}
C. Cohen-Tannoudji, J. Dupont-Roc, and G. Grynberg {\itshape Photons and Atoms:
  Introduction to Quantum Electrodynamics},  A Wiley-Interscience publication
  Wiley, 1989.

\bibitem[286]{Patil2014}
Y.S. Patil, L.M. Aycock, S. Chakram, and M. Vengalattore, {\itshape
  Nondestructive imaging of an ultracold lattice gas}, arXiv:1404.5583  (2014).

\bibitem[287]{Bernier2013}
J.S. Bernier, P. Barmettler, D. Poletti, and C. Kollath, {\itshape Emergence of
  spatially extended pair coherence through incoherent local environmental
  coupling}, Phys. Rev. A 87 (2013), p. 063608.

\bibitem[288]{Sarkar2014}
S. Sarkar, S. Langer, J. Schachenmayer, and A.J. Daley, {\itshape Light
  scattering and dissipative dynamics of many fermonic atoms in a lattice}, in
  preparation  (2014).

\bibitem[289]{Trotzky2010}
S. Trotzky, L. Pollet, F. Gerbier, U. Schnorrberger, I. Bloch, N.V. Prokof'ev,
  B. Svistunov, and M. Troyer, {\itshape Suppression of the critical
  temperature for superfluidity near the Mott transition}, Nature Phys. 6
  (2010), pp. 998--1004.

\bibitem[290]{Buchhold2014}
M. Buchhold and S. Diehl, {\itshape Non-Equilibrium Universality in the Heating
  Dynamics of Interacting Luttinger Liquids}, arXiv:1404.3740  (2014).

\bibitem[291]{Busch1998}
T. Busch, J.R. Anglin, J.I. Cirac, and P. Zoller, {\itshape Inhibition of
  spontaneous emission in Fermi gases}, EPL (Europhysics Letters) 44 (1998),
  p.~1.

\bibitem[292]{Misra1977}
B. Misra and E.C.G. Sudarshan, {\itshape The Zeno's paradox in quantum theory},
  Journal of Mathematical Physics 18 (1977), pp. 756--763.

\bibitem[293]{Fischer2001}
M.C. Fischer, B. Guti\'errez-Medina, and M.G. Raizen, {\itshape Observation of
  the Quantum Zeno and Anti-Zeno Effects in an Unstable System}, Phys. Rev.
  Lett. 87 (2001), p. 040402.

\bibitem[294]{Gagen1993}
M.J. Gagen, H.M. Wiseman, and G.J. Milburn, {\itshape Continuous position
  measurements and the quantum Zeno effect}, Phys. Rev. A 48 (1993), pp.
  132--142.

\bibitem[295]{Durr2009}
S. D\"urr, J.J. Garc\'\i{}a-Ripoll, N. Syassen, D.M. Bauer, M. Lettner, J.I.
  Cirac, and G. Rempe, {\itshape Lieb-Liniger model of a dissipation-induced
  Tonks-Girardeau gas}, Phys. Rev. A 79 (2009), p. 023614.

\bibitem[296]{Paredes2007}
B. Paredes, T. Keilmann, and J.I. Cirac, {\itshape Pfaffian-like ground state
  for three-body hard-core bosons in one-dimensional lattices}, Phys. Rev. A 75
  (2007), p. 053611.

\bibitem[297]{Buchler2007b}
H.P. Buchler, A. Micheli, and P. Zoller, {\itshape Three-body interactions with
  cold polar molecules}, Nat Phys 3 (2007), pp. 726--731.

\bibitem[298]{Ng2011}
K.K. Ng and M.F. Yang, {\itshape Thermal phase transitions in the attractive
  extended Bose-Hubbard model with three-body constraint}, Phys. Rev. B 83
  (2011), p. 100511.

\bibitem[299]{Mazza2010}
L. Mazza, M. Rizzi, M. Lewenstein, and J.I. Cirac, {\itshape Emerging bosons
  with three-body interactions from spin-1 atoms in optical lattices}, Phys.
  Rev. A 82 (2010), p. 043629.

\bibitem[300]{Silva-Valencia2011}
J. Silva-Valencia and A.M.C. Souza, {\itshape First Mott lobe of bosons with
  local two- and three-body interactions}, Phys. Rev. A 84 (2011), p. 065601.

\bibitem[301]{Safavi-Naini2012}
A. Safavi-Naini, J. Stechervon , B. Capogrosso-Sansone, and S.T. Rittenhouse,
  {\itshape First-Order Phase Transitions in Optical Lattices with Tunable
  Three-Body Onsite Interaction}, Phys. Rev. Lett. 109 (2012), p. 135302.

\bibitem[302]{singh2012}
M. Singh, A. Dhar, T. Mishra, R.V. Pai, and B.P. Das, {\itshape Three-body
  on-site interactions in ultracold bosonic atoms in optical lattices and
  superlattices}, Phys. Rev. A 85 (2012), p. 051604.

\bibitem[303]{Sowinski2012}
T. Sowi\'{n}ski, {\itshape Exact diagonalization of the one-dimensional
  Bose-Hubbard model with local three-body interactions}, Phys. Rev. A 85
  (2012), p. 065601.

\bibitem[304]{Daley2013}
A.J. Daley and J. Simon, {\itshape Effective three-body interactions via
  photon-assisted tunneling in an optical lattice}, arXiv:1311.1783  (2013).

\bibitem[305]{Vega2008}
I. Vegade~, D. Porras, and J. Ignacio~Cirac, {\itshape Matter-Wave Emission in
  Optical Lattices: Single Particle and Collective Effects}, Phys. Rev. Lett.
  101 (2008), p. 260404.

\bibitem[306]{Navarrete-Benlloch2011}
C. Navarrete-Benlloch, I. Vegade~, D. Porras, and J.I. Cirac, {\itshape
  Simulating quantum-optical phenomena with cold atoms in optical lattices},
  New Journal of Physics 13 (2011), p. 023024.

\bibitem[307]{Gericke2008}
T. Gericke, P. Wurtz, D. Reitz, T. Langen, and H. Ott, {\itshape
  High-resolution scanning electron microscopy of an ultracold quantum gas},
  Nat Phys 4 (2008), pp. 949--953.

\bibitem[308]{Weitenberg2011}
C. Weitenberg, M. Endres, J.F. Sherson, M. Cheneau, P. Schausz, T. Fukuhara, I.
  Bloch, and S. Kuhr, {\itshape Single-spin addressing in an atomic Mott
  insulator}, Nature 471 (2011), pp. 319--324.

\bibitem[309]{Bakr2009}
W.S. Bakr, J.I. Gillen, A. Peng, S. Folling, and M. Greiner, {\itshape A
  quantum gas microscope for detecting single atoms in a Hubbard-regime optical
  lattice}, Nature 462 (2009), pp. 74--77.

\bibitem[310]{Sherson2010}
J.F. Sherson, C. Weitenberg, M. Endres, M. Cheneau, I. Bloch, and S. Kuhr,
  {\itshape Single-atom-resolved fluorescence imaging of an atomic Mott
  insulator}, Nature 467 (2010), pp. 68--72.

\bibitem[311]{Witthaut2011}
D. Witthaut, F. Trimborn, H. Hennig, G. Kordas, T. Geisel, and S. Wimberger,
  {\itshape Beyond mean-field dynamics in open Bose-Hubbard chains}, Phys. Rev.
  A 83 (2011), p. 063608.

\bibitem[312]{Kordas2013}
G. Kordas, S. Wimberger, and D. Witthaut, {\itshape Decay and fragmentation in
  an open Bose-Hubbard chain}, Phys. Rev. A 87 (2013), p. 043618.

\bibitem[313]{Vidanovic2014}
I. Vidanovic, D. Cocks, and W. Hofstetter, {\itshape Dissipation through
  localised loss in bosonic systems with long-range interactions},
  arXiv:1402.0011  (2014).

\bibitem[314]{Witthaut2008}
D. Witthaut, F. Trimborn, and S. Wimberger, {\itshape Dissipation Induced
  Coherence of a Two-Mode Bose-Einstein Condensate}, Phys. Rev. Lett. 101
  (2008), p. 200402.

\bibitem[315]{Pawlowski2013}
K. Pawlowski, D. Spehner, A. Minguzzi, and G. Ferrini, {\itshape Macroscopic
  superpositions in Bose-Josephson junctions: Controlling decoherence due to
  atom losses}, arXiv:1302.2069  (2013).

\bibitem[316]{Spehner2014}
D. Spehner, K. Pawlowski, G. Ferrini, and A. Minguzzi, {\itshape Effect of
  one-, two-, and three-body atom loss processes on superpositions of phase
  states in Bose-Josephson junctions}, arXiv:1401.7238  (2014).

\bibitem[317]{Pitaevskii2003}
L. Pitaevskii and S. Stringari {\itshape Bose-Einstein Condensation},
  International Series of Monographs on Physics  Clarendon Press, 2003.

\bibitem[318]{Scelle2013}
R. Scelle, T. Rentrop, A. Trautmann, T. Schuster, and M.K. Oberthaler,
  {\itshape Motional Coherence of Fermions Immersed in a Bose Gas}, Phys. Rev.
  Lett. 111 (2013), p. 070401.

\bibitem[319]{Kasevich1992}
M. Kasevich and S. Chu, {\itshape Laser cooling below a photon recoil with
  three-level atoms}, Phys. Rev. Lett. 69 (1992), pp. 1741--1744.

\bibitem[320]{Aspect1988}
A. Aspect, E. Arimondo, R. Kaiser, N. Vansteenkiste, and C. Cohen-Tannoudji,
  {\itshape Laser Cooling below the One-Photon Recoil Energy by
  Velocity-Selective Coherent Population Trapping}, Phys. Rev. Lett. 61 (1988),
  pp. 826--829.

\bibitem[321]{Yang1989}
C.N. Yang, {\itshape $\eta$ pairing and off-diagonal long-range order in a
  Hubbard model}, Phys. Rev. Lett. 63 (1989), pp. 2144--2147.

\bibitem[322]{Porras2004}
D. Porras and J.I. Cirac, {\itshape Effective Quantum Spin Systems with Trapped
  Ions}, Phys. Rev. Lett. 92 (2004), p. 207901.

\bibitem[323]{Friedenauer2008}
A. Friedenauer, H. Schmitz, J.T. Glueckert, D. Porras, and T. Schaetz,
  {\itshape Simulating a quantum magnet with trapped ions}, Nat Phys 4 (2008),
  pp. 757--761.

\bibitem[324]{Kim2009}
K. Kim, M.S. Chang, R. Islam, S. Korenblit, L.M. Duan, and C. Monroe, {\itshape
  Entanglement and Tunable Spin-Spin Couplings between Trapped Ions Using
  Multiple Transverse Modes}, Phys. Rev. Lett. 103 (2009), p. 120502.

\bibitem[325]{Islam2013}
R. Islam, C. Senko, W.C. Campbell, S. Korenblit, J. Smith, A. Lee, E.E.
  Edwards, C.C.J. Wang, J.K. Freericks, and C. Monroe, {\itshape Emergence and
  Frustration of Magnetism with Variable-Range Interactions in a Quantum
  Simulator}, Science 340 (2013), pp. 583--587.

\bibitem[326]{Britton2012}
J.W. Britton, B.C. Sawyer, A.C. Keith, C.C.J. Wang, J.K. Freericks, H. Uys,
  M.J. Biercuk, and J.J. Bollinger, {\itshape Engineered two-dimensional Ising
  interactions in a trapped-ion quantum simulator with hundreds of spins},
  Nature 484 (2012), pp. 489--492.

\bibitem[327]{Mekhov2012}
I.B. Mekhov and H. Ritsch, {\itshape Quantum optics with ultracold quantum
  gases: towards the full quantum regime of the light--matter interaction},
  Journal of Physics B: Atomic, Molecular and Optical Physics 45 (2012), p.
  102001.

\bibitem[328]{Morrison2008}
S. Morrison and A.S. Parkins, {\itshape Dynamical Quantum Phase Transitions in
  the Dissipative Lipkin-Meshkov-Glick Model with Proposed Realization in
  Optical Cavity QED}, Phys. Rev. Lett. 100 (2008), p. 040403.

\bibitem[329]{Mekhov2007b}
I.B. Mekhov, C. Maschler, and H. Ritsch, {\itshape Probing quantum phases of
  ultracold atoms in optical lattices by transmission spectra in cavity quantum
  electrodynamics}, Nat Phys 3 (2007), pp. 319--323.

\bibitem[330]{Mekhov2007}
---{}---{}---, {\itshape Cavity-Enhanced Light Scattering in Optical Lattices
  to Probe Atomic Quantum Statistics}, Phys. Rev. Lett. 98 (2007), p. 100402.

\bibitem[331]{Mekhov2009}
I.B. Mekhov and H. Ritsch, {\itshape Quantum Nondemolition Measurements and
  State Preparation in Quantum Gases by Light Detection}, Phys. Rev. Lett. 102
  (2009), p. 020403.

\bibitem[332]{Maschler2008}
C. Maschler, I.B. Mekhov, and H. Ritsch, {\itshape Ultracold atoms in optical
  lattices generated by quantized light fields},  46 (2008), pp. 545--560.

\bibitem[333]{Niedenzu2010}
W. Niedenzu, R. Schulze, A. Vukics, and H. Ritsch, {\itshape Microscopic
  dynamics of ultracold particles in a ring-cavity optical lattice}, Phys. Rev.
  A 82 (2010), p. 043605.

\bibitem[334]{Marcos2012}
D. Marcos, A. Tomadin, S. Diehl, and P. Rabl, {\itshape Photon condensation in
  circuit quantum electrodynamics by engineered dissipation}, New Journal of
  Physics 14 (2012), p. 055005.

\bibitem[335]{Prosen2014}
T. Prosen, {\itshape Exact Nonequilibrium Steady State of an Open Hubbard
  Chain}, Phys. Rev. Lett. 112 (2014), p. 030603.

\bibitem[336]{Karevski2013}
D. Karevski, V. Popkov, and G.M. Sch\"utz, {\itshape Exact Matrix Product
  Solution for the Boundary-Driven Lindblad $XXZ$ Chain}, Phys. Rev. Lett. 110
  (2013), p. 047201.

\bibitem[337]{Prosen2011}
T. Prosen, {\itshape Open $XXZ$ Spin Chain: Nonequilibrium Steady State and a
  Strict Bound on Ballistic Transport}, Phys. Rev. Lett. 106 (2011), p. 217206.

\bibitem[338]{Prosen2010}
T. Prosen and B. \v{Z}unkovi\v{c}, {\itshape Exact solution of Markovian master
  equations for quadratic Fermi systems: thermal baths, open XY spin chains and
  non-equilibrium phase transition}, New Journal of Physics 12 (2010), p.
  025016.

\bibitem[339]{Znidaric2010}
M. \v{Z}nidari\v{c}, T. Prosen, G. Benenti, G. Casati, and D. Rossini,
  {\itshape Thermalization and ergodicity in one-dimensional many-body open
  quantum systems}, Phys. Rev. E 81 (2010), p. 051135.

\bibitem[340]{Prosen2008}
T. Prosen, {\itshape Third quantization: a general method to solve master
  equations for quadratic open Fermi systems}, New Journal of Physics 10
  (2008), p. 043026.

\bibitem[341]{Cai2013}
Z. Cai and T. Barthel, {\itshape Algebraic versus Exponential Decoherence in
  Dissipative Many-Particle Systems}, Phys. Rev. Lett. 111 (2013), p. 150403.

\bibitem[342]{Benenti2009}
G. Benenti, G. Casati, T. Prosen, D. Rossini, and M. \v{Z}nidari\v{c},
  {\itshape Charge and spin transport in strongly correlated one-dimensional
  quantum systems driven far from equilibrium}, Phys. Rev. B 80 (2009), p.
  035110.

\bibitem[343]{Prosen2009}
T. Prosen and M. {\v Z}nidari{\v c}, {\itshape Matrix product simulations of
  non-equilibrium steady states of quantum spin chains}, Journal of Statistical
  Mechanics: Theory and Experiment 2009 (2009), p. P02035.

\bibitem[344]{Minami2003}
T. Minami, C.O. Reinhold, and J. Burgd\"orfer, {\itshape Quantum-trajectory
  Monte Carlo method for internal-state evolution of fast ions traversing
  amorphous solids}, Phys. Rev. A 67 (2003), p. 022902.

\bibitem[345]{Seliger2007}
M. Seliger, C.O. Reinhold, T. Minami, D.R. Schultz, M.S. Pindzola, S. Yoshida,
  J. Burgd\"orfer, E. Lamour, J.P. Rozet, and D. Vernhet, {\itshape Electron
  capture and electron transport by fast ions penetrating solids: An open
  quantum system approach with sources and sinks}, Phys. Rev. A 75 (2007), p.
  032714.

\bibitem[346]{Seliger2005}
M. Seliger, C.O. Reinhold, T. Minami, and J. Burgd\"orfer, {\itshape Nonunitary
  quantum trajectory Monte Carlo method for open quantum systems}, Phys. Rev. A
  71 (2005), p. 062901.

\bibitem[347]{Breuer2012}
H.P. Breuer, {\itshape Foundations and measures of quantum non-Markovianity},
  Journal of Physics B: Atomic, Molecular and Optical Physics 45 (2012), p.
  154001.

\bibitem[348]{Breuer2004b}
---{}---{}---, {\itshape The non-Markovian quantum behavior of open systems},
  29 (2004), pp. 105--118.

\bibitem[349]{Breuer2007b}
---{}---{}---, {\itshape Non-Markovian generalization of the Lindblad theory of
  open quantum systems}, Phys. Rev. A 75 (2007), p. 022103.

\bibitem[350]{Breuer1999}
H.P. Breuer, B. Kappler, and F. Petruccione, {\itshape Stochastic wave-function
  method for non-Markovian quantum master equations}, Phys. Rev. A 59 (1999),
  pp. 1633--1643.

\bibitem[351]{Breuer2004}
H.P. Breuer, {\itshape Exact quantum jump approach to open systems in bosonic
  and spin baths}, Phys. Rev. A 69 (2004), p. 022115.

\bibitem[352]{Breuer2004c}
---{}---{}---, {\itshape Genuine quantum trajectories for non-Markovian
  processes}, Phys. Rev. A 70 (2004), p. 012106.

\bibitem[353]{Imamoglu1994}
A. Imamoglu, {\itshape Stochastic wave-function approach to non-Markovian
  systems}, Phys. Rev. A 50 (1994), pp. 3650--3653.

\bibitem[354]{Piilo2008}
J. Piilo, S. Maniscalco, K. H\"ark\"onen, and K.A. Suominen, {\itshape
  Non-Markovian Quantum Jumps}, Phys. Rev. Lett. 100 (2008), p. 180402.

\bibitem[355]{Piilo2009}
J. Piilo, K. H\"ark\"onen, S. Maniscalco, and K.A. Suominen, {\itshape Open
  system dynamics with non-Markovian quantum jumps}, Phys. Rev. A 79 (2009), p.
  062112.

\bibitem[356]{Haikka2013}
P. Haikka, S. McEndoo, and S. Maniscalco, {\itshape Non-Markovian probes in
  ultracold gases}, Phys. Rev. A 87 (2013), p. 012127.

\bibitem[357]{McEndoo2013}
S. McEndoo, P. Haikka, G.D. Chiara, G.M. Palma, and S. Maniscalco, {\itshape
  Entanglement control via reservoir engineering in ultracold atomic gases},
  EPL (Europhysics Letters) 101 (2013), p. 60005.

\bibitem[358]{Mollow1975}
B.R. Mollow, {\itshape Pure-state analysis of resonant light scattering:
  Radiative damping, saturation, and multiphoton effects}, Phys. Rev. A 12
  (1975), pp. 1919--1943.

\bibitem[359]{Barchielli1986}
A. Barchielli, {\itshape Measurement theory and stochastic differential
  equations in quantum mechanics}, Phys. Rev. A 34 (1986), pp. 1642--1649.

\bibitem[360]{Blatt1986}
R. Blatt, W. Ertmer, P. Zoller, and J.L. Hall, {\itshape Atomic-beam cooling: A
  simulation approach}, Phys. Rev. A 34 (1986), pp. 3022--3033.

\bibitem[361]{Zoller1987}
P. Zoller, M. Marte, and D.F. Walls, {\itshape Quantum jumps in atomic
  systems}, Phys. Rev. A 35 (1987), pp. 198--207.

\bibitem[362]{Endres2011}
M. Endres et~al., {\itshape Observation of Correlated Particle-Hole Pairs and
  String Order in Low-Dimensional Mott Insulators}, Science 334 (2011), pp.
  200--203.

\end{thebibliography}

\end{document}